\documentclass[conference]{IEEEtran}
\usepackage{graphicx}
\usepackage{balance}
\usepackage[hyphens]{url}
\usepackage[breaklinks]{hyperref}
\usepackage{color}
\usepackage{multirow,cite,xspace}
\usepackage{algorithm,algpseudocode}
\usepackage{amssymb}
\usepackage{amsmath}
\usepackage{amsfonts}

\newcommand{\oursystem}{{Omnivore}\xspace}
\newcommand{\mxnet}{{MXNet}\xspace}
\newcommand{\singa}{{SINGA}\xspace} 
\newcommand{\caffe}{{Caffe}\xspace}
\newcommand{\tensorflow}{{TensorFlow}\xspace}

\newcommand{\SE}{\textsf{SE}\xspace}
\newcommand{\HE}{\textsf{HE}\xspace}
\def\compactify{\itemsep=0pt \topsep=0pt \partopsep=0pt \parsep=0pt}

\newif\iflongversion

\longversiontrue

\iflongversion
	\newcommand{\extra}[1]{{\color{red} Appendix~\ref{#1}}}
\else
	\newcommand{\extra}[1]{{\color{red} the long version of this paper \cite{hadjis2016omnivore}}}
\fi

\ifCLASSINFOpdf
\else
\fi
\begin{document}

\newcommand{\R}{\mathbb{R}}
\newcommand{\cmr}[1]{{\color{blue}{#1}}}
\newcommand{\im}[1]{{\color{purple}{#1}}}
\newcommand{\imc}[1]{}
\newcommand{\ce}[1]{{\color{orange}{#1}}}
\newcommand{\yell}[1]{{\color{black}{#1}}}
\newcommand{\icde}[1]{{\color{black}{#1}}}
\newcommand{\red}[1]{{\color{red}{#1}}}
\newcommand{\magenta}[1]{{\color{magenta}{#1}}}

\newtheorem{theorem}{Theorem}
\newtheorem{definition}[theorem]{Definition}
\newtheorem{assumption}[theorem]{Assumption}
\newtheorem{proposition}[theorem]{Proposition}
\newtheorem{lemma}[theorem]{Lemma}
\newtheorem{corollary}[theorem]{Corollary}

%
\title{Omnivore: An Optimizer for Multi-device \\ Deep Learning on CPUs and GPUs}

\author{
\IEEEauthorblockN{Stefan Hadjis}
\IEEEauthorblockA{Dept. of Computer Science\\
Stanford University\\
Email: shadjis@stanford.edu}
\and
\IEEEauthorblockN{Ce Zhang}
\IEEEauthorblockA{Dept. of Computer Science\\
ETH Zurich\\
Email: ce.zhang@inf.ethz.ch} 
\and
\IEEEauthorblockN{Ioannis Mitliagkas, Dan Iter, Christopher R\'e }
\IEEEauthorblockA{Dept. of Computer Science\\
Stanford University\\
Email: \{imit, daniter, chrismre\}@stanford.edu}
}


%


\maketitle

\begin{abstract}
\boldmath
We study the factors affecting training time in multi-device deep learning systems. Given a specification of a convolutional neural network, our goal is to minimize the time to train this model on a cluster of commodity CPUs and GPUs.
We first focus on the single-node setting and show that by using standard batching and data-parallel techniques, throughput can be improved by at least $5.5\times$ over state-of-the-art systems on CPUs. This ensures an end-to-end training speed directly proportional to the throughput of a device regardless of its underlying hardware, allowing each node in the cluster to be treated as a black box.
Our second contribution is a theoretical and empirical study of the tradeoffs affecting end-to-end training time in a multiple-device setting. We identify the degree of asynchronous parallelization as a key factor affecting both hardware and statistical efficiency. We see that asynchrony can be viewed as introducing a momentum term. Our results imply that tuning momentum is critical in asynchronous parallel configurations, and suggest that published results that have not been fully tuned might report suboptimal performance for some configurations. 
For our third contribution, we use our novel understanding of the interaction between system and optimization dynamics to provide an efficient hyperparameter optimizer. Our optimizer involves a predictive model for the total time to convergence and selects an allocation of resources to minimize that time. We demonstrate that the most popular distributed deep learning systems fall within our tradeoff space, but do not optimize within the space. By doing this optimization, our prototype runs $1.9\times$ to $12\times$ faster than the fastest state-of-the-art systems.
\end{abstract}


%
\IEEEpeerreviewmaketitle

\section{Introduction}\label{sec:Introduction}
\label{sec:intro}

In recent years, deep learning has provided significant improvements 
in quality for a number of data-driven applications~\cite{Deng:2014:Book,Krizhevsky:2012:NIPS,Dean:2012:NIPS,LeCun:2015:Arxiv}. 
An important aspect of deep learning 
is that quality improves with the amount of data we can
process; therefore, advances in system efficiency and 
scalability directly improve quality. This 
has led to an arms race of distributed deep learning 
systems both in industry (e.g., Google's DistBelief~\cite{Dean:2012:NIPS}, Microsoft's Adam~\cite{Chilimbi:2014:OSDI}) and in 
academia~\cite{chen15mxnet,Wang:2015:TR,moritz15sparknet,iandola15firecaf}.

Despite this proliferation of deep learning systems, there 
have been few studies of deep learning from a data-systems
perspective. Each of these systems makes a set of design
decisions that may not work for other tasks or hardware settings. 
In our experience working with multiple Ph.D.-level users of 
these systems---including experts in pathology, radiology, 
computer vision, and the energy sector---it is often very 
difficult even for advanced users to make these design 
decisions themselves. It is not uncommon for a
suboptimal design choice to result in
an end-to-end runtime that is an order of magnitude slower
than what is possible. 
Moreover, most systems
provide no way of automatically selecting an optimal
configuration, placing this burden on the user. 
\icde{
This also contributes to
two debates about deep learning systems.

\vspace{0.15cm}
\noindent
{\bf Debate 1: CPU vs. GPU.}
There has been a long debate about CPUs vs. GPUs for 
deep learning. GPUs are popular for CNN systems because 
of the high throughput they provide, but they contain 
smaller off-chip memories. Microsoft's 
Adam argues that CPUs can deliver more cost-effective 
performance~\cite{Chilimbi:2014:OSDI}. For users who
cannot control their data-center hardware, Amazon's EC2 
provides GPUs but Google Compute does not.

\vspace{0.15cm}
\noindent
{\bf Debate 2: Synchronous vs. Asynchronous Training.}
Another debate is about the synchronization strategies
to use in multi-device deep learning. 
For example, Google's latest TensorFlow 
paper~\cite{abadi2016tensorflow} comes out in support of fully 
synchronous methods, citing recent papers~\cite{cui2016geeps,chen2016revisiting}. Other systems,
such as DistBelief \cite{Dean:2012:NIPS}, Project Adam~\cite{Chilimbi:2014:OSDI}, H2O \cite{candel2015deep}, and recent theoretical efforts~\cite{LiuJi2015}
 focus on asynchronous training and argue that
it is more scalable than fully synchronous strategies.

\vspace{0.1cm}
In this paper, we perform a
first study of the design space for deep learning systems.
We identify the key tradeoffs for
single-device and multi-device systems,
providing insights into the above
debates.
We find that the CPU
implementation of many state-of-the-art systems
misses textbook batching optimizations, which
can make the CPU implementation at 
least 5.5$\times$ faster.
We also see that when the momentum parameter~\cite{sutskever2013importance} is tuned,
there is {\em no penalty for asynchrony}---except in a short, cold-start training period.
This previously unknown connection between asynchrony and momentum was surprising even to leading groups in the area.
Recent work does not tune momentum~\cite{cui2016geeps,chen2016revisiting} and reports that asynchronous configurations are slower.
We see on different systems, including TensorFlow, that tuning {\em changes this outcome}: asynchronous configurations are faster when we tune. 
These theoretical and experimental insights 
help us understand the tradeoff  space better and build an automatic
optimizer to choose the optimal configuration.
As a result, 
our prototype system, Omnivore,
can be $1.9\times$ to $12\times$ faster 
than the fastest competitor systems.
Our results have attracted interest from major companies.
In collaboration with a major chip manufacturer, we are 
integrating our optimizers on new platforms of much larger scale.
}

\subsection*{Overview of Technical Contributions}

To conduct our study, we develop a prototype distributed 
system called Omnivore.\footnote{
\scriptsize{\url{https://github.com/HazyResearch/Omnivore}}
}
We make three contributions.

\vspace{0.05cm}
\paragraph*{Scope of Our Study}

We focus on perhaps the most popular deep learning models,
convolutional neural networks (CNNs), which are state-of-the-art for a
wide range of applications (e.g., image processing, video analysis,
drug discovery).
Our study answers the following question: {\it
  ``Given a cluster (e.g., X machines, Y GPUs, Z CPUs, etc.), how
  do I train my CNN as quickly as possible?''}. 
We assume that the following are given: (i) a deep
learning model (network architecture), (ii) a dataset for training this
model, and (iii) a set of computational resources (a number of devices
on machines, their throughput, and the network speed).\footnote{
\scriptsize{
We only do basic network optimization, and we assume 
that machines are connected by a uniform and fast 
topology, e.g., if they were housed on the same rack.
}
}
We then study how to minimize the total training time.  We build a
complete prototype capable of training the most popular deep learning
models. This allows us to hone in on two major choices: (i) how to use
hardware on each node and (ii) the degree to which
asynchrony can be tolerated.
Our work demystifies these factors
by identifying the key
tradeoffs that underlie all design decisions, providing theoretical
guidance about asynchrony, and quantifying the impact of those
tradeoffs experimentally. 


\paragraph*{Contribution 1: Single-Device Optimizations}
We show that it is possible to 
achieve throughput proportional to the maximum FLOPS of {\em both} CPUs and GPUs.
This is not trivial; while state-of-the-art systems achieve GPU speeds
proportional to the device throughput, existing CPU implementations
can be sped up significantly compared to what is reported in the
literature.
Our study builds on two key optimizations we reported in 
a workshop~\cite{hadjis15cct}. 
We use batching and data-parallel optimizations---not employed in
other systems---to achieve
end-to-end speedups of more than $5.5\times$ over state-of-the-art
systems on commodity CPUs.
Such optimizations are not always possible
on the GPU, but by selecting this strategy for the CPU, we now achieve
speeds proportional to the peak throughput.
This allows us to build a simpler optimizer
by modeling CPUs and GPUs as black boxes.

\paragraph*{Contribution 2: Multi-device Tradeoffs}
Our second contribution is an empirical study
of factors affecting
training time
for multi-device deep learning training.  We analyze the decisions made on
existing systems and find that while they are diverse, the strategies of
the most popular systems fall within a tradeoff space defined by two
fundamental dimensions~\cite{Chilimbi:2014:OSDI, iandola15firecaf,
  chen15mxnet, Dean:2012:NIPS, Wang:2015:TR, moritz15sparknet}: (i)
the server architecture, or how the layers of a CNN map to devices;
and (ii) the execution strategy, or how batches of data are mapped to
machines for processing.
We develop a simple framework that allows us to model each of these
approaches.
Devices are organized in {\em compute groups}.
Each compute group is responsible for a single batch of data per iteration.
Inside a group, the computation occurs in standard,
synchronous steps. Across groups, computation happens {\it
  asynchronously}. 

We study the impact of asynchrony on the end-to-end performance
of training deep learning systems. Not surprisingly, the more asynchrony there is,
the faster the system is for {\em each iteration},
which we call {\em hardware efficiency}.
This happens because there is less coordination between workers. 
The challenge is to understand how asynchrony affects
the {\em number of iterations to converge}, which
we call {\em statistical efficiency}. 
We provide novel understanding of the factors that affect 
statistical efficiency.
Empirically, we observe that 
the optimal value for the momentum parameter decreases as we increase the number of asynchronous workers.
\icde{
Indeed, our theory in~\cite{mitliagkas2016asynchrony} shows that asynchronous-parallel training
can be viewed as a synchronous update but with an increased, {\it implicit momentum} term.
Furthermore, if the
optimal momentum value for a problem is above the implicit momentum, 
then there is {\em no} penalty for running
asynchronously, as long as the momentum is properly tuned.
Although momentum is explicitly (i.e., algorithmically) introduced in almost every
system \cite{sutskever2013importance},
 we are the first to realize this connection between
asynchrony and momentum.
Here, we validate this experimentally:
we see that, by properly tuning
momentum, asynchronous methods can be at least $1.5$--$2.5\times$ faster
than using the standard momentum value of $0.9$, which is used
in most existing work. This understanding of
statistical efficiency, along with an analytical
hardware efficiency model, forms a novel system tradeoff space.
}


\paragraph*{Contribution 3: Simple Automatic Optimizer}
Based on our theory and empirical study, the
intuition behind our optimizer is very simple: pick 
the highest degree of asynchrony such that the 
implicit momentum induced by asynchrony is below 
the optimal momentum. \icde{Given a fixed number of
compute groups (which control the degree of asynchrony), we grid-search the parameters for
learning rate and momentum by 
{\em measuring} the statistical and hardware efficiency for minutes
(less than 10\% of the time to train a network).}
If the best-found momentum is non-zero, the optimizer chooses this configuration. 
If it is zero, we assume that there could be a better setting with fewer
compute groups.
Our optimizer is able to
choose a near-optimal point in the tradeoff space, and we demonstrate that our system achieves end-to-end
speedups of $1.9\times$ to $12\times$ on popular CNN workloads 
compared to state-of-the-art tools that choose suboptimal 
tradeoff points. 
We compare our simple
optimizer with a state-of-the-art Bayesian optimization
approach~\cite{Snoek:NIPS:2012}.  Both approaches are able to reach
the same final accuracy (within $1\%$),
but the Bayesian strategy takes almost $6\times$ as long. 
We can also apply our optimizer to other deep learning systems. 
In some cases, this prevents those other tools from
diverging, while in other cases, it speeds them up by $7\times$.



\paragraph*{Outline}
We present background in Section~\ref{sec:background}.
Section~\ref{sec:Omnivore}
and Section~\ref{sec:Distributed}
introduce the tradeoff space 
related to single-machine and
multi-device settings, respectively. 
Section~\ref{sec:Optimizer} describes the 
optimizer for making decisions in this tradeoff space.
We validate our results in Section~\ref{sec:Experiments}, 
discuss related work in Section~\ref{sec:RelatedWork}, and conclude in Section~\ref{sec:conclusion}.
\section{Background} \label{sec:background}


\subsection{Convolutional Neural Networks (CNNs)}

A convolutional neural network (CNN,~\cite{Krizhevsky:2012:NIPS}) consists of
layers $L_{1}, L_{2}, \ldots,  L_{P}$. Each layer is an operator which takes as input a 3D data tensor $D \in \mathbb{R}^{n\times n \times d_{in}}$
and transforms it to a resulting
3D data tensor 
$R \in \mathbb{R}^{m  \times m \times d_{out}}$,
i.e.\ $L^{FW}_{p}(D_{p}) = R_{p}$.
$FW$ indicates the layer running in the ``forward'' direction to transform $D$ into $R$. Layers have a second operation, backward or $BW$, described later.
Often, $D_{1}$, the input to the first layer $L_{1}$, is an 
image $I \in \mathbb{R}^{n\times n \times 3}$, where 3 represents the RGB color channels.


For layers after $L_{1}$ (the input layer), the input tensor $D_{p}$ comes from the output of a prior layer (usually $D_{p} = R_{p-1}$), such that the CNN layers are cascaded to define a composite operation  (boldface highlights inputs and outputs)
\begin{equation}
\boldsymbol{R_{P}} = L^{FW}_{P} \circ L^{FW}_{P-1} \circ ... \circ L^{FW}_{2} \circ L^{FW}_{1}(\mathbf{I}) 
\label{equation:fw_pass}
\end{equation}
The final result $R_{P}$ is the CNN's prediction for image $I$. For example, if the task is image classification with 1000 categories, the tensor $R_{P} \in \mathbb{R}^{1\times 1 \times 1000}$ is a  vector containing the probability of each category. This prediction is then compared to $C$, the true classification for $I$, using a \textit{loss function} $\ell(R_{P}, C)$ that evaluates the quality of the prediction. A lower loss indicates a better prediction.

Many types of layers exist in a CNN.
Some layers perform a pre-defined transformation such as downsampling while other layers contain a \textit{model, W,} and perform an operation parameterized by the model. Models are also known as \textit{weights} or \textit{parameters}.
The models of all layers constitute the entire set of weights or parameters of the CNN, i.e.,
\[
W = W_{\text{CNN}} = \{W_{L_{1}}, \ldots, W_{L_{P}}\}.
\]

\subsection{Stochastic Gradient Descent}
The goal of CNN \textit{training} is to optimize the model $W$ in order to minimize the loss function $\ell(R_{P}, C)$, also denoted as $\ell(W, I, C)$ to make the fact that
$R_P$ is a function of $W$ and $I$ explicit.
 Low loss is correlated with high prediction accuracy and in this work
we refer to both.
The most popular training algorithm for CNNs
is an iterative technique called \textit{stochastic gradient descent} (SGD).
Each SGD iteration consists of a \textit{forward} and \textit{backward} pass.

The input to each SGD iteration is an image-label tuple ($I$, $C$) as described above. The forward pass calculates the prediction $R_{P}$ of $I$ using equation~\eqref{equation:fw_pass}, and then the prediction error compared to $C$ is used to calculate the \textit{gradient} (or derivative) of $\ell$ with respect to $R_{P}$. We denote this gradient as $\nabla_{R_{P}} \ell$. Now the cascade of equation~\eqref{equation:fw_pass} runs in reverse by applying each layer in the ``backward'' direction:
\begin{equation}
L^{BW}_{1} \circ L^{BW}_{2} \circ \ldots \circ L^{BW}_{P-1} \circ L^{BW}_{P}(\mathbf{\nabla_{R_{P}} \ell})
\label{equation:bw_pass}
\end{equation}
equation~\eqref{equation:bw_pass} implements the chain rule of calculus.
The $BW$ operation of layer $p$ takes as input a data gradient $\nabla_{R_p} \ell$ and outputs a data gradient $\nabla_{D_p}\ell$. Internally, it also updates that layer's model $W_{L_p}$ by (i) calculating a gradient of the loss with respect to the model, $\nabla_{W_{L_p}} \ell$, and (ii) using an SGD update on the model.
SGD repeats for many, often millions of iterations, until the loss is sufficiently low, i.e.\ the model is sufficiently optimized.
The initial model $W^{(0)}$ is randomly initialized.
The SGD update at step $t$ takes the form
\begin{equation}
\label{eqn:update1}
W^{(t)} \gets W^{(t-1)} + V^{(t)},
\end{equation}
where the new step, $V^{(t)}$, consists of a scaled version of the previous step, $V^{(t-1)}$, plus a gradient calculated with $(I,C)$:
\begin{equation}
\label{eqn:update2}
V^{(t)} \gets \mu V^{(t-1)} 
	- \eta  \left[
		 \nabla_{W} \ell \left(W^{(t-1)}, I, C\right)
		 + \lambda W^{(t-1)}
	\right]
\end{equation}
In Section~\ref{sec:Distributed} we introduce the notion of {\em asynchronous updates}. The main change in equation~\eqref{eqn:update2} under asynchrony, is the use of an older model, $W^{(s)}$, when evaluating the gradient $\nabla_W \ell$.
This gradient (specifically, its negative) is the direction to ``step'' within the parameter space each SGD iteration.
The {\em learning rate}, $\eta$, is the scale factor applied to the magnitude of this gradient, i.e. the size of the step. 
$\lambda$ dictates the amount of regularization, which is an input to the training problem (part of the CNN model to train). 
$\mu$ is the amount of {\em explicit momentum} we add to the update.
Momentum is used to ``accelerate'' learning in the directions common to each gradient
by keeping a history of past gradients and adding this history to the
gradient of the current step, with past gradients decayed
exponentially.
Commonly, in order to produce more stable gradients, each SGD iteration does not process a single tuple ($I$, $C$), but a \textit{batch} of $b$ tuples, e.g., 256, in which case $D$ and $R$ become 4D.
The gradients from each tuple are summed to produce a single, combined gradient for that iteration.

Selecting the right values for {\em hyperparameters} $(\eta,\mu,b)$ is critical for performance.
We will describe a simple optimizer to pick these parameters.


\subsection{CNN Computation}

\begin{figure}
\centering
\includegraphics[width=0.5\textwidth]{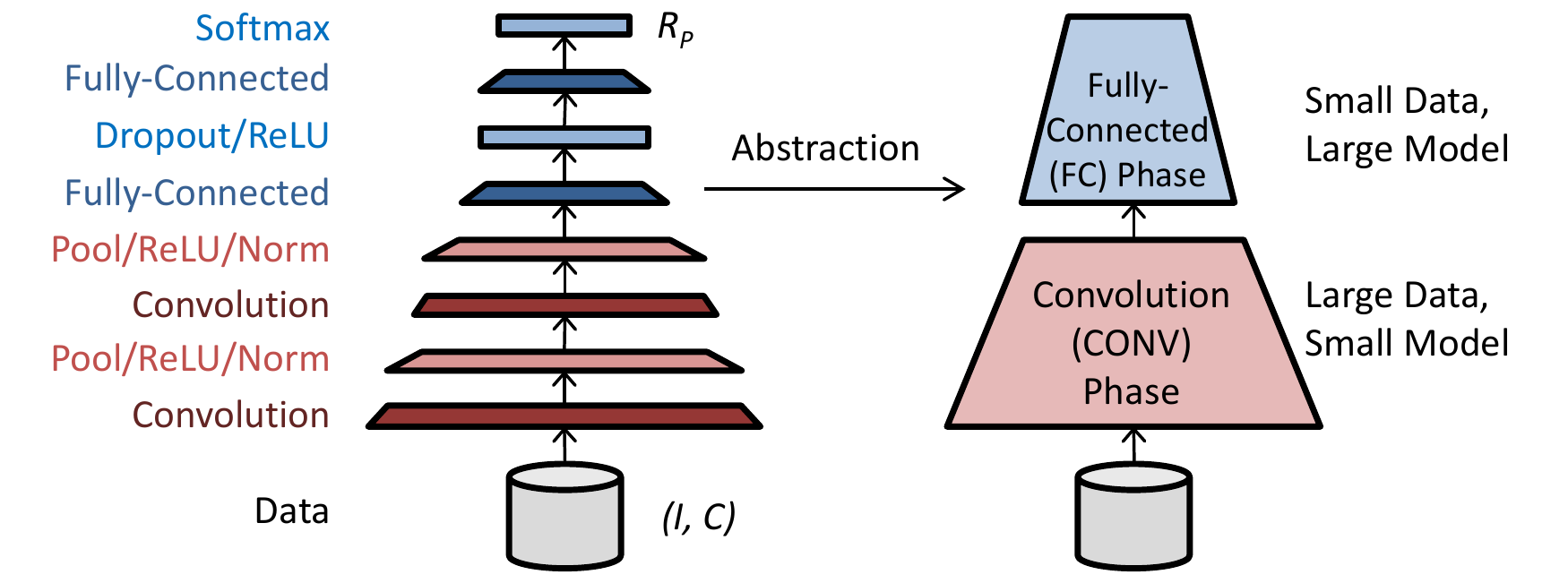}
\caption{Abstracting a CNN into two phases.}
\label{fig:CNN_Diagram}
\end{figure}


Of all the CNN layers, two
layers are the most computationally intensive,
\textit{convolutional} (conv) and \textit{fully-connected} (FC)
layers. A convolutional layer performs many
independent convolutions over an image, i.e., several sliding window
operations; an FC layer performs a dense matrix
multiplication. In many popular implementations, the bottleneck in both
layers is a matrix multiply implemented as a call to either a BLAS or
cuBLAS library. For our study, their data properties are more
important, and we refer to Chetlur et al.~\cite{Chetlur:2014:ArXvi}
for a more detailed description of
their use in machine learning.

Figure~\ref{fig:CNN_Diagram} illustrates a
state-of-the-art CNN, in which 
all convolutional layers always appear before all fully-connected layers.
In this work we introduce an abstraction which separates
a CNN into two phases, each consisting of a number of consecutive
layers: first the convolution phase (conv), whose layers have large
data tensors $D$ and $R$ (e.g., 100MB-1GB) and small models $W$
(e.g., 5-50 MB), followed by the fully-connected phase (FC), whose
layers have small $D$ and $R$ (e.g., 1-10MB) and large $W$ (e.g., 30-300
MB).  The reduction in data size is in part due to \textit{pooling}
layers in the convolution phase which perform down-sampling.
The increase in model size is due to the fact that convolution
layers repeat the same weights for the sliding window.
Note that our  two-phase categorization also applies to modern, non-linear CNNs~\cite{he2015resnet}. 

The computation for both the conv and FC phases is usually compute-bound although the conv phase contains significantly more computation (e.g., in AlexNet conv is 1.6 TFLOPs and FC is 80 GFLOPs, or 95\% of the computation is convolution). Within a machine, each layer's computation can be mapped to CPUs, GPUs, or a combination of both. In addition, this computation can be parallelized either using data parallelism (partitioning the data batch and replicating the model, which works well for the conv layers) or model parallelism (partitioning the model and replicating the data batch, which works well for the FC layers). 
In distributed settings, the layer computations are mapped across machines. We will show later this mapping is always done at the coarser granularity of entire phases (conv or FC) because it reduces network delays due to the distinct model and data sizes of the two phases. 
We study the choices of mapping and parallelization techniques
for both single and multiple devices 
in Section~\ref{sec:Omnivore} and~\ref{sec:Distributed}.


\subsection{Problem Definition}\label{sec:definition}

We study systems tradeoffs to build an optimizer for the most widely
used networks/algorithms. We focus on SGD due to its
popularity, although our optimizer applies to other algorithms as
well. More precisely, we are given as input the following: (i) a CNN architecture $\{L_1,...,L_P\}$, including regularization, 
(ii) a dataset $\mathcal{D}$ consisting of data batches, 
(iii) a device graph $\mathcal{G}$ in which vertices 
are hardware 
devices (specified by their throughput) and edges are
communication speeds between devices.  
Our goal is to design an
optimizer which 
creates a plan for 
\textit{physical mapping} and
\textit{execution strategy}
in order to train as quickly as possible.

A plan for physical mapping maps the computation
(both FW and BW) of each layer
to vertices (e.g., GPUs or CPU cores) of
the device graph. 
A plan for the execution strategy maps data batches 
to devices in order to parallelize SGD in
the multi-device case. The key decision 
here is the degree of asynchrony
in execution. 
Section~\ref{sec:Omnivore}
and Section~\ref{sec:Distributed} study how to do physical 
mapping within a device and across devices, respectively.
Section~\ref{sec:Distributed} also studies the impact of
asynchrony.
If the
cluster is homogeneous, we do not need
the explicit device graph -- instead, a few parameters,
such as the number of devices and the
throughput of each device, are enough
to specify the cluster for the optimizer.


\section{Single-Device Tradeoff} \label{sec:Omnivore}

We first study the systems tradeoffs within a single device.
We show
that for each device (GPU or CPU) we can achieve throughput that is
proportional to its peak FLOPS.  This enables the distributed
optimizer to treat each device as a black box in
Section~\ref{sec:Distributed}. 
This is not a trivial property for
deep learning: many existing systems~\cite{Jia:2014:arXiv,
  tensorflow, Bergstra:2010:Scipy} use either CPUs or GPUs,
but they often report that GPU implementations are an
order of magnitude faster than CPU
even when the devices offer
similar FLOPS. 
Therefore the challenge is to utilize the FLOPS on the
CPU. We study the key kernel in CNN implementations, which is compute
bound. We introduce a data batching technique which trades off memory
footprint for compute time and demonstrate that this tradeoff gives a
more than $5\times$ CPU speedup over existing systems. With this optimization
now both CPUs and GPUs give throughput proportional to the FLOPS
offered by the device.

\begin{figure}
\centering
\includegraphics[width=0.5\textwidth]{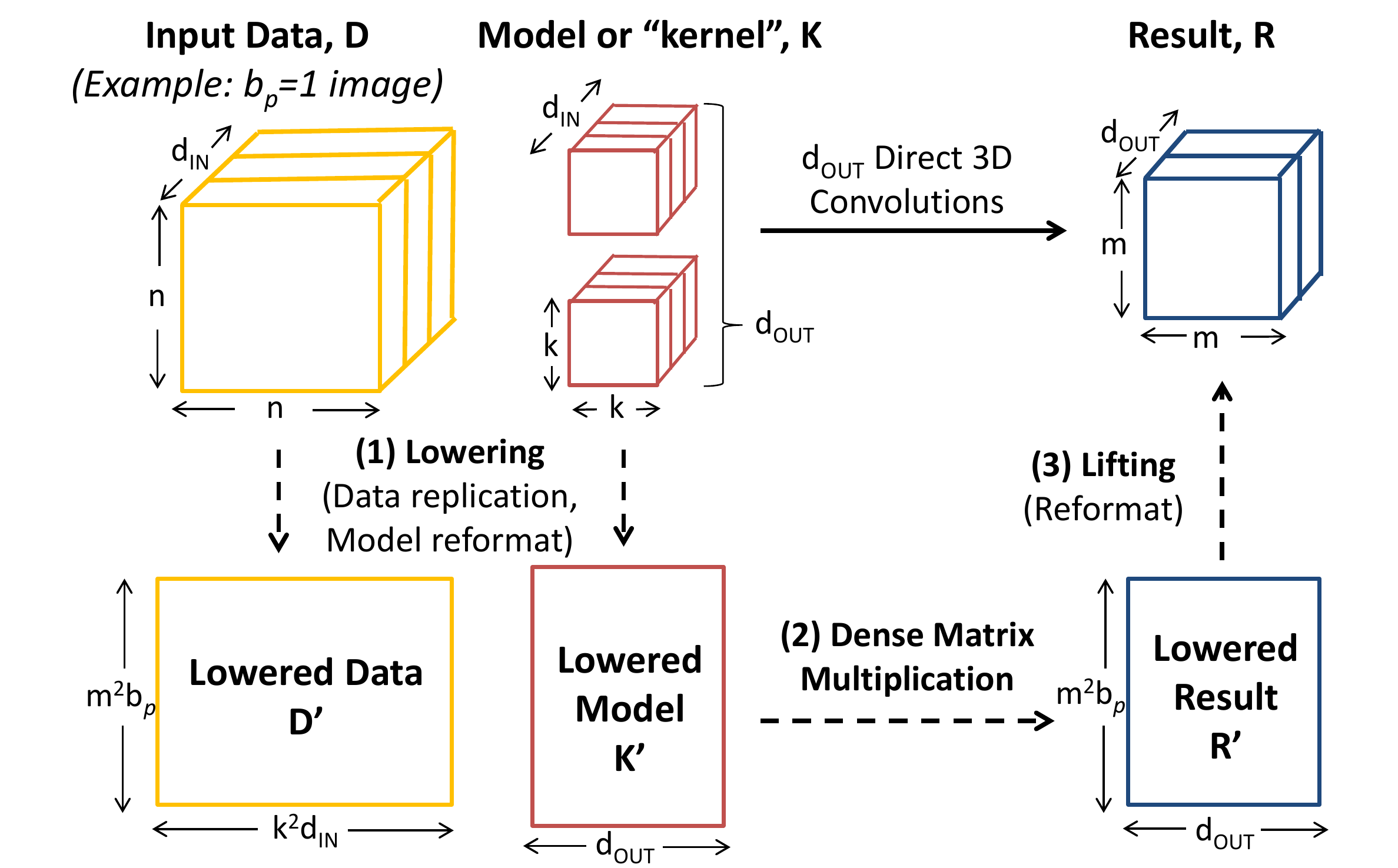} 
\vspace{-2.0em}
\caption{The computation performed by a convolutional layer which processes $b_{p}$ images at a time in parallel, where $1 \leq b_{p} \leq$ batch size. The convolutions can be implemented either directly (top) or using the lowering/GEMM method (bottom).}
\label{fig:conv_new_block}
\end{figure}

\subsection{Convolutional Layer Computation}

As reported in the literature and confirmed by our experiments,
the most computationally intensive layers in the CNN are the convolutional layers.
Together, all convolutional layers in a CNN often consume between 70-90\% of total execution time.
Recall that a convolutional layer contains a model, which we will also call its \textit{kernel} and denote as $K$.
A convolutional layer accepts as input a 4D data tensor $D \in \mathbb{R}^{n\times n \times d_{in} \times b}$, where recall $b$ is the batch size.
$K$ is also a 4D tensor, $K \in \mathbb{R}^{k \times k \times d_{in} \times d_{out}}$.
The output is a 4D data tensor $R \in \mathbb{R}^{m  \times m \times d_{out} \times b}$, where:
\begin{equation}
R_{x,y,z,w} = \sum_{d'=0}^{d_{in}-1} \sum_{x'=0}^{k-1} \sum_{y'=0}^{k-1}
D_{x-\frac{k}{2}+x',y-\frac{k}{2}+y',d',w} K_{x',y',d',z}
\label{eq:conv_full}
\end{equation}
Like most HPC kernels a straightforward implementation is suboptimal, and many optimized implementations of this convolution kernel
exist \cite{Vasilache:2014:arXiv, Jia:2014:arXiv, Chetlur:2014:ArXvi}. A popular implementation is to perform equation~\eqref{eq:conv_full} as a dense matrix multiplication (also known as GEMM, general matrix multiply), which \cite{Chetlur:2014:ArXvi} demonstrates to be versatile and fast for a range of size parameters, as GEMM kernels are highly optimized.

In order for equation~\eqref{eq:conv_full} to be carried out as a matrix multiplication an initial reformatting and replication step called \textit{lowering} is required to put the data and model into the correct format.
Figure~\ref{fig:conv_new_block} shows the three logical steps in the lowering process:
(i) lowering, which transforms 3D tensors $D$ and $K$ into 2D matrices $\hat{D}$ and $\hat{K}$;
(ii) {\em matrix multiply}, which multiplies $\hat{D}\hat{K}$ to get the result $\hat{R}$; and
(iii) {\em lifting}, which transforms $\hat{R}$ back to a tensor representation of $R$.
The lifting step is fast (reformatting), but the lowering step requires replication of the data, sometimes by a factor of 1 or 2 orders of magnitude. This blowup in the data size demands more off-chip memory and more computation in step (ii). 

\subsection{Batching and Data Parallelism}\label{sec:batching}

\begin{figure}[t!]
\centering
\scalebox{0.75}{
\begin{tabular}{c||rrrr}
\hline
{\bf Device Type} &
{\bf Device} &
{\bf \% Peak} &
{\bf \% Peak} &
{\bf \% Peak}\\
{\bf (EC2 Instance)} &
{\bf GFLOPS} &
{\bf Caffe} &
{\bf Omnivore} &
{\bf SGEMM}\\
\hline
1$\times$ CPU Xeon E5-2666   &
\multirow{2}{*}{742} & 
\multirow{2}{*}{18\%} & 
\multirow{2}{*}{56\%} & 
\multirow{2}{*}{81\%} \\
(c4.4xlarge) & & & & \\
\hline
2$\times$ CPU Xeon E5-2666   &
\multirow{2}{*}{1,670} & 
\multirow{2}{*}{8\%} & 
\multirow{2}{*}{40\%} & 
\multirow{2}{*}{71\%} \\
(c4.8xlarge) & & & & \\
\hline
1$\times$ GPU Grid K520   &
\multirow{2}{*}{1,229} & 
\multirow{2}{*}{53\%} & 
\multirow{2}{*}{54\%} & 
\multirow{2}{*}{99\%} \\
(g2.2xlarge) & & & & \\
\hline
Dual-GPU Grid K520   &
\multirow{2}{*}{2,458} & 
\multirow{2}{*}{26\%} & 
\multirow{2}{*}{52\%} & 
\multirow{2}{*}{99\%} \\
(g2.8xlarge) & & & & \\
\hline
\end{tabular}
}
\vspace{-0.6em}
\caption{Across several CPUs and GPUs we obtain throughput on
  convolution layers that is $\sim 50\%$ of the device peak
  (shown for AlexNet, total $FW+BW$ time for all conv layers). 
  We also show large GEMM as a reference of what the device can achieve.}
\label{fig:flops:prop}
\end{figure}

The design tradeoff between CPUs and GPUs arises as a result of this increase in data size.
Assume that we are given a fixed batch size of $b$ images (e.g.,\ $b=256$).  GPUs cannot fit an
entire batch of lowered data into off-chip memory, therefore many CNN
implementations on GPUs perform lowering/GEMM serially on one or few
images at a time until all $b$ have been processed. On the CPU
however, off-chip memory is larger which allows lowering/GEMM to be
performed on all $b$ images at once. This leads to significant CPU
speedups compared with state-of-the-art tools which do not explore
this tradeoff but use the same implementation suited to the GPU for
the CPU. In particular, as Figure~\ref{fig:flops:prop} shows, this allows us
to view a CPU or GPU as simply a device producing FLOPS; for
reference, we also include the FLOPS delivered by the most
popular CNN framework Caffe~\cite{Jia:2014:arXiv} and the call of an
optimized single-precision matrix multiply (SGEMM) for that hardware. 
The throughput obtained by Omnivore on all devices in Figure~\ref{fig:flops:prop} also matches the 
expected range for CNNs (on GPUs) reported by the manufacturer~\cite{Chetlur:2014:ArXvi}.

To achieve this throughput on the CPU, we use a simple \textit{convolution
  batching} technique in which we first lower all $b$ images in a
batch before performing any GEMM.  After this lowering, we perform a
\textit{single} GEMM for all $b$ images, rather than $b$ smaller
GEMMs.  This consumes $b\times$ the memory because the $\hat{D}$
matrix is $b\times$ larger than when lowering images one by
one. However, it has two speed benefits: (i) one large GEMM is faster
than $b$ smaller GEMMs because CPU caches and vector instructions are
fully utilized, and (ii) lowering all images in the batch at once
enables data parallelism to be used during lowering. Specifically for
(ii), given $n$ CPU cores, a batch is split into $n$ partitions, with
$b/n$ images per partition. A separate thread then performs lowering
and GEMM on each partition, i.e. each thread performs convolution on a
subset of the batch. 
This data parallelism can be applied to other layers.

\begin{figure}
\centering
\includegraphics[width=0.5\textwidth]{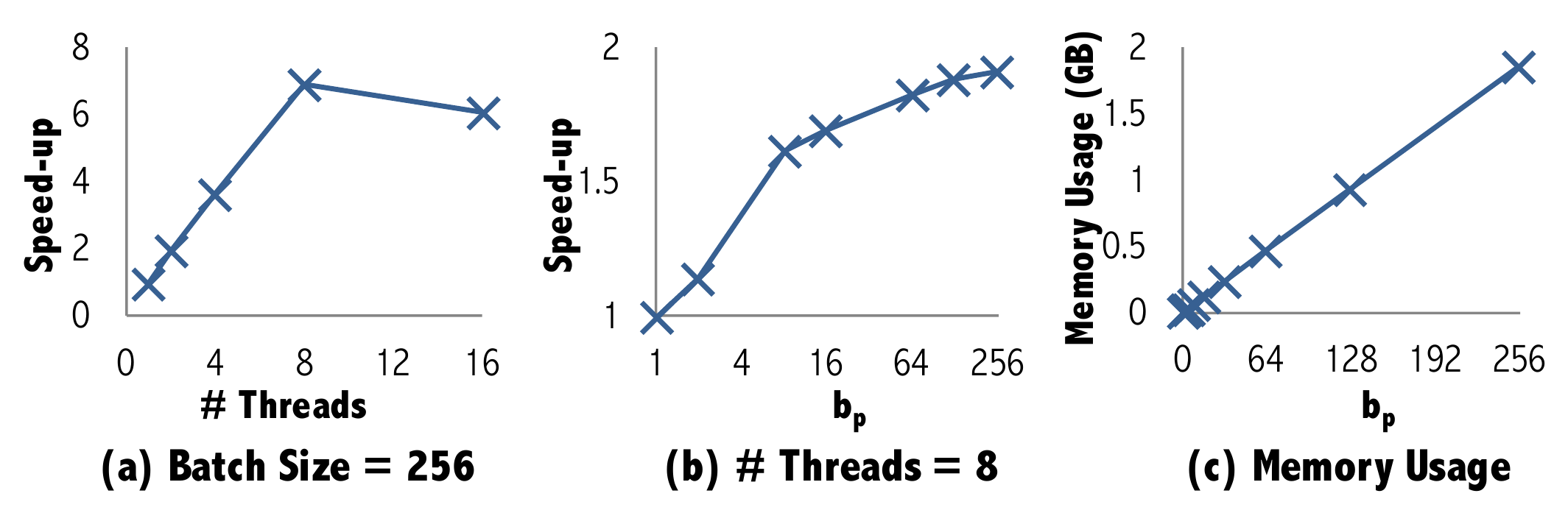}
\vspace{-2.0em}
\caption{The impact of batch size 
($b_{p}$)
and number of threads
on the GEMM kernel (total batch size 
$b$ 
= 256, 8 physical cores).
Shown for the GEMM in the largest convolutional layer of AlexNet.}
\label{fig:parallel}
\end{figure}

Generalizing this implementation tradeoff, $b_{p}$ images can be
processed in parallel by the convolution layer, where $1 \leq
b_{p} \leq b$. Increasing $b_{p}$ increases the memory footprint but
decreases the overall execution time.
Figure~\ref{fig:parallel} shows batching
experiments for a CPU GEMM kernel.  All points in
each graph perform GEMM on 256 images (i.e., $b=256$), but the number
of total GEMM calls depends on $b_{p}$ (e.g., if $b_{p}=256$, there is
one large GEMM).
We therefore advocate the
strategy of selecting $b_{p}$ as large as possible (up to $b$) such
that $\hat{D}$ fits into the off-chip memory. This can be predicted
because memory usage increases linearly with $b_{p}$ as seen in
Figure~\ref{fig:parallel} (c).
For a range of modern CPUs that we used in our experiments,
the optimal $b_{p}$ value is always $b$.

While this tradeoff is simple, it enables {\em FLOPS
  proportional scheduling} which allows us to abstract away the
details of devices in our distributed implementation.

\section{Multi-Device Tradeoff}\label{sec:Distributed}

In this section, we study the distributed setting. Given a CNN, an input set of data
and a set of devices, our goal is to map each layer of the CNN
and a subset of data to each device. 
Many distributed
CNN systems~\cite{Chilimbi:2014:OSDI, iandola15firecaf,
  chen15mxnet, Dean:2012:NIPS, Wang:2015:TR, moritz15sparknet}
  can be mapped to points within our
tradeoff space.

We show that a key tradeoff is the degree of
asynchrony. 
We build
models predicting how hardware
and statistical efficiency are affected
by asynchronous execution.  
By decoupling hardware efficiency and
statistical efficiency and creating separate models,
the optimizer can find a balance and minimize the total time to
convergence. Specifically, we argue for an analytic model for 
hardware efficiency and
we are able to give a new theoretical characterization of statistical
efficiency.


\begin{figure}
\centering
\includegraphics[width=0.5\textwidth]{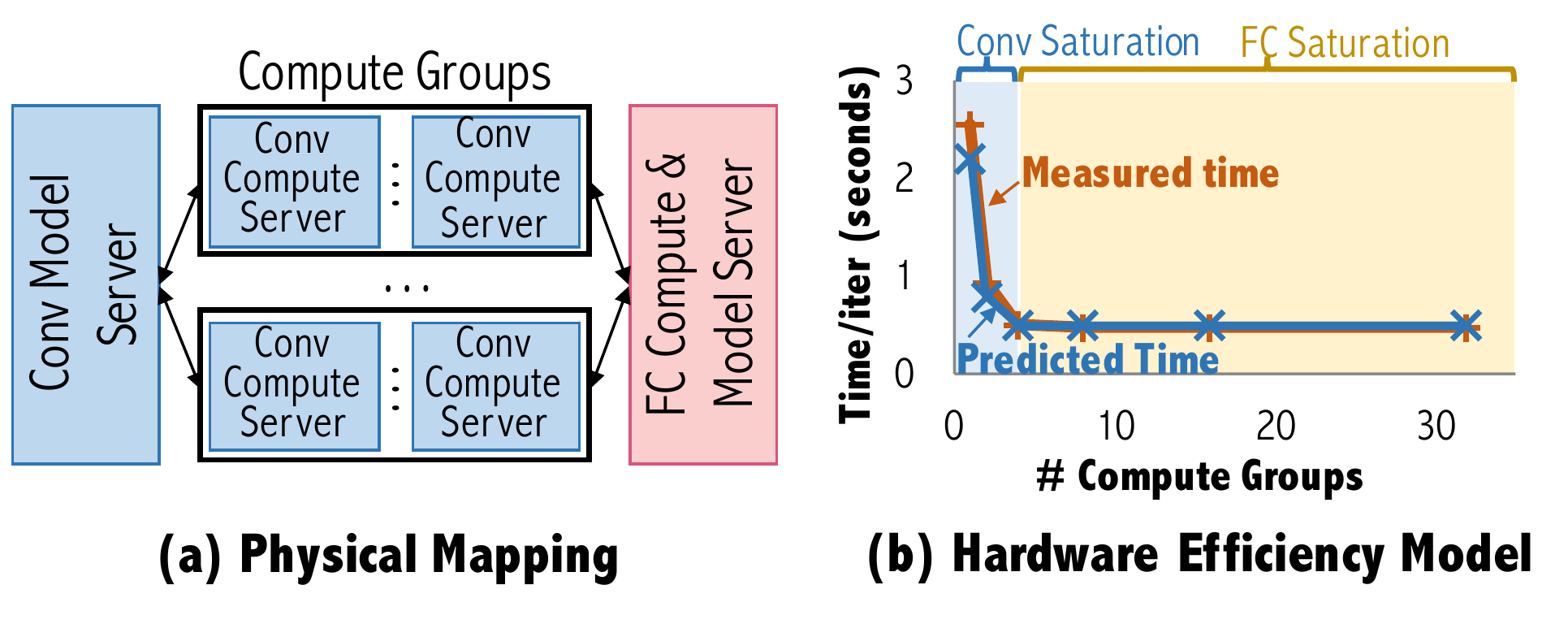}
\vspace{-2.0em}
\caption{(a) Default physical mapping used by Omnivore. (b) Predicted and measured iteration time as the number of devices (machines) per group changes (32 c4.4xlarge machine, AlexNet)}
\label{fig:HE_predict}
\vspace{-1.0em}
\end{figure}

\subsection{Setup and Assumptions}

We separate each layer into {\it compute
  servers}, responsible for computation, and a {\it model server}
responsible for handling reads from and writes to the model. These
``servers'' are virtual: many servers may be mapped to the same device
or an individual server may be mapped to several devices. While the
separation of compute and model servers is present in all systems,
only project Adam~\cite{Chilimbi:2014:OSDI} described mapping both
compute and model servers to the same device (or set of devices for parallelism), i.e.~\textit{merging}
the servers, which they did for FC layers. 
Figure~\ref{fig:HE_predict} (a) shows this mapping.
For concreteness, suppose there are $N+1$ devices: one device
handles the FC layers, and the remaining $N$ devices handle the conv
layers (motivated by $90-95\%$ of the computation being in the
convolutions). To simplify our 
exposition we refer to this reference architecture throughout Section~\ref{sec:Distributed}.
Section~\ref{sec:Optimizer} demonstrates the benefits of this architecture 
empirically both in terms of hardware and statistical efficiency. 

\paragraph*{Compute Groups}
The input data to a layer is divided into batches (whose size is
determined by the system). The main computational operation for each
layer is to (i) read the model; (ii) compute an operation on the data
(the forward or the backward pass) for the given a batch of data; and
(iii) update the model. We assign each compute server to a {\em
    compute group}. Within a compute group, many devices speed up the
  computation of an individual batch of data, e.g., all devices
  compute a single gradient for a batch. Across compute groups,
  different devices process distinct batches.

\paragraph*{Asynchrony and Staleness}
In most systems, compute groups communicate {\em asynchronously}
\cite{Dean:2012:NIPS,Chilimbi:2014:OSDI,hogwild}: the models are
updated without locking or barriers, and so forward and backward passes may be
computed with a stale model. Intuitively, the lack of coordination
allows one to make better use of the hardware (better hardware
efficiency) but the use of stale information may reduce 
statistical efficiency. If there are $g=S+1$ compute groups, we call $S$
the staleness parameter. This is justified as the computation within
each group in step (ii) above is very regular for CNNs (standard deviation of
runtime is less than 6\% of mean), hence these groups execute in a
nearly round-robin fashion.

Throughout this section, we make two simplifying assumptions: First,
we focus our description on the two phases described in
Section~\ref{sec:background}, i.e., we abstract networks into
convolutional layers (conv) which have a large amount of data but a
small model, and fully-connected layers (FC) which have small data but
a large model.  Practically, these two layers are the main bottlenecks
in current networks. Their differing characteristics give rise to
different points of the tradeoff space. Second, many popular networks
are simply a single chain, not a DAG (e.g.,\  AlexNet), and DAGs can be
serialized into a chain, hence we view the
input as a list of layers (without loss of generality).

\paragraph*{Execution}
Given a list of layers, the main computational loop is to move through
the list forward calling the forward operation and in reverse order
calling the backward operation at each step. As we have decomposed each
layer into a model server and compute servers and further mapped compute servers 
to compute groups over several devices, it is the responsibility of our 
execution engine to make sure that all data is on each device when needed. 
We use standard techniques to hide latency of device copies, 
e.g., double buffering.

\subsection{Hardware Efficiency Model}\label{sec:HE_Model}

The goal of this section is to create a predictive model $HE(S)$ for
hardware efficiency. The goal is to characterize
the impact of the amount of
staleness $S$ in the system (or equivalently the number of compute groups, $g$).
We derive a simple analytic model which
reasons about the bottlenecks.  An \textit{execution
  strategy} partitions the $N$ conv devices into $g$ compute groups.
Again for concreteness, assume there are $k$ devices per group
($k=N/g$).  Let $t_{conv}(k)$ be a function that returns the time that
a group of size $k$ needs to compute the convolution phase.
We make the assumption that FC only operates sequentially\footnote{ \scriptsize{For simplicity, we assume different groups
  (batches) cannot be executed in parallel on the FC server but
  our model can be extended to more general cases.}}.
Note that
the number of requests to the FC phase is a function of the number of
groups, and let $t_{fc}$ be the time that the FC phase needs to
serve one group.

Given $g$, $t_{conv}(k)$ and $t_{fc}$, our goal is to create a
hardware efficiency model which predicts the time per
iteration. There are two cases depending on which phase is the bottleneck.

\noindent
(1) When the FC phase is saturated, i.e., it starts to serve the next request immediately after the previous request finishes,
\begin{equation*}
\text{Time per iteration}_{\text{saturated fc}} = t_{fc};
\end{equation*}
(2) When the FC phase is not saturated, each conv group becomes the bottleneck. In this case, 
\begin{equation*}
\text{Time per iteration}_{\text{saturated conv}} = (t_{conv}(k) + t_{fc})/g
\end{equation*}
which is the total time for a single iteration divided by the number
of parallel groups.
Thus, our predicted model for iteration time, $HE(g)$, is:
\begin{equation*}
HE(g) = \max \{ t_{fc}, (t_{conv}(k) + t_{fc})/g \}
\end{equation*}
Given $t_{conv}(k)$, $t_{fc}$ and the number of groups, $g$, the model can now predict   what the mode of saturation will be and therefore the time per
iteration.



\paragraph*{Obtaining the Parameters}
The parameters above can be measured with high accuracy and low variance.  $t_{fc}$
can be measured by running an iteration on a single device, but
$t_{conv}(k)$, though still directly measurable, requires measurements for
each $k$. Instead, $t_{conv}(k)$ can be calculated from (i) the
throughput of each node; (ii) the network speed; and (iii) a
measurement of $t_{conv}(1)$ (which only needs to be measured for 
a single $k$, $k=1$, and on a single device). 
Figure~\ref{fig:HE_predict}(b) shows
that our hardware efficiency is accurate; detailed evaluation
\yell{
\iflongversion
in the appendix.
\fi
}


\subsection{Statistical Efficiency Model}\label{sec:SE_Model}

We now characterize the effect of staleness on statistical
efficiency.
\icde{
While working on \oursystem, we realized that 
the momentum value that yields the fastest convergence decreases as we increase the number of asynchronous workers. Using the right value can {\em significantly} reduce the number of iterations to reach a target loss.}
This motivated us to mathematically analyze this phenomenon in our companion theory paper \cite{mitliagkas2016asynchrony}.
The result is surprisingly elegant: under a simple model, asynchrony introduces an extra momentum term in the SGD update. This comes in agreement with our experimental findings in this paper.
\icde{
Here, we recap the essential result of our theory and validate it experimentally on different systems.}
Importantly, our result is {\em predictive}.
We use it in this work to complement our experimental findings on the importance of momentum tuning and to design the first {\em asynchrony-aware optimizer} for deep learning systems in Section~\ref{sec:Optimizer}.

We make the following assumptions, which are not necessary
but are helpful to capture the essence of our result.

\begin{itemize}\compactify
\item [(A0)] The batch for each step is drawn uniformly at
  random with replacement. This is a standard assumption of SGD.
  
\item [(A1)] Variations in the time to process a step are due
  to unmodeled system behavior. Also, variations are
  independent of the specific batch drawn. 
  This is justified by the fact that, for all batches, all computation involves
  dense operations.
  
\item [(A2)] The time it takes to process a step is exponentially
  distributed and independent from other steps. This is a simplifying
  but standard assumption from queuing theory~\cite{gross2008fundamentals}.
  A more
  general (and complex) version of Theorem~\ref{thm:queueing} below
  holds without this assumption.
\end{itemize}

\begin{theorem}
\label{thm:queueing}
Consider $g$ asynchronous groups and set explicit momentum to 
zero, i.e.\ $\mu=0$ in the update of equation~\eqref{eqn:update2}.
Under the above assumptions, the expected update becomes 
\begin{equation}
	\mathbb{E} V^{(t+1)}  
	= \left(1 - {1 \over g}\right) \mathbb{E}V^{(t)}
	- {\eta \over g} \mathbb{E}\nabla_W \ell(W^{(t)}).
\end{equation}
In which $\ell(W)$ denotes the expectation of $\ell(W,I,C)$ over the
random draw of possible batches $(I,C)$.
\end{theorem}

\begin{figure}[tbp]
\begin{center}
\includegraphics[height=1.045in]{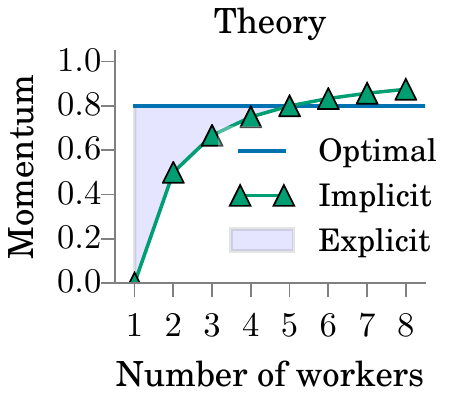}
\includegraphics[height=1.045in]{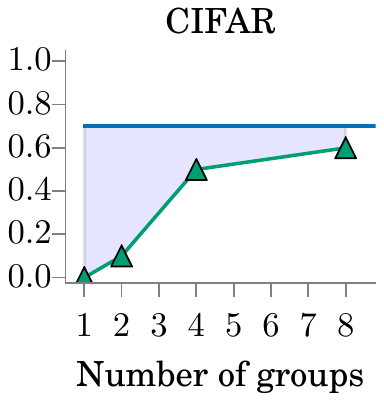}
\includegraphics[height=1.045in]{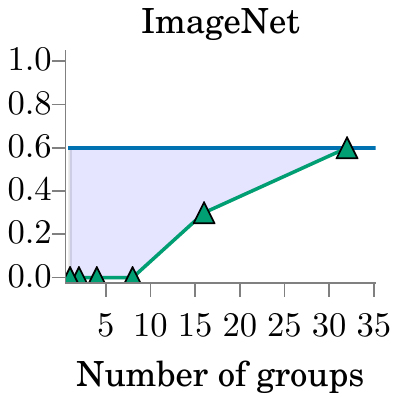}
\end{center}
\vspace{-1.0em}
\caption{Predicted and measured momentum moduli:
(Left) Predicted, Theorem~\ref{thm:queueing}
(Middle) Measured, CIFAR (Right) Measured, ImageNet} 
\label{fig:momentumstaleness}
\end{figure}

In plain English, {\it asynchrony increases momentum}--there is
implicit momentum of $1-1/g$. This not only
matches the standard form of momentum
used by deep learning practitioners in equation~\eqref{eqn:update2},
but also can predict measured system
behavior. Figure~\ref{fig:momentumstaleness} shows the predicted and
actual measured momentum for two datasets: 
As long as the asynchrony-induced implicit momentum is less than the optimal total momentum, we can algorithmically compensate with explicit momentum. When however, implicit momentum exceeds the optimal total momentum, we start incurring statistical inefficiency. We use this intuition as the basis for our optimizer.

\paragraph*{Cold-start}
We observe a phenomenon similar to {\it burn-in}~\cite{raftery1992many} in
Gibbs samplers. The model needs a few iterations to set
the appropriate scale of the model parameters. On Imagenet-1000, we find that 5
passes over the dataset ($<15\%$ of total execution) suffice to ``warm up'' the model.
As a result, the optimizer will start by running synchronously and then switches to
asynchronous (Section~\ref{sec:Optimizer}).

\section{Distributed Optimizer}\label{sec:Optimizer}
This section uses the models and tradeoff space characterization of the previous two sections to create (i) a plan for physical mapping which maps each server to a machine, and (ii) a plan for the execution strategy which defines the number of compute groups by allocating data batches to each server. As in previous sections we assume a fixed number of machines. We first discuss the process of physical mapping and then describe our optimizer. We conclude with theoretical and empirical justification for the optimizer, and in Section~\ref{sec:Experiments} compare it to state-of-the-art, Bayesian approaches. 

\subsection{Physical Mapping}
\label{sec:physical-mapping}

We observe that in all CNNs the architecture of Figure~\ref{fig:HE_predict} (a) works best, and
describe other physical maps\yell{
\iflongversion
in the appendix.
\fi}
Omnivore maps the FC compute and model servers to the same machine, an approach we call {\em merged FC}.
Merging the FC compute and model servers to the same devices
was shown in~\cite{Chilimbi:2014:OSDI} to reduce communication overhead in a CPU cluster (better hardware efficiency),
however it was not known that (1) this also benefits hardware efficiency for a GPU cluster,
and (2) this technique also benefits statistical efficiency by eliminating staleness in the FC model. 
These are both observations we make.
The remaining machines are used for the conv compute servers. The conv
model servers are mapped to one of the conv compute machines. These
optimizations are critical: on a cluster of 33 EC2 c4.4xlarge
machines, not merging the FC servers incurs an additional hardware
efficiency penalty of $1.2\times$ due to increased communication as
well as a statistical efficiency penalty of $2.5\times$ because of
staleness in the FC model.  The key tradeoff is therefore the number
of conv compute groups, which we describe next.

\subsection{Optimizer}
\label{sec:opt:opt}

\begin{algorithm}[t]
\small
\renewcommand{\algorithmicrequire}{\textbf{Input:}}
\renewcommand{\algorithmicensure}{\textbf{Output:}}

\begin{algorithmic}
\Require Time budget $T$ and possible choices of (1) \# compute groups $\mathcal{CG}$, (2) momentum 
$\mathcal{M}$, and (3) learning rate $\mathcal{H}$.
\Ensure Trained model $W$.
\end{algorithmic}

\hrulefill

\begin{algorithmic}[1]

\State $g = \mathcal{CG}$
\While {{\bf not reaching the termination criteria}}
\State $(\mu, \eta)$ $\leftarrow$ $\textbf{gridSearch}(\mathcal{M}, \mathcal{H} | W, g)$ 
\While {$\mu = 0$ and $g > 1$}
\State $g \leftarrow g/2$ 
\State $(\mu, \eta)$ $\leftarrow$ $\textbf{gridSearch}(\mathcal{M}, \mathcal{H} | W, g)$ 
\EndWhile
\State $W \leftarrow \textbf{train}(g, \mu, \eta, W)$ for $T$ minutes
\EndWhile
\State \Return $W$.
\end{algorithmic}
\caption{Automatic Optimizer for the Tradeoff}
\label{alg:optimizer}
\end{algorithm}

There are multiple {\em interdependent} factors
that have impact on the performance of the training procedure: 
(1) the number of compute groups; (2) the momentum; 
and (3) the learning rate.
The optimal setting of these parameters
might also change during training, so our optimizer runs periodically in
epochs (e.g., every hour).  Algorithm~\ref{alg:optimizer} shows the
end-to-end optimizer that runs after the initial cold-start period.

In each of the epochs, the key issue is to set the number of compute
groups, $g$. We perform a grid search over both the learning rate and
the momentum starting at a particular value of $g$.
This search determines the optimal explicit momentum for that $g$
by selecting the configuration with the lowest final loss.
  The key intuition is: {\it set the highest amount
  of asynchrony, $g$, such that this explicit momentum is non-zero.}
The reasoning is that,
when the optimal explicit momentum is $0$,
the implicit momentum is likely higher than the optimal value,
and a cause of statistical inefficiency
(c.f.\ Figure~\ref{fig:momentumstaleness}).
In this case, we reduce the amount of asynchrony by reducing $g$.
We provide an initial value for $g$ by leveraging the hardware efficiency model. In particular, we start with the smallest number of compute groups that saturate the FC server.
This can be determined analytically or through measurements
during the cold start phase.
Having selected a $(g,\eta,\mu)$, we run for the rest of the epoch. At the end of one hour, the epoch ends, the
model is checkpointed (written to disk), and the optimizer repeats.

\begin{figure}[t!]
\centering
\includegraphics[width=0.5\textwidth]{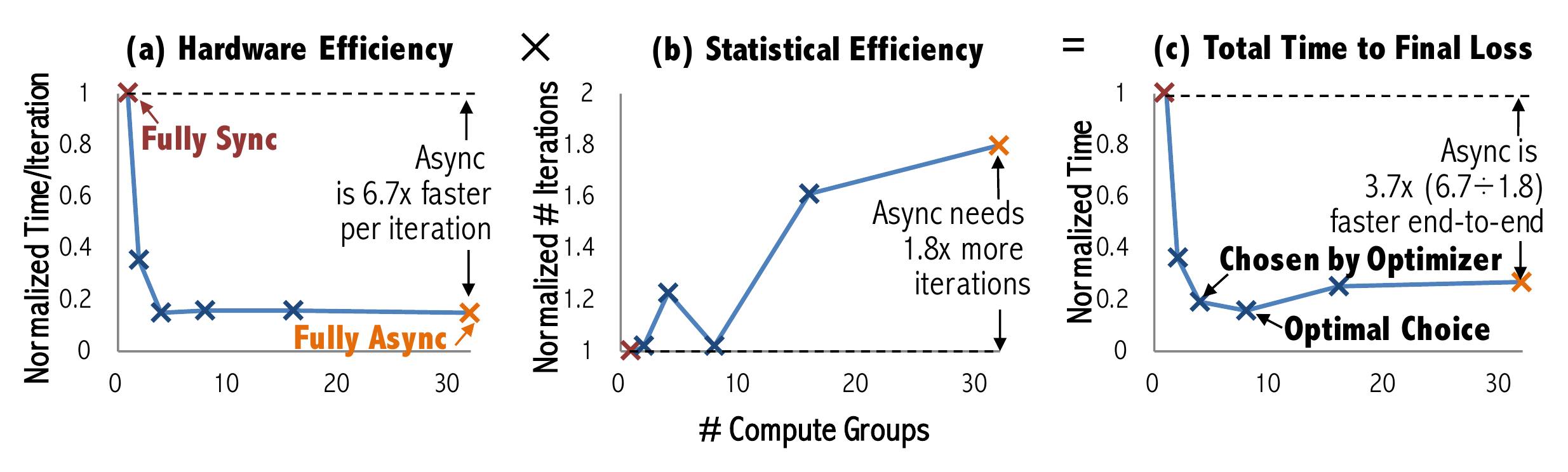}
\vspace{-2.5em}
\caption{Hardware efficiency, statistical efficiency, and total time for various execution strategies.}
\label{fig:sec4full}
\end{figure}

\paragraph*{Importance of Compute Groups}
We demonstrate that using the right number of compute groups has an impact on performance.
Fixing the total number of machines, we try different numbers of compute groups on \mbox{CPU-L~(Figure~\ref{tab:machine})} for the Imagenet 1000-class dataset, and AlexNet CNN. 
We grid-search a number of values for the learning rate and momentum and report the best result achieved in each case.
Figure~\ref{fig:sec4full} reports (a) the time
per iteration (hardware efficiency), (b) the number of iterations to reach a specific training loss (statistical
efficiency), and (c) their product, which
is the total time required to reach the final loss.
Note that the hardware efficiency curve in (a) is the same as in Figure~\ref{fig:HE_predict} (b).

We see in Figure~\ref{fig:sec4full}~(c) that $g=32$ (fully asynchronous) is $3.7\times$ faster than $g=1$ (synchronous) 
as measured by wall-clock time to final loss. 
This is due to its $6.7\times$ faster iteration time in (a), although it requires $1.8\times$ more iterations as shown in (b).
This matches the theory's prediction: increasing $g$ causes the optimal explicit momentum, $\mu^{*}$, to decrease.
We noted in 
Figure~\ref{fig:momentumstaleness} (right) that $\mu^*$ drops to $0$ at $g=32$, and consequently there is a penalty in statistical efficiency.
Running the optimizer of Algorithm~\ref{alg:optimizer} selects $g=4$, which is near-optimal: $5.3\times$ faster than sync and 
$1.4\times$ faster than async.
We repeat the same experiment for CIFAR and find the results are similar. In all cases, the optimal number of groups is more than $2\times$ faster compared to sync,
and that Algorithm~\ref{alg:optimizer} always picks a near-optimal point strictly better than sync.

\section{Experiments}\label{sec:Experiments}
We evaluate the runtime performance of our system. We show that, 
\oursystem outperforms state-of-the-art tools by $1.9\times$ to $12\times$
on a range of training tasks. Our result holds (i) across a diverse
range of hardware, including both CPUs and GPUs, (ii) in both single
device and multiple device/machine settings. Moreover, 
\oursystem contains an automatic optimizer and
therefore does not require users to input hyperparameter values.

Our experiments also validate that our optimizer
is up to $6\times$ faster compared to state-of-the-art Bayesian optimizers. 


\subsection{Experiment Setup}


\paragraph*{Datasets and Models}

We validate the performance of \oursystem on a diverse range of
datasets and models, as shown in Figure~\ref{tab:corpus}.  The largest corpus we use is
ImageNet~\cite{deng2009imagenet}, which contains 1.3M
images. ImageNet is the de facto benchmark for deep learning
systems~\cite{Krizhevsky:2012:NIPS,Dean:2012:NIPS,Chilimbi:2014:OSDI}.
Training a model on ImageNet can take tens of hours---even on the latest 7~TFLOPS Titan X GPU, NVIDIA reports that it took three days to train with a single GPU.\footnote{\scriptsize{\url{https://blogs.nvidia.com/blog/2015/03/17/digits-devbox/}}}
For some of our experiments, which require running many configurations
to convergence, we used a reduced version, ImageNet8, containing the
first $8$ classes. We train the standard
CaffeNet\footnote{\scriptsize{\url{https://github.com/BVLC/caffe/tree/master/models/bvlc_reference_caffenet}}}
on both data sets.  We also use smaller, but standard, datasets CIFAR
and MNIST, which are for object recognition and handwriting
recognition, respectively. We train the networks in Caffe's
tutorials for both.

\paragraph*{Metrics}
Our main metric of performance is the wall-clock time required to
achieve a given training accuracy. As in Section~\ref{sec:Optimizer}, 
we define statistical efficiency as the number of
iterations needed to achieve that training accuracy.

\paragraph*{Competitor Systems and Experiment Settings}

We compare \oursystem against a range of existing tools using both a
single-machine and multiple machines.  {\bf Single machine:} we
compare \oursystem to \caffe~\cite{Jia:2014:arXiv} and Google's
\tensorflow~\cite{tensorflow}, the two most popular CNN systems, using
Caffe's reference (CaffeNet) model.
{\bf Multiple machines:} we compare \oursystem to
\mxnet~\cite{chen15mxnet} and \singa~\cite{Wang:2015:TR}, two popular
distributed deep learning systems.\footnote{ \scriptsize{We also surveyed a range
  of distributed training systems including SparkNet, DL4J, and
  others.  We found that \mxnet is the
  fastest system on our datasets/tasks. At the time of writing, the
  official version of \caffe can only run on a single machine, and
  \tensorflow does not contain a distributed example for
  AlexNet/CaffeNet.}}  Both \mxnet and \singa support multiple
execution strategies, and we consider all of these strategies in our
experiments. We set up and tune both systems according to their
tutorials.  On data sets where SINGA performs strictly worse than
\mxnet, we omit its result from the figure.
To demonstrate the merits of momentum tuning and compute groups on other platforms, we also implemented these features in \tensorflow
\yell{
\iflongversion
	in the appendix.
\fi
}

\oursystem is implemented in C++.
\caffe,
\tensorflow, \mxnet, and \singa all implement
their core components in C++. We compile all 
systems with g++~4.8.2,
and use OpenBLAS 0.2.15, cuda 7.5, as well as both
cuDNN v3 and v4 for competitor systems and report the faster result.

\begin{figure}[t!]
\centering
\scriptsize
\begin{tabular}{crrr}
\hline
Data Set & \# Images & Image Size & \# Classes \\
\hline
{\bf ImageNet} &   1.3 M   & 256$\times$256$\times$3 & 1000 \\
{\bf ImageNet8} &  10 K    & 256$\times$256$\times$3 & 8    \\
{\bf CIFAR}    &   60 K    & 32$\times$32$\times$3   & 10   \\
{\bf MNIST}    &   60 K    & 28$\times$28$\times$1   & 10   \\
\hline
{\bf Shakespeare} & 162 K  & 25$\times$1$\times$128  & 128   \\
\hline
\end{tabular}
\caption{Statistics of Data Sets. All data sets are
image-related except {\bf Shakespeare}, a natural language corpus for text synthesization
used in our RNN experiments.}
\label{tab:corpus}
\end{figure}

\begin{figure}[t!]
\scriptsize
\centering
\begin{tabular}{c|rrrrr}
\hline
Name & Machines & TFLOPS & Network & \$/hour\\
\hline
{\bf 1xCPU} & 1 $\times$ \texttt{c4.4xlarge} & 0.74 & - & \$0.84\\
{\bf 2xCPU} & 1 $\times$ \texttt{c4.8xlarge} & 1.67 & - & \$1.68\\
{\bf 1xGPU} & 1 $\times$ \texttt{g2.2xlarge} & 1.23 & - & \$0.65\\
{\bf 4xGPU} & 1 $\times$ \texttt{g2.8xlarge} & 4.89 & - & \$2.60\\
\hline
{\bf CPU-S} & 9 $\times$ \texttt{c4.4xlarge} & 6.68 & 1 Gbits & \$7.56\\
{\bf CPU-L} & 33 $\times$ \texttt{c4.4xlarge}& 24.51& 1 Gbits & \$27.72\\
{\bf GPU-S} & 9 $\times$ \texttt{g2.8xlarge} & 44.24& 10 Gbits & \$23.40\\
\hline
\end{tabular}
\caption{Summary of Machines and Clusters. TFLOPS are
the total TFLOPS that the given machine/cluster
can provide. We are unable to open 33 4xGPU machines
due to limited EC2 availability and
therefore there is no GPU-L.
}
\label{tab:machine}
\end{figure}

\subsection{Performance Evaluation}\label{sec:small_cluster}

We validate that \oursystem has faster execution to the same quality as existing
systems in training deep learning models. More precisely, \oursystem
achieves the same training accuracy/loss {\em faster}, as measured by
wall-clock time.  We first present our main end-to-end performance
results, comparing \oursystem to the state-of-the-art.  Then, we 
validate our contributions by showing results (i) for a single machine
comparing to \caffe and \tensorflow and (ii) in the distributed setting
comparing to \mxnet and \singa. We evaluate the impact of our
tradeoff space and optimizer in Section~\ref{sec:detailed}.

\subsubsection{End-to-end Performance}
\label{sec:end-to-end}

For large datasets, \oursystem is faster than
state-of-the-art tools.  We validate this on ImageNet.
We compare \oursystem with \mxnet and \singa.
We train the standard CaffeNet on all systems
using both CPU-L and GPU-S clusters. We time out
each run after 8 hours and report the training 
accuracy at a given time. We tune
all competitor systems following the official 
performance tuning guideline\footnote{\scriptsize{\url{https://github.com/dmlc/mxnet/tree/db6f6a9418e5696b04be741a78a47ae877bb5505/example/image-classification}} and previous work~\cite{Bottou:2012:Tricks,Krizhevsky:2012:NIPS}}.\footnote{
The time for this tuning of other tools exceeds \oursystem's automatic tuning time for the cold start phase
so we omit these initial tuning times.}
For \oursystem, we use its automatic
optimizer that does not require any hyperparameters.
Because \singa is always slower than \mxnet 
in our experiments, we omit it from this figure.
but discuss it in Section~\ref{sec:distributed-experiments}.

\begin{figure}[t!]
\centering
\includegraphics[width=0.45\textwidth]{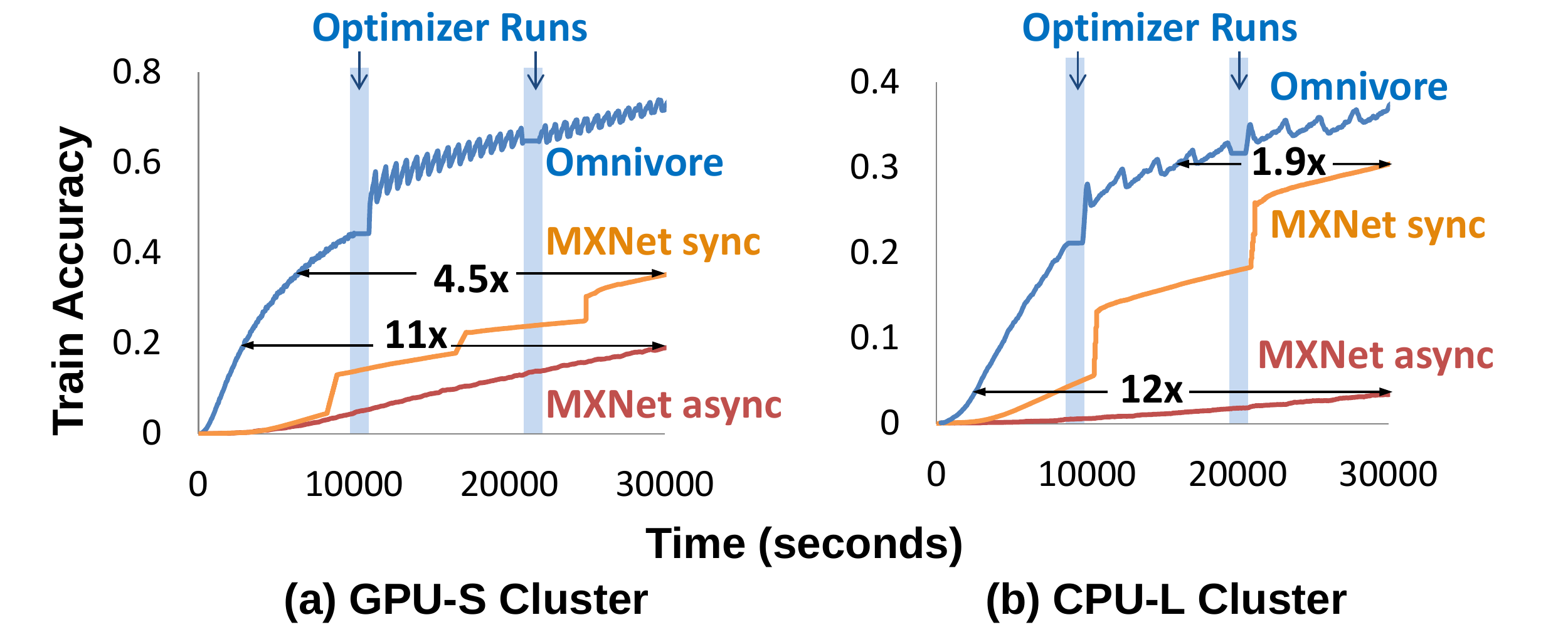}
\vspace{-0.5em}
\caption{End-to-end Performance of \oursystem and \mxnet on ImageNet.
We omit SINGA as it performs worse than \mxnet.
}
\label{fig:flagship}
\end{figure}

Figure~\ref{fig:flagship} shows the result. We report sync and
async for MXNet as their documentation suggests trying both.
\oursystem reaches the same accuracy up to  
$11\times$ faster than \mxnet on the GPU cluster
and $12\times$ faster
on the CPU cluster.
Compared to sync, \oursystem is $4.5\times$ and $1.9\times$ faster
respectively.
This includes the $10\%$ overhead of \oursystem's optimizer during the run.
The optimizer reduces momentum or learning rate each time it runs.
In the
remainder
of this section
we conduct detailed analysis of
\oursystem,
\mxnet, and \singa to understand this
improvement. As we will see,
\oursystem's optimizer, which searches within
the larger tradeoff space,
 is the key reason 
for our system's performance.

\subsubsection{Single Machine Experiments}
\label{sec:single-machine-experiments}

We validate our claim that \oursystem's performance is {\em
  FLOPS-proportional}, which means it scales with available FLOPS,
regardless of the number or type of devices available.  We first
explain the protocol of our experiments.  Then we report our results
and discuss the importance of FLOPS-proportional system performance.

\paragraph*{Protocol}
We compare \oursystem against Caffe and Tensorflow
on a single machine. We train CaffeNet on 
ImageNet on hardware described in Figure~\ref{tab:machine}.
We measure the time each system needs to finish 40 iterations
of training (following 10 iterations of warm-up), using the standard batch size in CaffeNet ($256$ images).
Our single-machine optimizations only affect hardware efficiency; the number of iterations needed for convergence does not change. Hence,
we can use time per iteration as a surrogate for performance.



\begin{figure}[t!]
\centering
\scriptsize
\begin{tabular}{c||r|r||r|r}

\hline
\multirow{2}{*}{\bf System} & \multicolumn{4}{c}{\bf Machines} \\
\cline{2-5}
                 & {1xCPU}     & {2xCPU}     & {1xGPU}              & {4xGPU} \\
\hline
{Caffe}          & 1.03$\times$   & 1$\times$   & {\bf 1.11$\times$}   & 1$\times$   \\
{TensorFlow}     & 1$\times$   & 1.05$\times$   & 1$\times$            & 1.26$\times$            \\
\hline
{Omnivore}       & {\bf 3.90$\times$}           
				 & {\bf 5.36$\times$}						    
				 & 1.04$\times$	
				 & {\bf 3.34$\times$}	           

											                \\
\hline
\end{tabular}
\caption{End-to-end single machine performance across 
different machines on
CaffeNet. Times are normalized as the speedup
over the slowest system on the same machine (the larger the better).
We run cuDNNv3, cuDNNv4 and no cuDNN
for \caffe and \tensorflow and report the
fastest one.
}
 \label{fig:e2e}
\end{figure}

\paragraph*{Results}

Figure~\ref{fig:e2e} shows the results. 
We normalize all execution times by the
slowest system 
in each column and report the resulting speedups.
We see that on a single CPU,
\oursystem is $3.9\times$ faster than both
\caffe and \tensorflow; on a single GPU, all systems show similar speed.
This is consistent with our observation in 
Section~\ref{sec:batching}: \tensorflow and \caffe 
use the same strategy for both CPU
and GPU, which is optimal for GPU but
suboptimal for CPU. One interesting
consequence of this speedup result is
that, although \caffe on 1xCPU 
is $7\times$ slower than on 1xGPU, 
\oursystem is only $1.8\times$ slower
on 1xCPU, which we will see matches the FLOPS ratio of these devices.
\oursystem's FLOPS-scaling extends to multiple devices, and
the gap with other systems increases for more CPU sockets
(2xCPU) or GPU cards (4xGPU).\footnote{\scriptsize{We also run experiments on
a 4-socket, 56-core Haswell 
CPU machine, and \oursystem is $13\times$ faster than \caffe.
}}



\paragraph*{FLOPS Proportionality}

The training performance of CPUs is commonly believed to be an order of magnitude slower than GPU performance.
Literature often reports this, and we showed that it is the case for \caffe and \tensorflow on 1xCPU and 1xGPU.
We validate that
\oursystem delivers performance {\em proportional
to the FLOPS that a device can provide}.
As shown in Figure~\ref{tab:machine},
1xGPU provides 1.7$\times$ more FLOPS
than 1xCPU, and 
\oursystem has a $1.8\times$ gap
between 1xCPU and 1xGPU. 
In other words, 
regardless of the type of device, \oursystem performs 
proportionally to the number of FLOPS available.
We also observe that proportionality holds for all machines in Table~\ref{tab:machine}.
%
FLOPS-proportionality means that,
using both CPUs and GPUs on the same machine,
we should be able to construct an even faster system.
We validate this by using both CPUs and GPUs on
4xGPU, whose CPU and GPUs provide 0.67 TFLOPS
and 4.89 TFLOPS, respectively.
By using data parallelism across the CPU and
a single GPU, \oursystem achieves an $18\%$ speedup
over just using the GPU.

\subsubsection{Distributed Experiments}
\label{sec:distributed-experiments}

We conduct experiments to understand our end-to-end improvement
across three different clusters described in
Figure~\ref{tab:machine}. As we will show, our tradeoff
characterization leads to the performance gains of \oursystem.
 We first describe the settings and the performance metric used in the
experiments.  Then we discuss the optimizer's contribution and analyze
its decisions across different clusters.

\paragraph*{Protocol}

\begin{figure}[t!]
\includegraphics[width=0.48\textwidth]{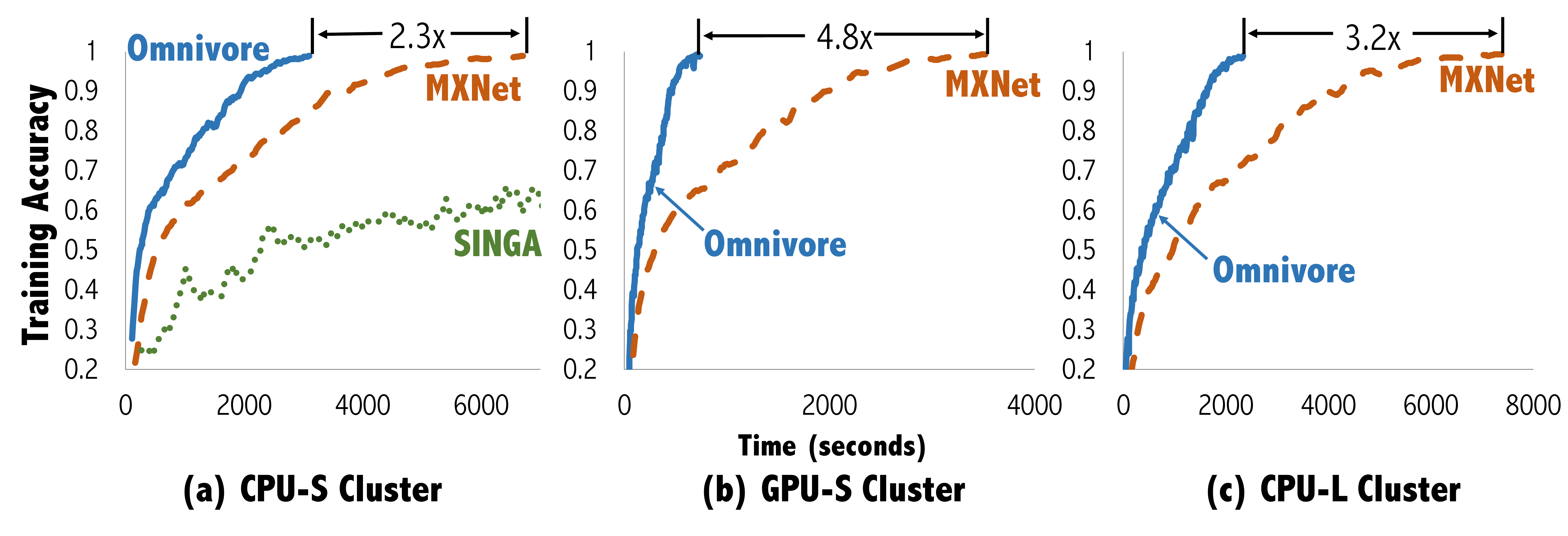}
\vspace{-0.5em}
\caption{Comparison of \oursystem with \mxnet and \singa on ImageNet8. We omit the SINGA
for CPU-L because it performs worse than \mxnet.
}
\label{tab:e2e_imagenet8}
\end{figure}

We tune \oursystem, \mxnet,
and \singa and run them
to convergence under multiple settings.
Thus, we use ImageNet8 
that contains the first eight classes of ImageNet.
This allows us to grid search all parameters
in \mxnet and \singa, including synchronization 
strategies and learning rate, and pick the best
run. 
We do not include the time of our optimizer,
which takes significantly less time
compared with the grid search we did for
\mxnet and \singa. We run all systems for 2 hours 
and measure the training accuracy at a given time.
Figure~\ref{tab:e2e_imagenet8} shows the results.

\paragraph*{Results}

{\bf (Small CPU Cluster: CPU-S)}
CPU-S is a small CPU cluster that contains
$9$ 1xCPU machines. Because each machine
is slow and the network is fast, we expect that 
a fully synchronous strategy will be fast in terms
of hardware efficiency while having the best
statistical efficiency.
Figure~\ref{tab:e2e_imagenet8}(a) shows the results.
All systems reach accuracy 60\% within
2 hours, and \oursystem reaches 60\% the fastest.
\oursystem is $2.3\times$ faster than \mxnet (388 seconds
vs. 907 seconds). At $3000$ seconds, \oursystem already 
achieves an accuracy $>99\%$,
while \mxnet achieves the same accuracy after $7000$ seconds.
The speed up here is also $2.3\times$.
As expected, both systems chose a fully synchronous strategy,
so statistical efficiency is not the cause of this performance gap.
The $2.3\times$ speed up is due to our CPU-based
optimization (Section~\ref{sec:Omnivore})
and merging FC servers (Section~\ref{sec:physical-mapping}).\footnote{
\scriptsize{
\singa does not converge to 99\% in 2 hours
and \oursystem is 11$\times$ faster than \singa
to reach 60\% accuracy.}}

{\bf (Small GPU Cluster: GPU-S)}
GPU-S is a small GPU cluster that contains
$9$ 4xGPU machines (36 GPUs). Because
each node is significantly (7$\times$) faster than 1xCPU,
we expect the optimal strategy to be more
asynchronous, and thus, statistical 
efficiency to come into play.
Figure~\ref{tab:e2e_imagenet8}(b) shows the result.\footnote{
\scriptsize{As of the time of writting, SINGA does not support GPUs and is omitted from Figure~\ref{tab:e2e_imagenet8}(b).}
}
Similar to the CPU-S cluster, \oursystem
outperforms \mxnet: it reaches $99\%$ accuracy $4.8\times$ faster.
\mxnet only supports completely synchronous or asynchronous execution, and its optimal run uses the completely synchronous strategy.
On the other hand, \oursystem's optimizer chooses to run with 
two compute groups.
Had \oursystem chosen the same strategy
as \mxnet, it would be $1.7\times$ slower
than \oursystem's actual choice
due to a different choice of the
synchronization strategy.
The remainder of the $4.8\times$ gap is
due to the physical mapping (merging FC) 
used by \oursystem, and this improves both hardware and statistical efficiency:
while originally described 
as a mechanism to reduce network communication~\cite{Chilimbi:2014:OSDI}, 
we find that merged FC also improves statistical efficiency by reducing 
staleness in the FC model.

{\bf (Large CPU Cluster: CPU-L)}
CPU-L is a CPU cluster with
33 1xCPU machines. Because the number of machines
is large, we expect the synchronization across
all machines would incur a large penalty in terms
of hardware efficiency---thus, we expect the optimal strategy to be more
asynchronous. Figure~\ref{tab:e2e_imagenet8}(c)
shows the result.
We see that \oursystem is $3.2\times$ faster than \mxnet
to reach 99\% accuracy.
The best \mxnet strategy was to 
train completely synchronously; Omnivore's optimizer now 
chose four compute groups. 
Had \oursystem chosen the same strategy as \mxnet,
it would incur $5\times$ overhead for hardware efficiency 
but only gain
a $2\times$ benefit for statistical efficiency.
Also, had \oursystem simply
chosen a fully asynchronous configuration,
it would be $3\times$ slower.
This shows the importance of choosing the
right number of groups to balance
statistical and hardware efficiency.

\paragraph*{Impact of Optimizer}

In the experiments above, we use grid search to find the optimal
strategy for both \mxnet and \singa. On the other hand, \oursystem
relies on the optimizer to automatically choose the best strategy.
Had we not used the grid search for \mxnet and relied on default
parameters the performance gap would be $20\times$ on
ImageNet8. This is
compared to \mxnet's completely asynchronous strategy, which is 
recommended in their performance tuning guideline for networks like AlexNet. 

\paragraph*{Comparison across Clusters}

\oursystem's optimizer makes different choices on different clusters.
It is interesting to compare them.
As we can see, given the same amount of machines (CPU-S vs.\ GPU-S),
as devices get faster, \oursystem tends to choose
more asynchronous strategies.
Intuitively, the faster
compute nodes get, the easier for the network to get congested.
The fully synchronous approach incurs higher penalty in that case.
On the other hand, given the same
speed of each compute node (CPU-S vs.\ CPU-L), when
the number of machines gets larger, \oursystem also tends to choose
a strategy that is between a fully synchronous 
and a fully asynchronous strategy: (1) when
the staleness gets very large, even a properly tuned, fully asynchronous
strategy incurs a penalty in terms of statistical efficiency,
and (2) when the number of machines that need to be synchronized
gets larger, a fully synchronous strategy incurs a penalty
in terms of hardware efficiency.
\oursystem's optimizer makes it possible for us
to be robust across different devices and cluster sizes.

\subsection{Tradeoff and Optimizer of \oursystem}\label{sec:detailed}

We validate the hypothesis that 
(1) the tradeoffs studied in this paper and (2) 
the automatic optimizer have a significant impact
on the performance of \oursystem. 
We study the importance of compute groups, as well as 
compute-group-specific momentum tuning.
We also study the effectiveness of \oursystem's 
automatic optimizer for this tradeoff space by comparing 
it against a standard Bayesian optimizer.


\subsubsection{The Tradeoff Space}

In this section we demonstrate that the various dimensions of the tradeoff space have a significant impact on performance.
Throughout this work we already illustrated some
of these tradeoffs, so here we only summarize them and leave
the detailed discussion \yell{\iflongversion
for the appendix.
\fi}
We see that tuning the learning rate is necessary for convergence
and the physical mapping has both \HE and \SE benefits.
We also showed that using the optimal number of compute groups 
can yield $6.7\times$ speedups compared to fully synchronous 
and $1.8\times$ compared to fully asynchronous execution.
This holds on a range of datasets and clusters.
We also implement compute groups within 
\tensorflow and demonstrate the tradeoffs for 
the Inception-v3 network \yell{\iflongversion
in the appendix.
\fi}
We now focus on the importance of properly tuned momentum, which 
as we showed in Section~\ref{sec:SE_Model} is a function of the level of asynchrony.

\begin{figure}
\centering
\includegraphics[width=0.25\textwidth]{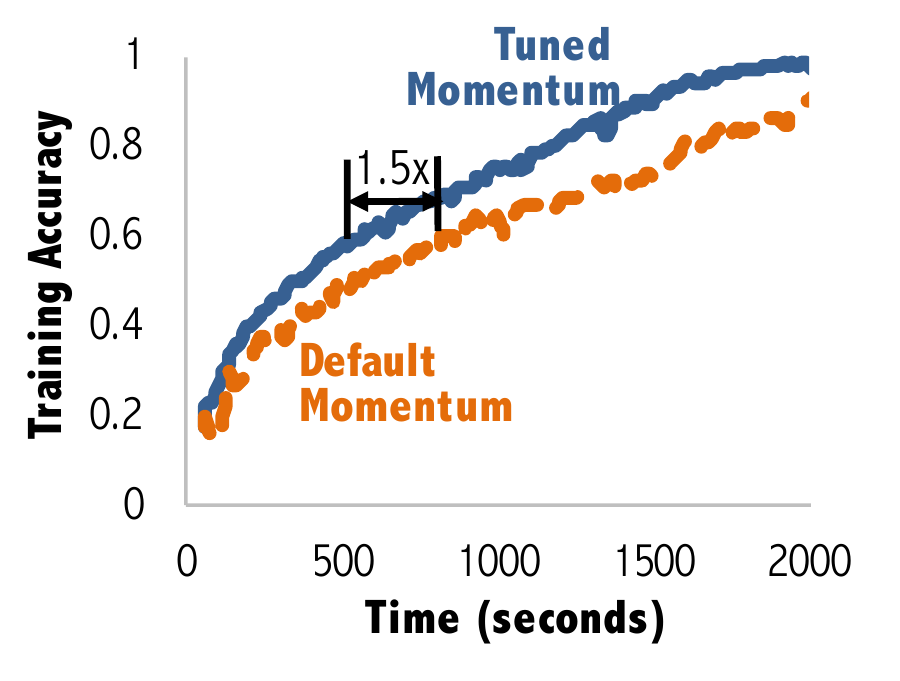}
\vspace{-1.5em}
\caption{Lesion study of momentum. Default momentum = 0.9, which is also
the optimal momentum for the fully synchronous strategy.}
\label{fig:lesion_momentum}
\vspace{-1.0em}
\end{figure}

\paragraph*{Importance of Momentum Tuning}
We validate that the correct value of momentum depends on the number
of groups. We expect that properly tuned momentum would outperform a
momentum tuned agnostically to the number of compute groups.
Therefore, we compare different methods of momentum tuning on the
optimal number groups for ImageNet8 on CPU-L, which was 4 (recall 
Figure~\ref{tab:e2e_imagenet8}(c)).  We fix that number of groups and
(i) set momentum to $0.9$ (as reported in
AlexNet~\cite{Krizhevsky:2012:NIPS}); (ii) use the momentum tuned for
a synchronous system; (iii) tune the momentum using \oursystem's
optimizer for 4 compute groups.  As we see in
Figure~\ref{fig:lesion_momentum}, tuning for the right amount of
asynchrony is important: if \oursystem did not tune momentum, it would
be 1.5$\times$ slower. Further experiments show 
that tuning momentum can yield speedups of 2$\times$.
On TensorFlow, we observe similar speedups: 
When momentum is set to $0.9$, synchronous training is faster. 
When, however, we perform momentum tuning, the asynchronous configuration wins with its performance relative to sync improved by a factor of $2.4\times$ 
This both verifies our expectations for this
experiment and provides further support for our theory in
Section~\ref{sec:SE_Model}.

\paragraph*{Discussion: Other Models}
We find that these tradeoffs are impactful when applied to other
models. We find that for Recurrent Neural Network models and LSTM
models (e.g., \cite{Graves:2013}), the same choices affect
performance---for example, choosing a completely synchronous or
asynchronous configuration can be up to 2$\times$ slower than the
optimal configuration.  
This could imply speedups for applications in which RNNs are widely
used, such as handwriting, speech recognition, and general sequence or time-series
data. 

\subsubsection{Optimizer}\label{sec:exp-optimizer}

We validate that our optimizer outperforms
state-of-the-art Bayesian optimization algorithms.
We compare our optimizer with the optimizer
proposed by Snoek et al.~\cite{Snoek:NIPS:2012}.
We measure both the number of configurations and the
total number of epochs that the Bayesian optimizer needs to achieve an
accuracy within $1\%$ of the highest accuracy \oursystem achieves. 
In our experiments, the Bayesian optimizer
  never discovers a configuration which outperforms the
  configuration  \oursystem obtains by grid search. 
  We found that the Bayesian
optimizer takes on average $12$ runs to find a near-optimal strategy, which on average
is $6\times$ more epochs than just running that strategy. Because of
this search overhead it was not feasible to use the full ImageNet 1000 dataset
so we used ImageNet8 (whereas recall from Figure~\ref{fig:flagship}
that \oursystem had an overhead of only 10\% on ImageNet 1000).
 We used the GPU-S cluster.  Typically Bayesian
optimizers can amortize this cost by running in parallel, but
here that is not possible as the parameters depend on the hardware
configuration and so the optimizer needs complete access to the
entire cluster.

\section{Related Work}\label{sec:RelatedWork}

{\bf Single Node.} Optimizing CNN performance has become a well-studied problem in recent years. Popular libraries include Caffe~\cite{Jia:2014:arXiv}, cuDNN~\cite{Chetlur:2014:ArXvi}, TensorFlow~\cite{tensorflow}, Theano~\cite{Bergstra:2010:Scipy}, and Torch. 
To compute convolutions, many of these frameworks use lowering, an idea proposed by Chellapilla et al.~\cite{Chellapilla:2006:ICFHR} that takes advantage of highly-optimized BLAS libraries. Our work follows from this line of research and demonstrates how to optimize lowering for CPUs in order to build a system which is robust to different types of hardware.

{\bf Distributed Deep Learning.} Distributed systems for Deep Learning popular, with SINGA~\cite{Wang:2015:TR}, MXNET~\cite{chen15mxnet}, FireCaffe~\cite{iandola15firecaf}, SparkNet~\cite{moritz15sparknet}, DL4J, 
DistBelief~\cite{Dean:2012:NIPS}, and Project Adam~\cite{Chilimbi:2014:OSDI} selecting different execution strategies and other optimizations. Our study shows a combined tradeoff space on the union of all these techniques. We did this by decoupling the hardware and statistical efficiency for each technique and optimizing them separately.
Our work is the first to provide a theoretical characterization for statistical efficiency and to show that hyper-parameters need to be tuned to compensate for asynchrony.

The idea of decoupling the number of iterations and time per iteration and analyzing each separately is not new for distributed CNN systems. MXNet reported hardware efficiency and statistical efficiency separately.
SINGA
went deeper into the tradeoff, identifying the compute group size as a tunable parameter. They advocate combining both 
synchronous and asynchronous training and offer a flexible training architecture which enables trading off the convergence rate with the time per iteration in order to to minimize training time. However, while SINGA identifies this tradeoff and provides experimental evidence of its importance similar to the curves we showed, 
the user still needs to manually choose a configuration.

SparkNet also separated the time per iteration and number of iterations by building models of each. They did not explore the same tradeoff of machines per group, but rather a similar tradeoff related to staleness. Because SparkNet uses a MapReduce framework they implement \textit{model averaging}. Within this technique they explore, in isolation, how the staleness (their $\tau$ parameter) impacts the number of iterations to convergence and the time per iteration. Their hardware efficiency model was measured (both network speed and compute time) and their statistical efficiency model was also empirical (they varied staleness and measured the statistical efficiency penalty).

\section{Conclusions} \label{sec:conclusion}

We described the first explicit study of the tradeoff space for deep
learning systems, a popular, high-value type of industrially deployed
learning systems. We identified critical issues in how one maps layers
to devices and are the first to systematically study widely used
techniques like asynchrony. We designed a new optimizer and showed that it
has excellent end-to-end performance and is independent of our
particular implementation substructure.
We are collaborating with a major chip manufacturer to apply asynchrony-aware tuning and compute groups onto new platforms of much larger scale.

\bibliographystyle{IEEEtran}
\bibliography{omnivore_full}
%
%
%

\pagebreak

\iflongversion
\section{Acknowledgements}

The authors would like to thank Chris Aberger and the rest of the Hazy Group for their feedback and help, as well as HyoukJoong Lee, Nadathur Rajagopalan Satish, Peter Bailis and Benjamin Recht for their thoughtful comments.
We would like to thank Intel, Toshiba and the Moore Foundation for support along with DARPA through MEMEX (FA8750-14-2-0240), SIMPLEX (N66001-15-C-4043), and XDATA (FA8750-12-2-0335) programs, and the Office of Naval Research (N000141210041 and N000141310129). Any opinions, findings, and conclusions or recommendations expressed in this material are those of the authors and do not necessarily reflect the views of DARPA, ONR, or the U.S. government.

\begin{appendices}

\vspace{0.2in}
\section*{Appendix}

The appendix sections mirror the section structure of the main paper, carrying  corresponding supplementary material.

Appendix~\ref{sec:app:intro} includes supplementary information for the Introduction, most importantly more discussion on the CPU vs GPU debate.

Appendix~\ref{sec:app:background} includes a discussion of CNN trends.

Appendix~\ref{sec:app:singlenode} includes a full tradeoff analysis on lowering strategies---first reported in a workshop paper~\cite{hadjis15cct}---that shows that the strategy we use in this paper works best for most CNN kernels.
It also includes a more comprehensive treatment of batching and data parallelism for the GPU.
Finally, it shows that FLOPS-proportionality holds for all machines in Table~\ref{tab:machine}.

Appendix~\ref{sec:app:distributed} includes a full discussion of the tradeoff space, including terminology, a survey (Appendix~\ref{sec:app:distributed:existing}) of many diverse distributed CNN systems~\cite{Chilimbi:2014:OSDI, iandola15firecaf,chen15mxnet, Dean:2012:NIPS, Wang:2015:TR, moritz15sparknet}
and shows that they can be mapped to points within our trade-off space.
Appendix~\ref{sec:app:distributed:hemodel} includes a proof for the hardware efficiency model we use, describes how to measure necessary quantities from the system, and shows that it works across a range of datasets.

Appendix~\ref{sec:app:optimizer} gives more details on our distributed optimizer:
Appendix~\ref{sec:app:optimizer:batchsize} includes discussion and experiments on selecting the batch size;
Appendix~\ref{sec:app:optimizer:pm} describes other physical mappings we studied and related analysis;
Appendix~\ref{sec:app:optimizer:coldstart} discusses in more detail the {\em cold-start period} of optimization.

Appendix~\ref{sec:app:distexperiments} includes the full setup of our distributed experiments:
Appendix~\ref{sec:app:distexperiments:smallcluster} contains full details for our small cluster experiments;
Appendix~\ref{sec:app:distexperiments:detailed} contains a full trade-off space analysis for ImageNet and CIFAR10;
Appendix~\ref{sec:app:distexperiments:largecluster} shows that tuning momentum can yield speedups of 2$\times$;
Appendices~\ref{sec:app:distexperiments:largecluster} and \ref{sec:app:distexperiments:scalability} contain full details on our larger cluster experiments and discuss the impact of optimizing the hyper-parameters of competitor systems;
Appendix~\ref{sec:app:distexperiments:endtoend} includes details on the end-to-end experiment on ImageNet1000, as well as the choices of our optimizer.
Appendix~\ref{sec:app:distexperiments:rnn} describes preliminary experiments on RNNs;
Appendix~\ref{sec:app:distexperiments:standardschedules} compares our optimizer to standard schedules and; Appendix~\ref{sec:app:distexperiments:bayesian} compares to Bayesian hyper-parameter optimization.

Appendix~\ref{sec:app:tensorflow}, includes our TensorFlow results. We show that on Inception-v3, momentum tuning can be result-changing:
when momentum is set to $0.9$, synchronous training is faster. 
When, however, we perform momentum tuning, the asynchronous configuration wins with its performance relative to sync improved by a factor of $2.4\times$. We also study the effect of compute groups.

Appendix~\ref{sec:app:tco} gives a total cost of ownership analysis.

Appendix~\ref{sec:app:conclusions} includes supplementary discussion from the Conclusions section.

\section{Appendix for Introduction (Section~\ref{sec:intro})}
\label{sec:app:intro}

\begin{table}[]
\centering
\caption{Tradeoff space when designing distributed deep learning systems.}
\label{table:full_tradeoff}
\begin{tabular}{|l|l|}
\hline
\textbf{Tradeoff}                                                                        & \textbf{Some Examples}                                                                                                                               \\ \hline
Single Node Hardware                                                                     & \begin{tabular}[c]{@{}l@{}}CPUs, GPUs, both together\end{tabular}                                                                                  \\ \hline
Type of Parallelism                                                                      & Model, Data                                                                                                                               \\ \hline
\begin{tabular}[c]{@{}l@{}}Batch size\end{tabular} & \begin{tabular}[c]{@{}l@{}}Large (few accurate updates), \\Small (many parallel updates)\end{tabular}                                                                 \\ \hline
\begin{tabular}[c]{@{}l@{}}Batch allocation \\ within node\end{tabular}                  & \begin{tabular}[c]{@{}l@{}}1 CPU or GPU per batch,\\ many CPUs or GPUs per batch\end{tabular} \\ \hline
\begin{tabular}[c]{@{}l@{}}Batch allocation\\ across nodes\end{tabular}                  & \begin{tabular}[c]{@{}l@{}}1 batch across all machines,\\ 1 parallel batch per machine,\\groups of machines per batch\end{tabular}                                                               \\ \hline
\begin{tabular}[c]{@{}l@{}}Combining Model\\ Replicas\end{tabular}                       & \begin{tabular}[c]{@{}l@{}}Atomic gradient updates,\\Model averaging, Ensembles,\\Race conditions (Hogwild!)\end{tabular}                                                               \\ \hline
Server Architecture                                                                       & \begin{tabular}[c]{@{}l@{}}Separate parameter/compute,\\ Merged parameter/compute\end{tabular}                                                     \\ \hline
Network Architecture                                                                      & \begin{tabular}[c]{@{}l@{}}Model Size (impacts\\communication),\\ Model Depth (impacts memory\\/ batch size)\end{tabular}                                                                  \\ \hline
Optimization Algorithm                                                                   & SGD, Adagrad, momentum\\ \hline
\begin{tabular}[c]{@{}l@{}}Hyperparameter\\ Optimization\end{tabular}                    & \begin{tabular}[c]{@{}l@{}}Grid/Random search, Bayesian,\\Plateau (e.g., ResNet), Decay\\ Schedule (e.g., Inception-v4)\end{tabular}                                              \\ \hline
\end{tabular}
\vspace{-1.0em}
\end{table}

\paragraph*{Tradeoff Table}
While many distributed deep learning systems exist, each of these makes design decisions suited for
a particular type of (1) compute cluster, (2) deep learning model and (3) dataset, although these same
decisions may not work for other problem settings or hardware resources.  This is because deep learning
is known to be complex both from a computational and machine learning perspective,
and designing a highly efficient and scalable deep learning engine involves a number of interrelated design decisions. 
Table~\ref{table:full_tradeoff} shows a number of factors to consider when designing distributed deep learning systems. Given a fixed number of machines, a number of tradeoffs exist from how to use hardware on each node, to how to allocate batches of data to machines, to the type of deep learning model to use in order to minimize communication. In addition, these decisions impact one another -- for instance the batch size influences the number of machines which can be used to effectively parallelize within a batch and therefore influences the total number of parallel gradients being computed to update the model.
Our work, which is a study, demystifies these
factors by identifying the key tradeoffs which underlie all
design decisions and quantifying the impact of those tradeoffs
experimentally.

\subsubsection*{Contribution 1: Single-Node Optimization}

Even focusing on just a single node, there has been a long debate
about CPUs vs GPUs for deep learning.
GPUs are popular for CNN systems because of the high throughput they provide.
Modern GPUs offer between 1.2 TFLOPS (NVIDIA GRID K520, per GPU) and 8 TFLOPS (NVIDIA Titan Z).
However, GPUs contain smaller off-chip memories than CPUs, and GPUs are connected to host memory by a slow PCI-e interconnect.
On the other hand, Microsoft's Project Adam argues that CPUs can deliver more cost-effective performance~\cite{Chilimbi:2014:OSDI}.\footnote{{\scriptsize\url{http://www.wired.com/2014/07/microsoft-adam/}}} This debate is only going to get more interesting, as modern GPUs offer high-speed interconnect with host memory\footnote{{\scriptsize\url{http://nvidianews.nvidia.com/news/nvidia-launches-world-s-first-high-speed-gpu-interconnect-helping-pave-the-way-to-exascale-computing}}} while Intel's current Haswell CPU can achieve 1.4 TFLOPS on a single chip.\footnote{\scriptsize{Xeon E5-2698 v3, \url{http://ark.intel.com/products/81060}}}. Moreover, SIMD parallelism has doubled in each of the last four Intel CPU generations and is likely to continue.\footnote{\scriptsize  SIMD scales linearly in power and area (whereas   frequency scaling  is cubic)   \url{http://parasol.tamu.edu/lcpc2014/keynote-tian.pdf}.}
Our work is the first to conduct a systematic study to understand the relative
performance of CPUs and GPUs for deep learning.

\section{Appendix for Background  (Section~\ref{sec:background})}
\label{sec:app:background}

\subsection{CNN Computation}

\paragraph*{AlexNet FLOPS}
We approximate the FLOPs (\# floating point operations) in AlexNet by the sum of all the GEMM operations with batch size 256. Specifically, we add 1 GEMM in the forward pass for each Conv and FC layer, plus two GEMMs in the backward pass for each Conv and FC layer (although Conv1 backward has only 1 GEMM because no gradient is needed with respect to the data).

\paragraph*{Terminology}
This section introduces model and data parallelism as two techniques to parallelize CNNs. For a full description of these concepts, see the Terminology in Appendix~\ref{sec:terminology}.

\paragraph*{CNN Trends}
This section viewed CNNs as two phases, Conv and FC. Recent CNNs, e.g., Residual Networks (ResNets) and Inception Networks, can also be categorized into this partitioning. For instance early Inception variants contained multiple FC layers at different parts of the network, but from a computational point of view these are all considered to be part of the FC phase.

In particular, CNNs have undergone a number of changes in the past few years\footnote{\scriptsize{\url{http://cs231n.github.io/convolutional-networks/\#case}}}. We summarize a few here:
\begin{itemize}
  \item Multiple FC layers replaced with average pooling, leaving a single fully-connected layer (for the softmax). This leads to a reduction in the overall model size (e.g., 60 million parameters for AlexNet compared to 4 million for GoogleNet)
  \item Increase in network depth (e.g., AlexNet with 5 conv layers, compared to ResNets\footnote{\scriptsize{\url{https://github.com/KaimingHe/deep-residual-networks}}} with $>150$). This increases memory requirements of networks, and makes multi-device training necessary.
\end{itemize}

As we will see, the optimizer presented in this paper considers the impacts of each of these points when making decisions for physical mapping and execution strategy.\\

\subsection{Problem Definition}

\paragraph*{Scope of Work}
Our work does not focus on improving machine learning techniques but rather studies systems tradeoffs in order to build an optimizer that is robust to the most widely used networks/algorithms. Our study uses the SGD algorithm due to its popularity, although our optimizer applies to other algorithms as well. Similarly we do not modify the CNN architecture but assume that this is provided by the user.

\paragraph*{Terminology}
The physical mapping maps the layer computation to vertices in $\mathcal{G}$.
Vertices in $\mathcal{G}$ may contain other vertices, e.g., GPUs or CPU cores within a machine. 
Section~\ref{sec:Omnivore} first studies how to map the CNN to hardware within a machine, and concludes
that with proper optimization only the throughput of each vertex matters, not its underlying hardware.
Section~\ref{sec:Distributed} then studies how to map the CNN to all of $\mathcal{G}$, i.e. across machines.

\section{Appendix for Single-Node Tradeoffs (Section~\ref{sec:Omnivore})}
\label{sec:app:singlenode}

This section describes the optimizer's physical plan (how to map the CNN to hardware) at the level of a single machine.
Given a machine containing devices of various throughput (CPUs, GPUs), our goal is to run the CNN computation as quickly as possible. We do this by identifying two tradeoffs.
Our first tradeoff introduces a data batching technique which trades off memory footprint for compute time. We demonstrate that this tradeoff gives a $>5\times$ CPU speedup over existing systems. With this optimization now both CPUs and GPUs give throughput proportional to the FLOPS offered by the device, and our second tradeoff partitions the CNN computation across both the CPU and GPU to combine their FLOPS -- a known HPC trick, but one which has not been applied to CNNs.

\subsection{Convolutional Layer Computation}

A {\em 3D convolution} consumes a pair of order $3$ tensors--the data $D \in \mathbb{R}^{n\times n \times d_{in}}$ and the kernel $K \in \mathbb{R}^{k \times k \times d_{in}}$. For example, if $D$ is a color image with 3 (RGB) color channels, $d_{in}=3$. In AlexNet~\cite{Krizhevsky:2012:NIPS}, $n \in [13,227]$, $k \in [3,11]$, and $d_{in} \in [3, 384]$, The output is a 2D matrix $R \in \mathbb{R}^{m  \times m}$ where $m = n-k+1$ and each element $R_{x,y}$ is defined as:
\begin{equation}
R_{x,y} = \sum_{d'=0}^{d_{in}-1} \sum_{x'=0}^{k-1} \sum_{y'=0}^{k-1}
D_{x-\frac{k}{2}+x',y-\frac{k}{2}+y',d'} K_{x',y',d'}
\label{eq:conv}\end{equation}
The kernel also supports parameters called \textit{padding} and \textit{stride}, which affect the size of $m$. For details on stride and padding see \url{http://cs231n.github.io/convolutional-networks/\#conv}.

This is the standard image 3D discrete convolution.
A \textit{convolution layer} in the CNN contains a number of kernels $\{K_j\}$, not just one, where we call $d_{out}=|{K_j}|$ the {\em number of output channels}.
These kernels $\{K_j\}$ constitute the \textit{model} of the convolutional layer, and
the reason for computing multiple kernels rather than just 1 in the convolutional layer is to
a more powerful machine learning model.
The convolutional layer takes as input the 3D data tensor $D$ and performs $d_{out}$ 3D convolutions, one per $\{K_j\}$, such that the output of the convolutional layer is now not a 2D matrix $R$ but a 3D tensor $R \in \mathbb{R}^{m  \times m \times d_{out}}$. Similarly the model of the CNN can be viewed as a 4D tensor $K \in \mathbb{R}^{k \times k \times d_{in} \times d_{out}}$.

Finally, recall that often rather than process a single data example the CNN processes a batch of $b$ examples simultaneously.
The motivation for doing this is that gradient computations during learning are less noisy.
Therefore in the most general case, the input $D$ to a 
convolutional layer is not 1 but $b$ 3D data tensors, or equivalently a 4D data tensor $D \in \mathbb{R}^{n\times n \times d_{in} \times b}$.
The model is unchanged, but the convolutional layer now performs the $d_{out}$ 3D convolutions on each example in the batch, i.e.
the batched convolution layer performs $b\cdot d_{out}$ 3D convolutions.
The output of the convolutional layer is therefore also a 4D tensor, $R \in \mathbb{R}^{m  \times m \times d_{out} \times b}$.

To summarize, a convolutional layer accepts as input a 4D data tensor $D \in \mathbb{R}^{n\times n \times d \times b}$,
performs a total of $b\cdot d_{out}$ 3D discrete convolutions using $D$ and its model  $K \in \mathbb{R}^{k \times k \times d_{in} \times d_{out}}$,
and outputs a 4D output data tensor $R \in \mathbb{R}^{m  \times m \times d_{out} \times b}$.
The full formula is:
\begin{equation*}
R_{x,y,z,w} = \sum_{d'=0}^{d_{in}-1} \sum_{x'=0}^{k-1} \sum_{y'=0}^{k-1}
D_{x-\frac{k}{2}+x',y-\frac{k}{2}+y',d',w} K_{x',y',d',z}
\end{equation*}

Many implementations of this convolutional layer exist. Like most other HPC kernels, a straightforward implementation of this
operation is suboptimal. Optimized implementations include directly computing
the convolutions, as in cuda-convnet2,\footnote{\scriptsize{\url{https://code.google.com/p/cuda-convnet2/}}}, computing the convolution as a discrete Fourier transform as in~\cite{Vasilache:2014:arXiv}, or implementing the convolution as a matrix multiply, as in Caffe~\cite{Jia:2014:arXiv} or cuDNN~\cite{Chetlur:2014:ArXvi}.

While the studies in these papers conclude that different strategies perform best for different kernel sizes, cuDNN~\cite{Chetlur:2014:ArXvi} demonstrates that the third technique of performing convolution as a matrix multiplication is versatile and fast for a range of size parameters, as matrix multiplication kernels are often highly optimized.

In order for the convolution to be carried out as a matrix multiplication, an initial reformatting phase called \textit{lowering} is required to put the data and kernel into the correct format, which we discuss in the next subsection of this appendix.

\subsubsection{Convolution by Lowering and GEMM}

{\em Lowering} followed by a general dense matrix multiplication (GEMM) is a popular way to implement the convolution operation.
Figure~\ref{fig:conv_new_block} shows the three logical steps in the lowering process:
(1) lowering, which transforms 4D tensors $D$ and $K$ into 2D matrices $\hat{D}$ and $\hat{K}$;
(2) {\em matrix multiply (GEMM)}, in which we multiply $\hat{D}\hat{K}$ to get the result $\hat{R}$; and
(3) {\em lifting}, which transforms $\hat{R}$ back to a tensor representation of $R$.

\begin{description}
\item {\bf Lowering Phase} in which we construct the matrix $\hat{D}$ and
  $\hat{K}$. A value of $D$ will appear more than once in the lowered
  matrices.

\item {\bf Multiply Phase (GEMM)} in which we multiply $\hat{D}$ and
  $\hat{K}$ to create $\hat{R}=\hat{D}\hat{K}$.

\item {\bf Lifting Phase} in which we map $\hat{R}$ back to $R$.
\end{description}

Three techniques exist to perform this process, each corresponding to a different way to group the sum in Equation~\ref{eq:conv}. Each of these techniques requires replicating the data or the kernel in order to allow the convolution to be implemented as a GEMM, but the amount of replication and the size of the GEMM depend on whether the replication happens in the lowering phase, lifting phase, or partly in each. The tradeoff is studied in detail by~\cite{hadjis15cct}, which concludes that the best choice is determined entirely by the ratio $d_{in}/d_{out}$ (the ratio of the number of input channels to the number of output channels of the convolution), and that for modern CNNs these ratios suggest that the data replication should be done during the lowering phase. Therefore in the lowering used by this work, there are two components to convolution: lowering (which requires replication of data), and the GEMM. The lifting does not require any memory copies or computation in the optimal technique. Note also that in this technique, the kernel does not require any replication, only the data. CNNs are continuously evolving however, and so it is possible that future CNN architectures will benefit from other lowering strategies. For a full study of the lowering tradeoff, refer to~\cite{hadjis15cct}.

The amount of data replication required by the lowering in this work is $m^{2}k^{2}/n^{2}$, where $m < n$ and $m$ depends on the stride and padding of the convolution. The replication can be on the order of 1 to 2 orders of magnitude (i.e. $10-100\times$ more data). In turn, this blowup in the data size requires more memory and computation in step 2 (GEMM). The benefit however is that a sparse computation has become dense, which is important for hardware implementations because the direct computation of the 3D convolution is usually memory bandwidth-bound (due to the small convolution window size, $k \in [3,11]$). A GEMM implementation on the other hand, while performing more computation as a result of lowering, receives hardware acceleration which eclipses the increase in data size.

\subsection{Batching and Data Parallelism}

\subsubsection{Batching Analysis}

\textit{Batching} is the implementation tradeoff that arises between the CPU and GPU as a result of available off-chip memory. It concerns how many images of the batch to process in parallel by the convolution layer. 
Recall that $b_{p}$ images are processed in parallel, where
where $1 \leq b_{p} \leq b$. $b$ is the batch size, i.e. the total number of images
that need to be processed.
The value of $b_{p}$ (how much to batch the convolution computation) is determined by how many lowered images can be stored in off-chip memory.

Modern GPU cards cannot store entire batches of lowered data into off-chip memory and implementations of CNNs on GPUs perform lowering and GEMM serially on one or few images at a time until all $b$ have been processed, i.e. $b_{p} = 1$.
On the CPU, off-chip memory is larger which allows for batching techniques that perform lowering and GEMM on all images in parallel and therefore allow CPU caches and vector instructions to be fully utilized. This tradeoff is continuing to evolve however, as newer GPU cards contain more off-chip memory (e.g., 12 GB in the Titan X), and also use new implementations which perform lowering and GEMM without having to materialize the intermediate lowered representation, as described by~\cite{Chetlur:2014:ArXvi}. Therefore we believe that this tradeoff will grow in importance.

For example $b_{p} = 1$ in Caffe~\cite{Jia:2014:arXiv}:
in order to process $b$ images in a batch, Caffe performs convolution on one image at a time, i.e. lowering/GEMM 
is done serially for each image in the batch (lower one image, run GEMM, lower
the next image, etc.) This has the smallest possible memory footprint as it only maintains the 
lowered matrix of a single image in memory, and is the preferred strategy for 
devices with limited memory.
Figure~\ref{fig:parallel}(c) showed that the memory footprint for the convolution is
directly proportional to $b_{p}$.

Computationally however, Figure~\ref{fig:parallel} (b) showed that $b_{p}=1$ suffers from lower hardware
utilization. 
This figure was run for the GEMM in the Conv2 layer of Alexnet (although we observed similar trends
for other Conv layers). Specifically the matrix $\hat{D}$ is ``$B$'' in the GEMM operation $A \times B$, and increasing $b_{p}$
increased the number of columns in $\hat{D}$.

In Figure~\ref{fig:parallel} (b) we fix the number of threads to 8,
vary $b_{p}$, and plot the speedup normalized to $b_{p} = 1$.
Increasing $b_{p}$ reduces the number of total GEMM calls, and this gives a $2\times$ overall
speedup for $b_{p} = 256$ compared to $b_{p} = 1$.
This is because a small $b_{p}$ means that $\hat{D}$ becomes thinner.
For thinner matrices, possible partition sizes of the underlying GEMM algorithm are
smaller and so the kernel cannot run optimally, for example the L2 and L3
caches cannot be filled during blocking optimizations. As a result $b_{p}=1$ is
more likely memory-bandwidth-bound than higher batch sizes (and this
phenomenon is likely more severe when the GEMM kernel is
executed with multiple threads.)
Also note that Figure~\ref{fig:parallel} was run on a CPU machine with 8 physical cores.
Regarding Figure~\ref{fig:parallel} (a),
the reason that 16 threads was slightly slower than 8 is that 
we hit the memory bandwidth bottleneck.

For modern CPUs, memory is large and so $b = b_{p}$. We use $b = b_{p}$ for the remaining
CPU experiments in this section.

\subsubsection{Data Parallelism in non-GEMM Kernels}

On the CPU, recall that the batching technique above makes a single matrix that is $b$ times (full batch size) larger
than it would be for a single image, and then performs a single GEMM on this large matrix.
A related strategy is to split a batch into multiple partitions, and process
each partition in parallel using a separate thread.

For GEMM, processing an entire batch of size $b$ with $n$
threads is equivalent to partitioning the batch into $p$ partitions of size $b/p$ with
$n/p$ threads used in each GEMM. 
For example, batching $b$ images and performing
a single GEMM with 8 threads is equivalent to creating 8 matrices, each with $b/8$ images, and
performing 8 parallel GEMM kernels with 1 thread each.
These are equivalent as this is exactly how BLAS
parallelizes GEMM: by partitioning partition columns of B
in $A \times B$ and allocating 1 thread per partition.

While partitioning and then performing the GEMM kernel is the same as simply performing
the GEMM kernel, this is not true for other kernels which are not multi-threaded
For non-GEMM kernels such as lowering, or other layers such as pooling,
the second technique of partitioning the batch and processing each partition
using a separate thread gives significant speedups (this is simply data parallelism)
across all cores. For example, it can be used to lower images in parallel using
all cores by assigning a subset of the images in the batch to each core.

\begin{figure}
\centering
\includegraphics[width=0.5\textwidth]{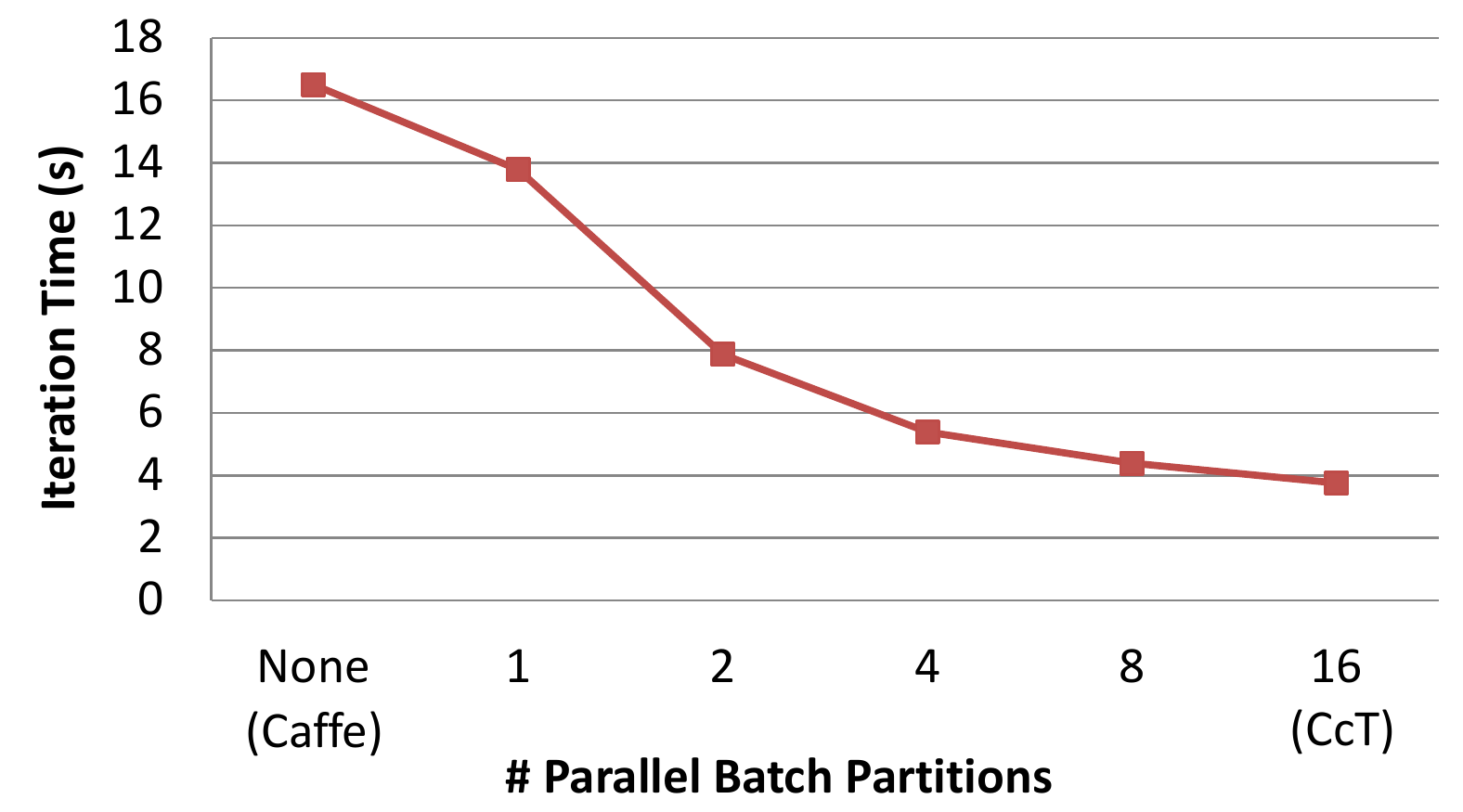}
\caption{The impact of data parallelism on the end-to-end
execution time of CaffeNet, run with 256 images per
mini-batch on an Amazon EC2 c4.4xlarge instance.}
\vspace{-1.5em}
\label{fig:CcT_Batch}
\end{figure}

Figure~\ref{fig:CcT_Batch} shows the impact
of data parallelism on a full end-to-end CaffeNet on the
EC2 c4.4xlarge instance with 8 physical cores (CPU only). The batch
size used is 256 images and the horizontal axis represents
into how many parallel partitions we partitioned these 256
images for each layer of the CNN. I.e. the horizontal axis is the number of threads used
for data parallelism. Note that the GEMM kernel is always parallelized (OpenBLAS was used)
and uses the maximum number of threads (16).

``None'' indicates the default Caffe implementation.
For all layers, each image is processed serially.
For example, one image is lowered, then the convolution GEMM
is performed, and then the next image is lowered, etc.
The only multi-threaded kernels are the BLAS GEMM kernels used in the convolution and FC.

``1'' is identical to the Caffe implementation except that lowering is 
first done for all images (serially, i.e. one image lowered at a time), and then
a single, large GEMM is done for the convolution.
All other layers are the same.

For all other number of parallel partitions
p, the 256 images were equally split into p partitions.
For example if p = 2, two partitions of size 128 images each were created.
Then, threads processed each partition in parallel, one thread per partition.
For example if p = 2, lowering was done with two threads, each lowering
128 images.
Following the lowering, a GEMM was performed for each partition. The total
number of threads used for these GEMM kernels was always 16, i.e. the 
GEMM is performed
on each partition with 16/p threads per
GEMM.

Figure~\ref{fig:CcT_Batch} (None vs 1) shows that batching the GEMM kernels (one large GEMM as opposed to 256 smaller ones) saves $\sim 2.2$s of convolution time. Then, data parallelism provides another $\sim 10$ s of reduction. The final time is then $4$s, where $\sim3$s is spend in convolution layers. Therefore the batching of the GEMM made the convolution roughly $\sim2\times$ faster (in the optimized execution), and the remaining speedups were due to data parallelism. 

Finally, studying more closely the speedups from data parallelism, the time reduction from 14 seconds to 4 seconds from data parallelism was roughly $80\%$ due to speeding up the lowering, and $20\%$ due to speeding up the other layers. I.e. the final iteration time would be $\sim6$s if data parallelism was only used for the Conv layers. The remaining $20\%$ is for data parallelism in pooling, normalization, dropout  and ReLU layers. Fully-Connected layers are simply a GEMM which always uses 16 threads, so data parallelism and model parallelism do not apply on the CPU, although as we also saw previously this GEMM can be made slightly faster by using 8 threads instead for the FC layers (because there are 8 physical cores).

Overall, 
batching combined with data parallelism gives more than a $4\times$ speedup
end-to-end
when running Caffe's AlexNet on 8 physical cores. Importantly, this end-to-end speed is now
proportional to the peak throughput of the CPU.\\

In summary, we show two sources of speedup on the CPU. First, by batching the lowering
and GEMM, we perform a single GEMM which is $b\times$ larger, as opposed to $b$ smaller GEMMs, which
as described above has better hardware utilization on the CPU. Second, we apply the
batch partition (data parallel) technique above to parallelize non-GEMM kernels such as lowering.
These optimizations are possible for the CPU because it has more memory to store the lowered
matrices. As a result the CPU performance is proportional to the device FLOPS, which allows
partitioning computation across both the CPU and GPU proportional to the device throughput.

\subsection{Device Throughput}
\label{sec:app:singlenode::throughput}

\paragraph*{FLOPS Experiments}
Figure~\ref{fig:flops:prop} showed throughput for the CNN when using
Caffe, Omnivore and also for reference a single-precision GEMM kernel.
For the CPU the GEMM experiment used OpenBLAS and matrices of size $16384 \times 16384$.
For the GPU GEMM we used a GEMM kernel from NVIDIA's CUDA examples.
For Caffe and Omnivore we focus on only the Convolution layers, specifically the time
to run forwards and backwards on all 5 layers of CaffeNet.

\paragraph*{FLOPS calculations}

The c4.4xlarge instance contains a single-socket Haswell CPU with 8 physical cores.
The c4.4xlarge instance CPU FLOPS are calculated as: $8$ physical cores $\times$ $2.9$ GHz $\times 32 = 0.742$ TFLOPs, where 32 is the single-precision Haswell instructions per cycle (8-float SIMD $\times$ 2 FMA per cycle, and FMA is fused-multiply-add).

Each g2.2xlarge instance provides a \textit{single} Grid K520 GPU, i.e. $1536$ cores $\times 800$ MHz = $1.23$ TFLOPS (the Grid K520 contains a total of 3072 cores, 1536/GPU). 

The c4.8xlarge instance contains a dual-socket Haswell CPU with 18 physical cores.
The FLOPS are calculated as: $18$ physical cores $\times$ $2.9$ GHz $\times 32 = 1.670$ TFLOPs.

\subsection{FLOPS-Proportional Scheduling}

Given that both CPU and GPU speeds are now proportional to the device FLOPS,
we next consider whether the CPU and GPU can be used
simultaneously to process each layer.
We do this using data parallelism
(batch is partitioned, model is replicated)
for all layers in the convolution phase, which is compute-bound
and has small data.
The tradeoff is what fraction of the batch to give to each device.
We select the simple but optimal choice 
that a device should process a fraction $p$ of
the input where $p$ is the proportion of total FLOPS which that device contributes.
e.g., if a CPU provides 1 TFLOPS and a GPU 4 TFLOPS,
$1/5$ of the batch is processed by the CPU for an ideal speedup of $20\%$ over the GPU alone.

\subsection{Single-Node Experiments}
\label{sec:app:singlenode::experiments}

Omnivore matches Caffe's output on each layer. It accepts the same input files as Caffe and produces the same outputs.
Our experiments compare against Caffe and use the CaffeNet\footnote{{\scriptsize\url{https://github.com/BVLC/caffe/tree/master/models/bvlc_reference_caffenet}}} CNN, which is Caffe's implementation of the popular AlexNet (the default architecture for benchmarking),
as well as the ImageNet dataset.

Both systems take as input the same network configuration files that Caffe provides.  We remove grouping for convolution layers because the full AlexNet fits in the memory of a single  K520 (g2.2xlarge) GPU.
We use the same external libraries for both Caffe and Omnivore: GCC-4.8, CUDA-7.5, and OpenBLAS.
For Caffe we report both cuDNN v3 and v4.

We built the same timer into Caffe and Omnivore (measuring wall-clock time, \texttt{clock\_gettime} in C). We run for 50 iterations and omit the first 10 in case there are disk or other delays in the first few iterations not representative of steady-state execution. 
Beyond these first 10 iterations we noticed that all iterations were consistent in terms of time for both tools and had a coefficient of variation less than 5\%.

We also ran Tensorflow using the same protocol as above (40 iterations, burn-in of 10, identical network).

\begin{figure}[t!] \centering
  \includegraphics[width=0.5\textwidth]{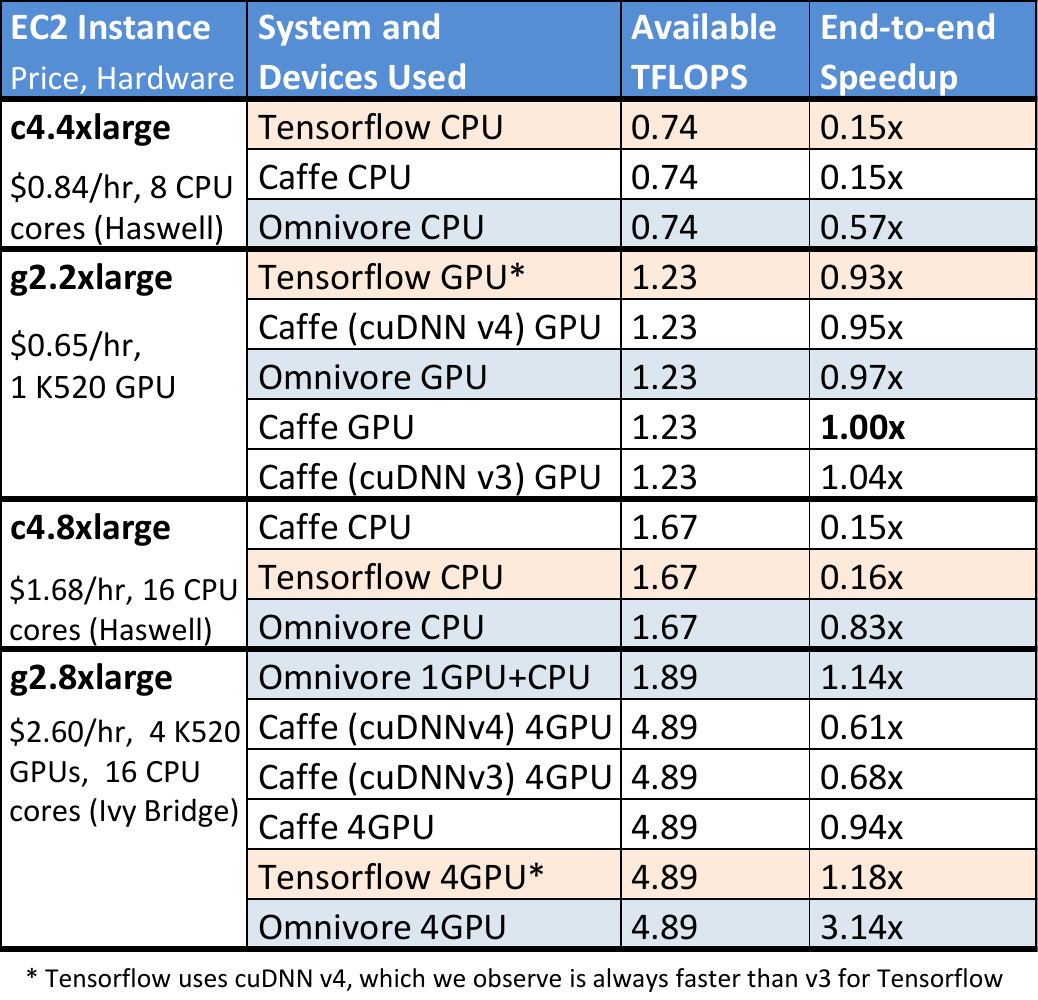}
  \caption{End-to-end performance comparison across EC2 machines on
    CaffeNet. All numbers are normalized as the speedup over running Caffe's
    GPU version on g2.2xlarge instance (\$0.65/hour).} \label{fig:e2e_full}
\vspace{-1.0em}
\end{figure}

\subsubsection{End-to-end Performance}

Figure~\ref{fig:e2e_full} shows the results of running Omnivore and Caffe  on various EC2 instances. 
Given that Omnivore and Caffe generate the same outputs, we concentrate on throughput. We ran both Omnivore and Caffe on each EC2 instance for 50 iterations, and counted the last 40.
On the c4.4xlarge CPU instance  Omnivore outperforms Caffe by nearly $4\times$ due to batching and data parallelism.  On the g2.2xlarge GPU instance Omnivore and Caffe achieve the same speed (within $5\%$). Note that the c4.4xlarge instance offers $60\%$ of the FLOPS of the g2.2xlarge instance, and the ratio of Omnivore speeds on these instances is $59\%$, i.e. Omnivore delivers speed proportional to the FLOPS. On the two-socket c4.8xlarge CPU instance Omnivore is now $5.5\times$ faster than Caffe. Caffe does not speed up given the additional cores, but Omnivore does. The speedup is not linear because the extra cores are distributed across sockets and not all layers are compute-bound.  However, these results show that  given similar device throughput, CPU CNN performance is not an order of magnitude slower than GPU performance as the literature often reports and as is the case in Caffe. Lastly we compared Omnivore to Caffe on a 4-socket, 56-core Haswell (non-EC2) CPU machine, and Omnivore is $13\times$ faster than Caffe.

\paragraph*{FLOPS calculations} See Appendix~\ref{sec:app:singlenode::throughput} for the FLOPS calculations in Figure~\ref{fig:e2e_full}.

\subsubsection{CPU/GPU Hybrid and Multi-GPU}

Figure~\ref{fig:e2e_full}  also shows that using the CPU in a GPU instance can accelerate purely GPU training. The g2.8xlarge instance's CPU provides 0.67 TFLOPS, and by using data parallelism across the CPU and a single GPU we achieve an $18\%$ speedup in Omnivore end-to-end over just using the GPU. This is faster than all other single GPU results.

Finally, we also apply this data parallel partitioning across multiple GPUs. We ran both Omnivore and Caffe using the 4 GPUs on the  g2.8xlarge instance and show that while Caffe actually slows down compared to the 1 GPU case, Omnivore becomes $3.1\times$ faster.\\

For CPU + GPU, each g2.8xlarge GPU provides 1.23 TFLOPS as shown above, and the CPU provides 665.6 TFLOPS (Sandy/Ivy bridge, i.e. $16$ SP instructions per cycle $\times 16$ physical cores $\times 2.6$ GHz). The ratio of CPU:GPU FLOPS is therefore 1:2, i.e. we should partition roughly 1/3 of the data on the CPU and 2/3 on the GPU. Since a batch size is 256 images, we rounded $67\%$ on the GPU to $75\%$, such that the CPU processes 64 images, because this partition is better suited to hardware (although we see only a $5\%$ speedup compared to using the exact ratio). This partitioning gives a $18\%$ speedup over just using the GPU.

For parallelization across 4 GPUs, we use data parallelism for all layers (each GPU given 1/4 of the batch and a model replica) except for the FC layers, which use model parallelism (each GPU given 1/4 of the model and a replica of the batch). 

We ran Caffe on 4 GPUs with cuDNN v4, cuDNN v3 and no cuDNN, and found that Caffe was fastest but neither gave a speedup compared to 1 GPU.\\

Therefore while CNN parallelization is challenging even for state-of-the-art systems, we've shown that FLOPS-proportional partitioning is possible across a range of hardware devices.
We now extend this technique to multiple machines, where the added challenge of network delay motivates re-thinking the SGD algorithm.

\section{Appendix for Distributed Tradeoffs (Section~\ref{sec:Distributed})}
\label{sec:app:distributed}

Having studied the tradeoffs for a single machine, this section now studies the distributed setting.
The goal of this section is to build an optimizer that creates
(1) a physical plan $P(A,G)$ which maps the CNN architecture $A$ to machines in the device graph $G$, and
(2) an execution plan $E(G,D)$ which parallelizes SGD by allocating data batches from the dataset $D$ to each machine.
This section begins by describing why these two are the most important tasks for the optimizer.
While many distributed CNN systems exist~\cite{Chilimbi:2014:OSDI, iandola15firecaf, chen15mxnet, Dean:2012:NIPS, Wang:2015:TR, moritz15sparknet}
and each describes their own distribution techniques,
upon analyzing these strategies we discover that, though diverse,
they all describe either (1) or (2) above.
Given these two fundamental design dimensions, we then arrange existing strategies into a 
tradeoff space and restate our optimizer's goal precisely within this space.
The remainder of the section then quantifies the impact of these tradeoffs
to allow optimization within the space.

\subsection{Distributed CNN Tradeoff Space}\label{sec:tradeoff}

\begin{figure}
\centering
\includegraphics[width=0.5\textwidth]{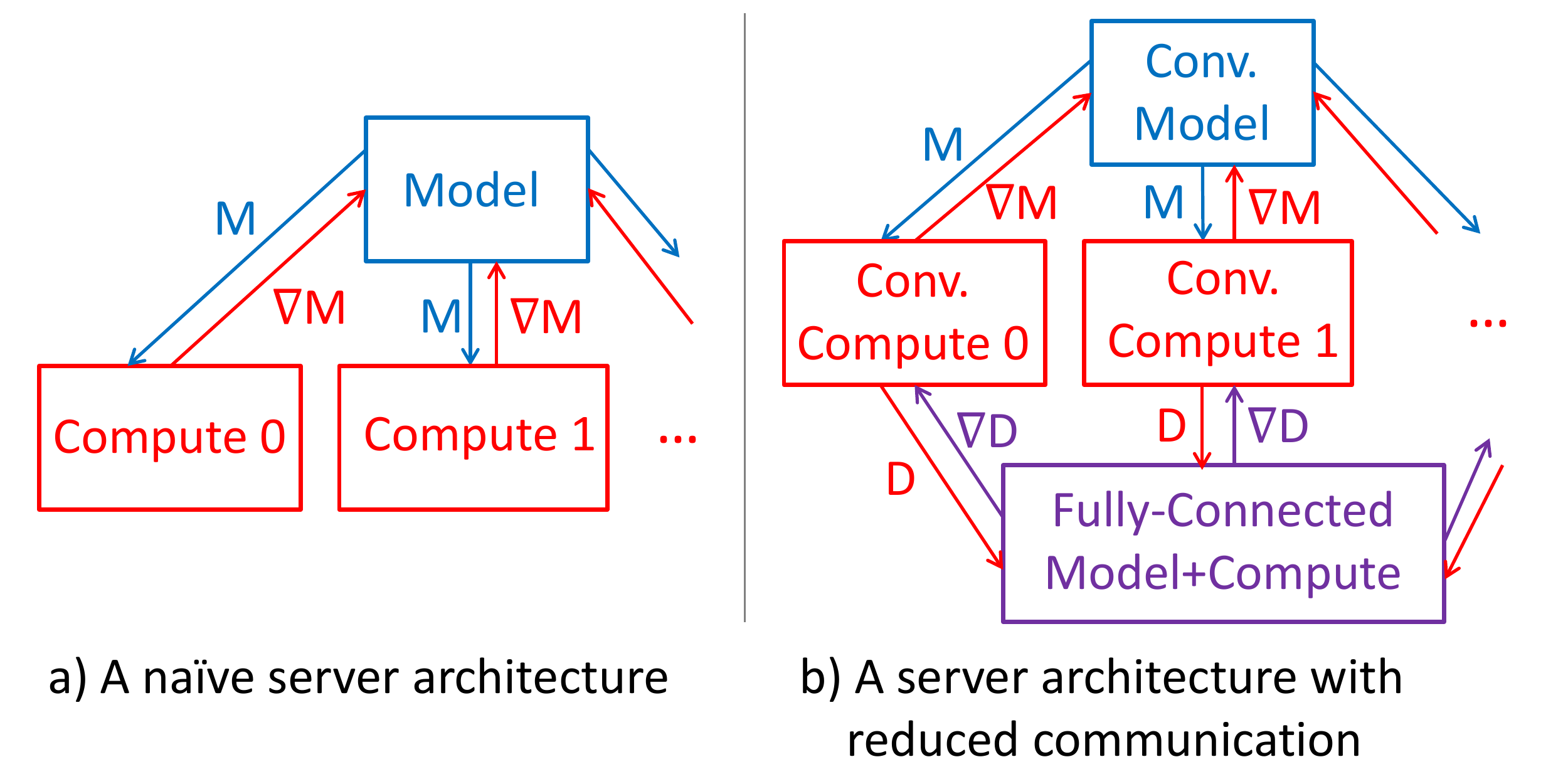}
\caption{SGD server architectures.}
\vspace{-1.0em}
\label{fig:arch_simple}
\end{figure}

This section describes popular techniques for distributed CNN training and refines them into a tradeoff space.

\subsubsection{Distributed Stochastic Gradient Descent}

Recall that SGD iteration $i$ (1) reads a batch of data $B_{i}$, (2) uses the current model $M_{i}$ to compute the model gradient $\nabla M(M_{i})$, and (3) subtracts $\nabla M(M_{i})$ from $M_{i}$ to obtain $M_{i+1}$. Iteration $i+1$ reads a new batch $B_{i+1}$ and the algorithm repeats until convergence.

Figure~\ref{fig:arch_simple} (a) shows a common distributed implementation of SGD in which the entire CNN model is stored on a server called a \textit{model server} or \textit{parameter server}. There are also a number of \textit{compute servers} which each perform the CNN calculation (Equations~\ref{equation:fw_pass} and~\ref{equation:bw_pass}). Each iteration, every compute server reads $M$ over the network from the parameter server, as well as a batch of data $B$ from a local database. Each compute server calculates a gradient $\nabla M$ which is sent back to the parameter server and used to update the model. In this example, compute servers operate in parallel and do not synchronize or communicate with one another.

In Figure~\ref{fig:arch_simple} (a), each server physically maps to a single machine (node). Generally, multiple servers can map to a single machine or a single server to multiple machines. For example, parameter and compute servers can map to the same node to reduce network communication.
Figure~\ref{fig:arch_simple} (b) shows a more complex \textit{server architecture} for SGD in which rather than two types of servers (compute and model), there are now 4 types: (1) conv compute, (2) conv model, (3) FC compute, and (4) FC model. 1 and 2 are for layers in the conv phase of the CNN while 3 and 4 are for layers in the FC phase. In Figure~\ref{fig:arch_simple} (b), the FC compute and model servers map physically to the same machine, i.e. the computation of the FC phase happens on the same machine as where the FC model is stored. There are many benefits to Figure~\ref{fig:arch_simple} (b) vs. (a): Recall that FC has small data, large model whereas conv has large data, small model. This server architecture has the benefit of only needing to communicate the conv model (and its gradients) and the FC data (and its gradients) across the network. Second, computation is offloaded from the compute machines to the FC model server, which otherwise is majorly idle. These benefits both improve hardware efficiency, and were described in~\cite{Chilimbi:2014:OSDI}. However, yet another benefit is that by having only a single FC compute machine, the FC model does not experience any \textit{staleness} -- a term we define next. This improves statistical efficiency.

\subsubsection{Staleness}

\begin{figure}[t]
\centering
\includegraphics[width=0.5\textwidth]{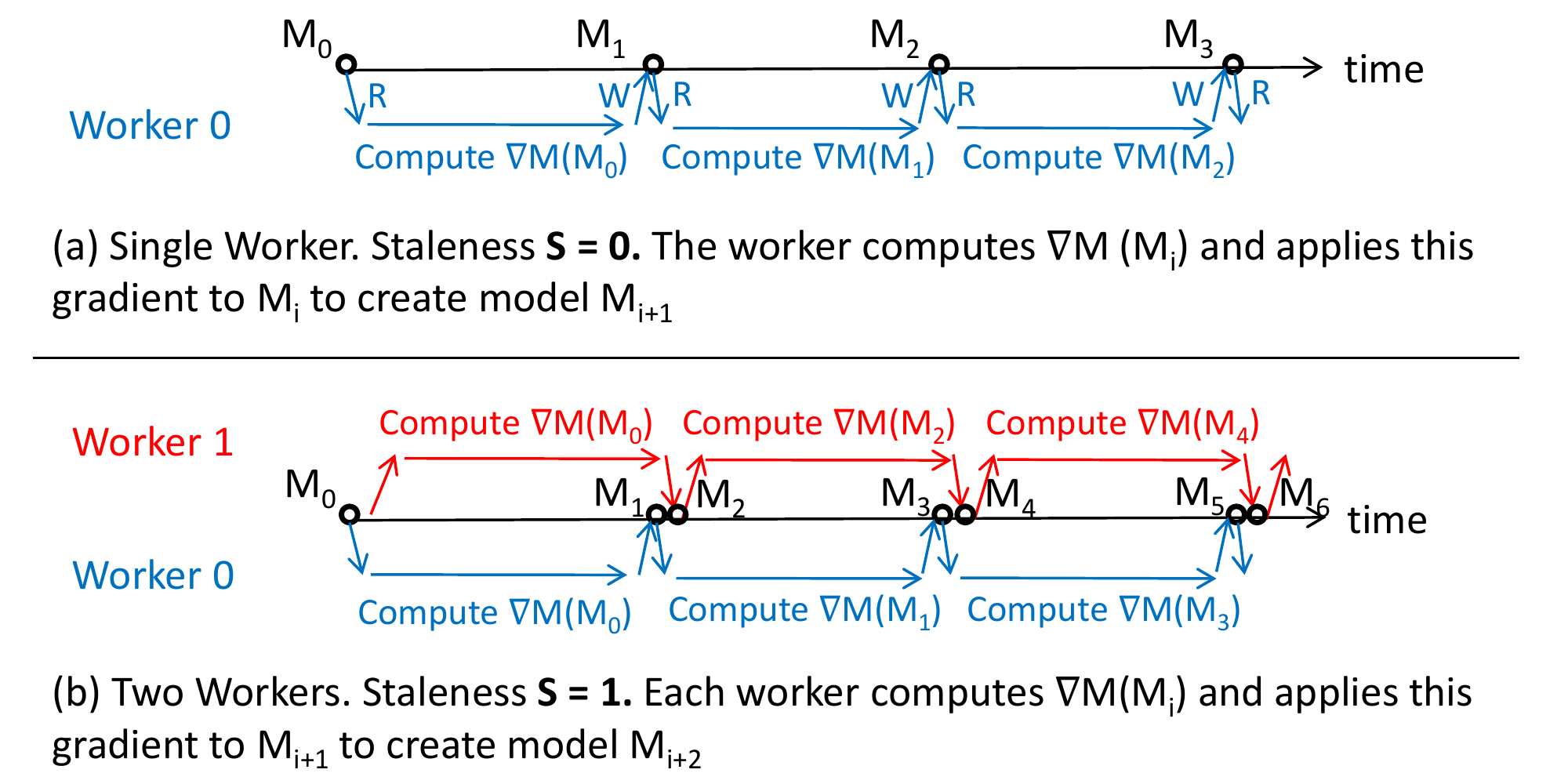} 
\caption{A graphical illustration of (a) one worker (S=0) and (b) two workers (S=1).}
\label{fig:stale1}
\vspace{-1.0em}
\end{figure}

\textit{Staleness} is a metric used to quantify the statistical efficiency (\# iterations to convergence). Figure~\ref{fig:stale1} shows staleness graphically for the simple server architecture of Figure~\ref{fig:arch_simple} (a). In Figure~\ref{fig:stale1} (a) there is a single worker, and so that worker always computes the gradient using the
current model. In this diagram, we assume that once a worker sends the updates back to the model server, it is immediately sent a new model, i.e. in this diagram, write/read is an atomic operation for each worker (the write to the model and read from the model happen together).

In Figure~\ref{fig:stale1} (b), there are now two concurrent workers. Assume for now that (1) these workers update the model in round-robin order, (2) write/read is again atomic, and (3) the server architecture is again as in Figure~\ref{fig:arch_simple} (a). We notice that now each worker computes the gradient using a copy of the model which is \textit{stale} by 1 iteration, e.g., updating model 2 to produce model 3, but doing so using a gradient update which was calculated using model 1. The reason for this staleness is that while worker 0 is computing its next update, worker 1 updates the model.
This staleness is bad for statistical efficiency (i.e. more iterations are required to converge) because now each gradient used to update the model no longer points in the direction of steepest descent for the model it is applied to, but rather the direction of steepest direction for a model from an earlier iteration.

Precisely, we define \textit{staleness} as follows: given N workers, the staleness is the number of updates to the model between a worker's read from the model and subsequent write back to the model. Because the updates are round-robin, this staleness is the same for all workers, and is equal to $S = N-1$. e.g., if there are 100 parallel workers, $S=99$. Intuitively, the staleness is just equal to the number of parallel workers sending gradient updates (minus one, although this can often be ignored because the number of workers is large).


The three assumptions above are useful to give a precise definition but are not necessary in practice. 

\paragraph*{Assumption 1:}
In practice the workers do not proceed in round-robin order due to inherent variance across machines, but we observe empirically that for dense models the updates are nearly round-robin. This is because dense model computations like those used in deep learning have roughly constant time per iteration (this is not true for sparse models).

\paragraph*{Assumption 2:}
Writes and reads do not need to be atomic, and in fact this can be beneficial for statistical efficiency. Rather than have a conv compute server request an entire updated model as soon as it publishes all of its gradients to the conv model server, it may instead publish gradients layer-by-layer in a backwards fashion during the backward pass of iteration $i$, and then lazily request the updated model in the forwards pass of the next iteration $i+1$. For example, AlexNet may update the model with gradients from conv5, conv4, conv3, conv2, and conv1, and then begin its next forwards pass and request conv1, conv2, conv3, conv4 and conv5. These requests can happen asynchronously from the computation to hide latency and overlap the network delays with the computation. As a result the write/read for conv1 may be almost atomic, but there would be some delay between the write/read for conv5. This delay in fact reduces staleness slightly because it reduces the number of intermediate writes by other workers between a worker's read and subsequent write (intuitively, in the extreme case, if the delay was very large, then every worker would write before any of them read the new model. Then this is just equal to mini-batch, except with a larger batch size, although that would of course make each iteration slower, i.e. harm hardware efficiency). In practice we observed a roughly 20\% reduction in the number of iterations to converge by requesting models in this lazy fashion (which does not harm hardware efficiency).

\paragraph*{Assumption 3:}
Staleness also applies to the server architecture of Figure~\ref{fig:arch_simple} (b), which recall reduces network communication by merging the FC model and FC compute servers, i.e. mapping them to the same machine which does both the gradient computation and model updates for the FC phase. This merged FC server processes only one batch at a time, and thus produces only one FC model gradient at a time. This means that the staleness for the FC model is 0 which is good for statistical efficiency. The conv compute servers on the other hand still calculate the updates to the conv model in parallel (once they receive their data gradients from the FC machine),
therefore the merged architecture in Figure~\ref{fig:arch_simple} (b) still contains staleness, but only for the conv model.

\subsubsection{Compute Groups}

\begin{figure}
\centering
\includegraphics[width=0.5\textwidth]{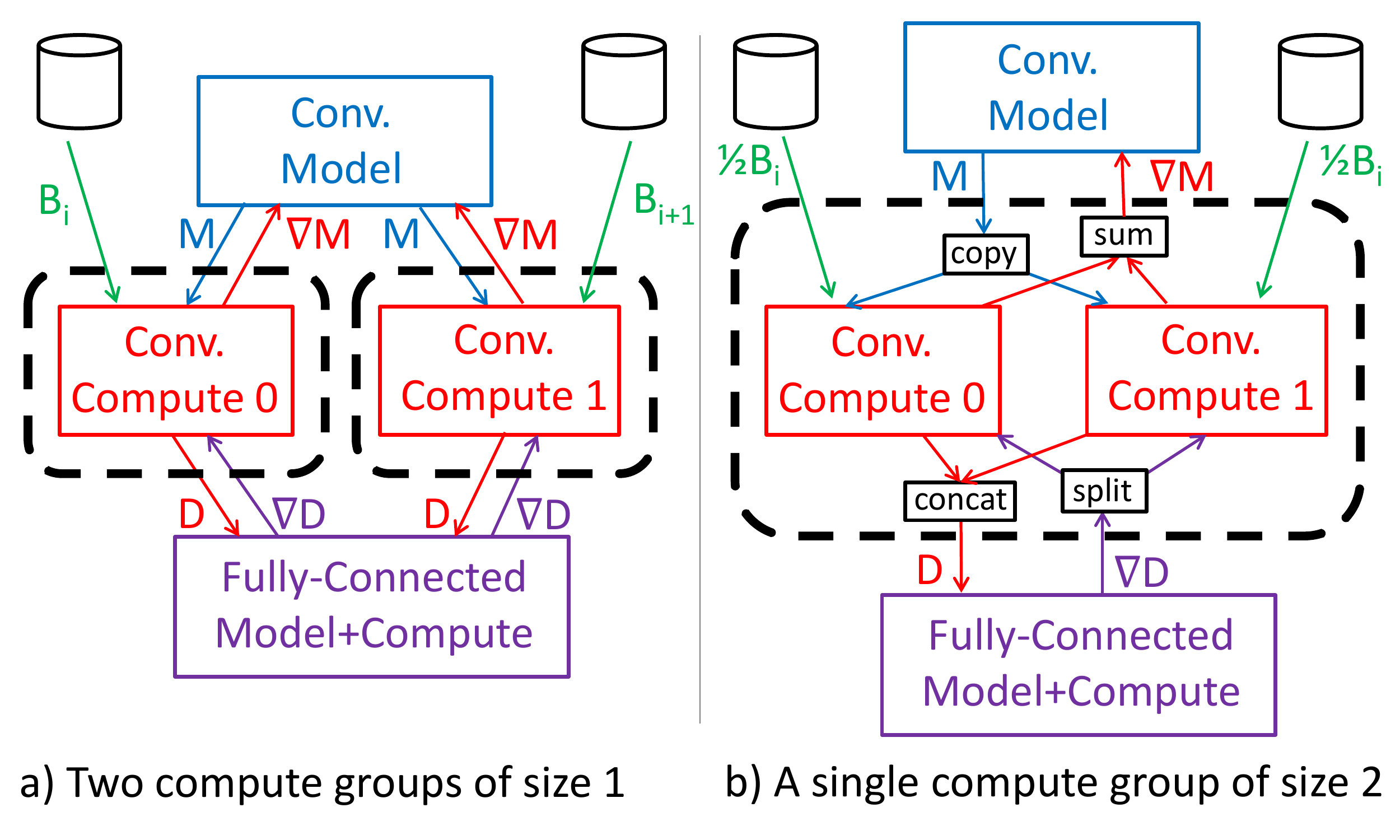}
\caption{Two different execution strategies given a fixed machine budget.}
\vspace{-1.0em}
\label{fig:arch_cg}
\end{figure}

The final concept common to all systems is the \textit{compute group}. Consider the example of two conv compute machines in Figure~\ref{fig:arch_cg} (a). This configuration has a conv staleness of $S=1$, as there are 2 workers independently updating the model. However, it is not necessary to introduce staleness in order for both compute machines to be utilized, and Figure~\ref{fig:arch_cg} (b) shows a second configuration in which data parallelism is used across the machines: each conv compute worker processes \textit{half} the data of a single batch, using the same model replica, and produces half of the final gradient. A barrier is then introduced in which the gradients are summed, and a single, final gradient is applied to the model. 
Figure~\ref{fig:arch_cg} (b) seemingly solves both problems: all the hardware is being utilized 
(good for hardware efficiency), and $S=0$
(good for statistical efficiency).
However the price we pay is in hardware efficiency, specifically the cost of synchronization across the machines. Indeed, we will show that the hardware efficiency of Figure~\ref{fig:arch_cg} (b) is poorer than that of Figure~\ref{fig:arch_cg} (a).

We define a compute group as a group of machines working together to process a single batch of data. A compute group is characterized by processing a single data batch at a time, with all machines in the group using the same model replica, and returning a single gradient to the model server. The compute groups in Figure~\ref{fig:arch_cg} are shown with black dotted lines. Figure~\ref{fig:arch_cg} (a) has 2 compute groups, and since there are 2 machines, the \textit{compute group size} is 1 machine. Figure~\ref{fig:arch_cg} (b) has 1 compute group of size 2. Note that this allows us to simplify our definition of staleness: because the number of parallel gradients being computed in the system is equal to the number of compute groups, \textit{the staleness is just equal to the number of compute groups} (minus one).

Generally, if we have $N$ machines used as conv compute servers, Figure~\ref{fig:arch_cg} (a) and (b) show two extreme cases. Figure~\ref{fig:arch_cg} (a) is the extreme case of 1 machine per compute group, and $N$ groups. This technique is often called \textit{asynchronous SGD}, or ``async'' for short.
In async, workers do not communicate and each worker updates the model independently. Every worker computes a separate gradient using a separate batch and separate model replica, and then sends these gradients to the parameter server in order to update the model. Figure~\ref{fig:arch_cg} (b) is the other extreme case of 1 group, and all $N$ machines in that single compute group. This technique is often called \textit{synchronous SGD}, or ``sync''. In sync, all machines work synchronously and in parallel on a single batch of data and using a single model replica to compute a single gradient. In this case the gradients over all workers are aggregated each iteration (or batch) before updating the model. An intermediate configuration could also exist, for example one which has 4 groups each of size $N/4$. There could even be groups of different sizes if different machines have different throughput (some GPUs, some CPUs, etc.).
Notice that because the compute group in Figure~\ref{fig:arch_cg} (b) parallelizes the conv phase, it uses data parallelism,
i.e. the batch is partitioned across the machines in the group.

\subsubsection{Precise Problem Definition}

This section has described two key tradeoffs: (1) the server architecture, concerned with physically mapping servers to machines, and (2) the execution strategy (the number of compute groups vs. their size), concerned with mapping batches to servers. 
We can now restate the goals of the optimizer from~\ref{sec:definition} in terms of our tradeoff space.
Given (1) $A$, the CNN architecture, (2) $D$, the dataset, and (3) $G$, the device graph, our optimizer transforms $A$ into $S$, an abstracted network of logical server types ($\text{Conv}_{\text{model}}$, $\text{Conv}_{\text{compute}}$, $\text{FC}_{\text{model}}$, and $\text{FC}_{\text{compute}}$), and creates (1) a physical plan $P(S,G)$ mapping each server in $S$ to machines in $G$
(note that we also refer to this distributed portion of the physical plan as the server architecture), and (2) an execution plan $E(S,D)$ which defines the number of compute groups by allocating batches of $D$ to each server in $S$.
Both of these choices impact hardware and statistical efficiency, and our next task is to quantify this impact.
First, we present terminology and then finish this section by describing where existing systems fall within this tradeoff space.

\begin{figure}
\centering
\includegraphics[width=0.5\textwidth]{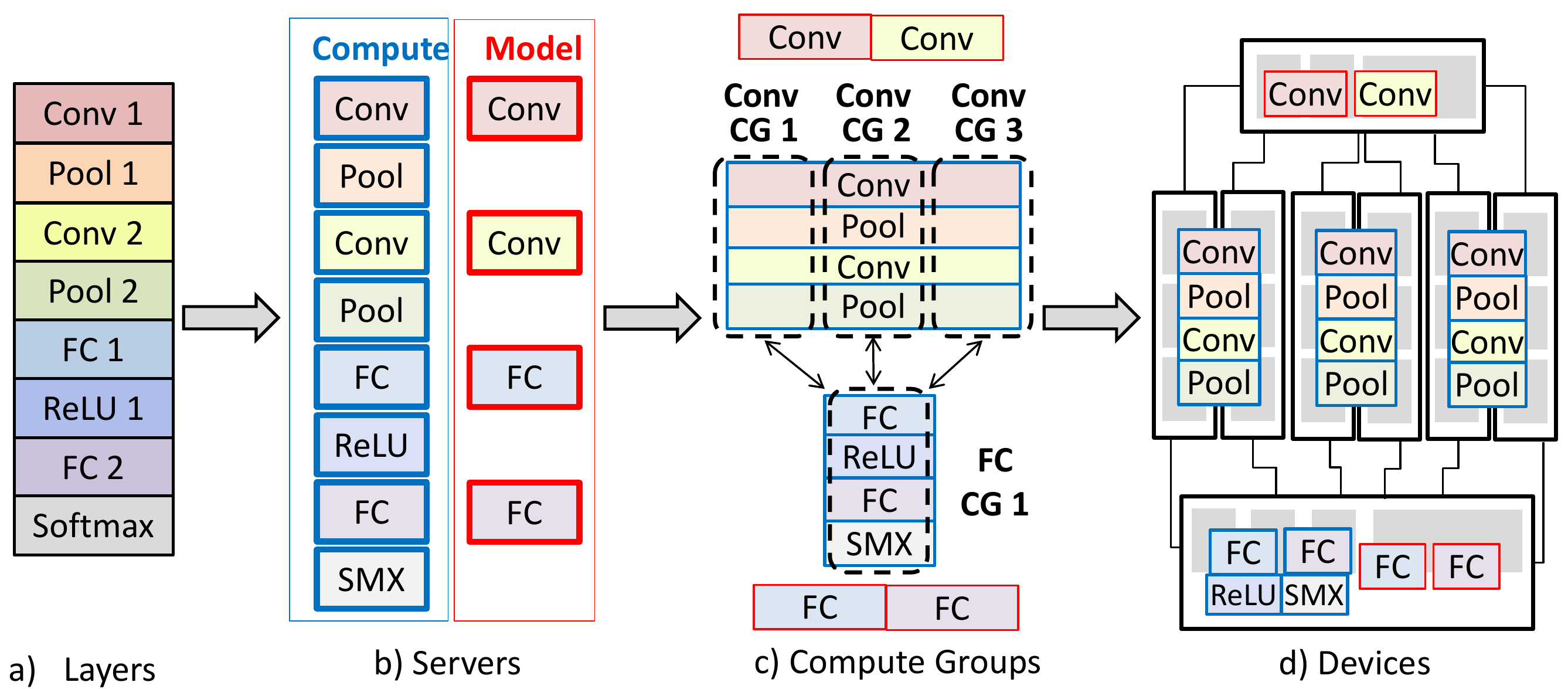}
  \caption{An Overview of our Mapping.}
  \label{Fig:Mapping:Stages}
\vspace{-1.0em}
\end{figure}

Our mapping from CNN layers to devices is shown in Figure~\ref{Fig:Mapping:Stages}.

We are given: (1) a network, which can be viewed as a labeled directed
acyclic graph in which each node is a type of layer (e.g., convolution,
fully connected, max pooling) and edges indicate dataflow
dependencies, and (2) a set of devices grouped into machines. Our goal
is to devise a mapping from the network to the machines.

\subsection{Terminology}\label{sec:terminology}

In this work we interchangeably uses the terms \textit{node} and \textit{machine} to refer to a single box, connected to other nodes or machines over a network. We also interchangeably use the terms \textit{device} and (when describing a device graph) \textit{vertex} to refer to a discrete unit of compute hardware (e.g., a GPU or CPU core. When discussing a cluster, a device can also be a machine in that cluster.). Finally, we also use the terms group and compute group interchangeably, as defined in Section~\ref{sec:background}.

Existing work also often contains a lot of terminology, which we summarize here. Consider for these examples the simple case of \textit{N identical compute devices} (e.g., N machines, N GPUs, N CPU cores, etc.)

\subsubsection{Synchronous Batches} 

These first two definitions apply to a single batch of data, i.e. there are not multiple parallel batches in the system used to compute asynchronous gradients, but rather a single batch used to compute a single gradient. As a result, this gradient is applied to the model at once to produce a new model, i.e. there only ever a single model in the system (although it may be replicated, in which case all models are identical). This is also known as the synchronous case above.

\textbf{Model Parallelism:} A single replica of the model and $N$ replicas of the data batch are created. Each device is given $1/N$ of the model replica and 1 replica of the data batch. This is useful for parallelizing fully-connected layers, which contain small data but large models: each device receives $1/N$ of the model and a replica of the data, and uses the data to compute a gradient for that portion of the model. These gradients in total combine to a single gradient with respect to that entire model, using the data batch.

Note that logically, a single gradient is produced by all devices together and that single gradient is used to update the model and produce a new model (physically the updates may occur locally on each device communication, i.e. the device updates its portion of the model). 

\textbf{Data Parallelism:} A single replica of the data batch and $N$ (identical) replicas of the model are created. Each device is given $1/N$ of the data replica and 1 replica of the model. This is useful for parallelizing convolutional layers, which contain small models but large data: each device receives $1/N$ of the data and a replica of the model, and together they calculate a single gradient with respect to that entire model using the data batch.

Note that, as mentioned above, here the model is \textit{always the same for each device}, i.e. the model is replicated and a single gradient is produced. That gradient is then used to update the model and the model is then re-broadcast to each device. So while model replicas exist, they are identical model replicas.

(Note: model replica is a logical term and does not always correspond to a physical replica. In particular, when two devices share memory, e.g., 2 CPU threads running on 2 cores, no physical replica is necessary as the same model in memory can be read by both.)

\subsubsection{Asynchronous Batches} 

The above definitions apply to a single data batch in the system (although it may have been replicated for the purpose of parallelization). Consequently, there was always one gradient being computed at once, and so each model replica was identical. 

This definition focuses on the asynchronous case mentioned previously, i.e. $N_{B}$ parallel batches in the system being used to compute $N_{B}$ parallel gradients. Each of these $N_{B}$ batches of data in the system is different. Each batch is allocated to a compute group (a group of devices), and each batch (i.e. each group of devices) is given a replica of the model. For example if $N_{B} = N/4$, then there will be 4 devices assigned to each batch, and one model replica assigned to each of these $N/4$ batches (i.e. each group of 4 devices). Each compute group of 4 devices will then compute a gradient given that model and that batch, and the devices in the group may do so using either model or data parallelism (i.e. choosing to create additional model or data replicas within the group, as described above, e.g., if the group uses data parallelism it will create 4 additional model replicas, but each of these models will be identical).

Note that here the number of batches in the system, the number of compute groups, and the number of parallel gradients being computed by the system are all the same number. Within a compute group (group of devices), additional model replicas will be the same. However across compute groups, the model replicas may be out of sync, because of asynchronous gradient computations. This is discussed in the final definition.

To make all the definitions above concrete, consider the example of 2 devices, e.g., 2 GPUs. There are 3 possible scenarios: (a) 2 parallel asynchronous batches (``async''), (b) 1 batch with data parallelism across the 2 devices (``sync''), or (c) 1 batch replicated twice with model parallelism across the 2 devices (also sync).

\subsubsection{Combining Model Replicas}

If there are G asynchronous compute groups, each using a separate data batch and producing a separate gradient each iteration, then each group will also have a separate version of the model as a result of the asynchronous gradient computations.

\textbf{Parameter server} is a technique in which these separate compute groups will not perform their model updates locally, but rather each publish their gradients to a global parameter server which will then broadcast the updated model to a group upon receiving the update for that group. In this way the groups always have models which are almost the same (they will still be slightly out of sync because gradients are published asynchronously and so models are returned to the groups asynchronously as well). DistBelief, Adam, SINGA, MXNet use this technique, and it is the focus of our study.

Note that if the parameter server waits for each parallel group to publish a gradient and then broadcasts the model to all groups (i.e. introduces a barrier), this is no longer asynchronous, and is now equivalent to the synchronous case above (data parallelism) where the batch size is $N$ times larger than it is per individual compute group.

Finally, the updates to the server from parallel groups can happen with or without race conditions. The case of race conditions is known as Hogwild!.

\textbf{Model Averaging} is a technique in which there is also a global parameter server, but model updates happen locally within each group. Then periodically, every $\tau$ iterations, the groups will publish not their gradients but their entire models to the parameter server. The parameter server will average the models (reduce step) and then broadcast them to each group (map step). This averaging does not have theoretical guarantees because neural networks are non-convex, but works in practice. This technique works well for map/reduce frameworks, e.g., Spark/Hadoop, and is used by SparkNet and DL4J. Here the models are more different than they are in the parameter server case.

The choice of the $\tau$ parameter is similar to the tradeoff of multiple groups of varying size, except that here staleness comes not from multiple asynchronous workers updating a single model, but multiple workers with a separate model combining models. In the case where $\tau = 1$, this is identical to the synchronous case of parameter server (all machines in a single group, i.e. all computing a single batch/gradient).

\textbf{Ensembles} are used by AlexNet. Here each group trains an entirely separate model to convergence and then predictions of these models are combined (e.g., through voting). The gradients or models themselves are never combined.

In this study we focus on the parameter server approach, which is the most widely used by distributed CNN systems.

\subsection{Existing Systems}
\label{sec:app:distributed:existing}

\begin{table}[]
\centering
\caption{Points in the distributed CNN tradeoff space chosen by popular systems. g is the number of compute groups, N is the number of machines.}
\label{table:other_systems_1}
\scalebox{0.9}{
\begin{tabular}{|l|c|c|c|c|c|}
\hline
\textbf{Tool}                                          & \multicolumn{1}{l|}{\textbf{\begin{tabular}[c]{@{}l@{}}sync\\(g=1)\end{tabular}}} & \multicolumn{1}{l|}{\textbf{\begin{tabular}[c]{@{}l@{}}1$<$g$<$N\end{tabular}}}      & \multicolumn{1}{l|}{\textbf{\begin{tabular}[c]{@{}l@{}}async\\(g=N)\end{tabular}}} & \multicolumn{1}{l|}{\textbf{\begin{tabular}[c]{@{}l@{}}Model\\ Avg.\end{tabular}}} & \multicolumn{1}{l|}{\textbf{\begin{tabular}[c]{@{}l@{}}Merge\\ FC\end{tabular}}} \\ \hline
DistBelief~\cite{Dean:2012:NIPS}                                             & $\bullet$                                                                         & $\bullet$                                                                            & $\bullet$                                                                          &                                                                                    &                                                                                    \\ \hline
Adam~\cite{Chilimbi:2014:OSDI}                                                    &                                                                                   & $\bullet$                                                                            & $\bullet$                                                                          &                                                                                    & $\bullet$                                                                          \\ \hline
FireCaffe~\cite{iandola15firecaf}                                              & $\bullet$                                                                         &                                                                                      &                                                                                    &                                                                                    &                                                                                    \\ \hline
MXNet~\cite{chen15mxnet}                                                   & $\bullet$                                                                         &                                                                                      & $\bullet$                                                                          &                                                                                    &                                                                                    \\ \hline
SINGA~\cite{Wang:2015:TR}                            & $\bullet$                                                                         & $\bullet$                                                                            & $\bullet$                                                                          &                                                                                    &                                                                                    \\ \hline
SparkNet~\cite{moritz15sparknet}                                               &                                                                                   &                                                                                      &                                                                                    & $\bullet$                                                                          &                                                                                    \\ \hline
DL4J                             &                                                                                   &                                                                                      &                                                                                    & $\bullet$                                                                          &                                                                                    \\ \hline
\end{tabular}
}
\vspace{-1.0em}
\end{table}

Using the terminology above, we discuss design decisions made by CNN
systems in Table~\ref{table:other_systems_1}. In our review of the
literature, these are the tradeoffs which we identified as most impactful to
minimizing convergence time.

\paragraph*{Execution Strategy}
In terms of execution strategies, Microsoft's Project Adam~\cite{Chilimbi:2014:OSDI} reports that the async strategy is effective and uses one or a few (e.g., 4) machines per compute group. Moreover, they report that using a technique called Hogwild!~\cite{hogwild}, which introduces race conditions, is an effective source of additional hardware speedups. On the other end of the spectrum, FireCaffe~\cite{iandola15firecaf} implements the sync strategy and notices that high scalability can be achieved by having all machines work synchronously and in parallel on a single, large batch of data, and by reducing communication using reduction trees. MXNet~\cite{chen15mxnet} implements both the sync and async strategies, and allows the user to select which to use. They also call the sync strategy the \textit{Bulk Synchronous Parallel} (BSP) protocol. Finally, Google DistBelief~\cite{Dean:2012:NIPS} and Apache SINGA~\cite{Wang:2015:TR} implement both sync and async as well as the compromise of multiple, larger compute groups. They also call sync Sandblaster, and cases of more than one group Downpour (e.g., Downpour with group size of 1 machine is equivalent to the async). SINGA calls the intermediate case of multiple, larger compute groups \textit{hybrid} (SINGA also has hybrid parallelism, which is different. That is a combination of model and data parallelism.)

All these systems implement the \textit{Parameter Server} technique, described above in the Terminology section (Appendix~\ref{sec:terminology}). It is the most widely used technique by distributed CNN systems and the technique we focus on in this study. Another technique known as model averaging, which works well within a map/reduce framework, is used by SparkNet~\cite{moritz15sparknet} and DL4J (\url{http://deeplearning4j.org/}). Model averaging is also described above. The key difference between model averaging and parameter server is the way in which model replicas are combined. 

\paragraph*{Physical Map, Modern Networks}

The second point in the distributed tradeoff space is the server architecture (how servers map to hardware), specifically whether the FC compute and FC model servers are mapped to the same physical machine (or machines, for multi-machine model parallelism).

This is a technique introduced by Microsoft's Project Adam~\cite{Chilimbi:2014:OSDI} to avoid communicating the large FC model and its gradient. 
The method was reported for older networks with large, fully-connected layers (AlexNet, VGG), however it also is useful for modern networks (Inception, ResNets). In traditional networks, the fully-connected layers contained the majority of the model parameters~\cite{Krizhevsky:2012:NIPS} ($>90\%$). Newer networks instead use average pooling and have only a single FC layer (for softmax regression)~\cite{he2015resnet}. Therefore newer networks contain fewer parameters in the FC phase, and fewer parameters overall, however this single FC layer can still be very large (e.g., when predicting among 22,000 classes on ImageNet 22k) and still benefits from reduced FC communication (because the number of FC weights will always be less than the number of inputs to the FC phase).
Therefore while newer networks contain only a single FC layer, the merged FC optimization of Project Adam is still relevant, and while a characteristic of newer networks is that their overall model size is smaller due to the elimination of multiple FC layers, ultimately this does not translate to reduced communication overall because the cost of communicating the FC layers has been eliminated in prior work.

In addition, as we show, the benefit of merging the FC servers is not only improving \HE due to reduced network communication, as~\cite{Chilimbi:2014:OSDI} noted, but also improving \SE because staleness in the FC model is eliminated (i.e. the device or devices which compute the FC model gradient updates also store that subset of the model). Moreover there is no consequence of merging the FC servers for small FC models because little computation in the FC phase means it is less likely for the FC to saturate (become the bottleneck), but merging still provides the benefit of (1) improving \SE, (2) reducing communication and (3) offloading computation to the parameter server machines. 

As our goal is to be a complete study, we study both cases (many large FC layers, few small FC layers) in order to build a system which is robust to any application. For example future CNN architectures may employ multiple FC layers again to support transfer learning tasks, or may need to predict among many object classes (e.g., hundreds of thousands or millions), further increasing the communication bottleneck for the FC. In addition, FC layers are also used for RNNs and other architectures.\\

This is the subset of the tradeoff which we study in this work, summarized in Table~\ref{table:other_systems_1}. Our goal is to find best point in the tradeoff space given the model, data, and hardware specifications from the user. Specifically we do not change the neural network architecture, but assume that this model is given to us. I.e. we do not focus on machine learning algorithms or techniques in this work, but rather study systems tradeoffs which exist for the most widely used networks/algorithms. We focus on the SGD algorithm for learning, as it has been and continues to be used along with momentum by annual ImageNet winners~\cite{Krizhevsky:2012:NIPS, he2015resnet}. Other algorithms exist for training deep learning models and can also be parallelized, for example Google's deep learning system uses Adagrad~\cite{Dean:2012:NIPS}. Microsoft on the other hand uses SGD~\cite{Chilimbi:2014:OSDI}. Because the systems tradeoffs we study in this work are orthogonal to the choice of update algorithm, in this work we focus on SGD, although the same tradeoff applies to other algorithms as well.

\subsection{Hardware Efficiency Model}
\label{sec:app:distributed:hemodel}

The goal of this section is to create a predictive model for how the
hardware efficiency (\HE) varies with the amount of staleness
$S$ in the system, given a fixed number of machines and batch size.
Recall that the staleness $S$ is equal to the number of compute groups (minus one). This is because a compute group is characterized by processing a unique data batch and returning a unique gradient to the model server, so the number of parallel gradients being computed is equal to the number of compute groups.
$S=0$ is the case of 1 compute group, also called the synchronous case. 

\begin{figure}
\centering
\includegraphics[width=0.5\textwidth]{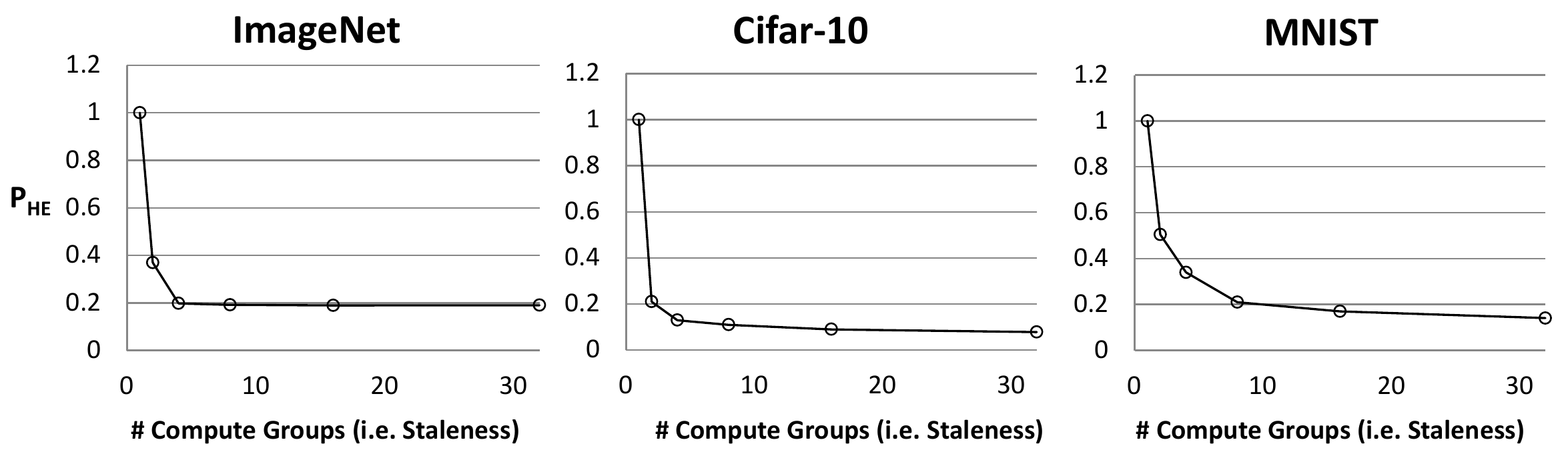}
\caption{Hardware efficiency penalty for various numbers of compute groups. The number of machines is fixed to 32. Each machine is an EC2 c4.4xlarge (CPU) machine.}
\vspace{-1.0em}
\label{fig:Phe}
\end{figure}

Figure~\ref{fig:Phe} shows a plot of staleness vs. \textit{hardware efficiency penalty} for three datasets. The standard networks from Caffe's tutorials are used for each dataset.
The hardware efficiency penalty, or $P_{HE}$, is defined as the ratio of the time per iteration relative to the time per iteration for this synchronous case,
\begin{equation*}
P_{HE}(S) = \frac{HE(S)}{HE(0)}
\end{equation*}
A higher $P_{HE}$ is worse (more time per iteration).
$P_{HE}$ decreases (iterations become faster) as the number of compute groups increases.
This is because, if we fix the number of machines to be $N$, the smallest time per iteration occurs when there is no synchronization, i.e. when the compute group size is 1 and there are $N$ compute groups (asynchronous case). In this case $S=N-1$. 
As the compute group sizes increase, and hence the number of groups decreases (because the number of machines is fixed to $N$), $P_{HE}$ will increase due to synchronization delays within the groups. 
When the number of compute groups is 1, that group contains all $N$ machines and requires the most synchronization. In this case $S=0$, and this has the highest penalty $P_{HE}$.

The hardware efficiency penalty is dimensionless. It is the ratio of hardware efficiencies (time per iter / time per iter). Because $P_{HE}$ is normalized to the synchronous ($S=0$) case, $P_{HE} \leq 1$. Note also that the hardware efficiency penalty is only comparable across different staleness values if the number of machines is fixed.

The hardware efficiency penalty can be predicted analytically. 
However, as we will see in a later section, the cold-start phase of the optimizer performs a short adaptive grid search across different staleness values.
Because a few iterations are run for various staleness values, the execution time for these iterations can be used to provide a precise hardware efficiency model
(deep learning layer computations are dense, not sparse, so there is little variation in iteration time).
Nevertheless understanding the hardware efficiency precisely is important:
\begin{enumerate}
  \item to understand the execution bottlenecks and either hard-code or allow the optimizer to make physical mapping choices
  \item because it may be too time-consuming to obtain static information for every staleness value of interest, and
  \item because our work is a study meant to inform future systems which may not use a static optimizer
\end{enumerate}

Figure~\ref{fig:Phe} was run on 33 EC2 CPU machines. The
server architecture shown in Figure~\ref{fig:arch_cg} was used, i.e. one machine contains the merged FC compute and FC model servers, and the other 32 machines contain Conv Compute servers. The Conv Model server is mapped to one of the Conv Compute machines. 
We make two observations, which are true for all datasets in Figure~\ref{fig:Phe}:

\textbf{Observation 1:} As the number of groups decreases (and hence as their sizes increases), the hardware efficiency becomes poorer. There are two reasons for this: (1) machine variance, which causes synchronization delays, and (2) network congestion, because the convolution model needs to be sent simultaneously to all machines in the group (and gradients need to be send back simultaneously).
Our analysis below shows that while machine variance exists, it is insignificant compared to increased network congestion.

\textbf{Observation 2:} As the number of groups increases, the speedup is not linear (it saturates). This is because the FC phase processes only a single batch at once, or equivalently because the FC compute and FC model server map to the same physical machine (or machines, as the FC compute / model server may use multi-machine model parallelism). 
Recall that this has benefits for both hardware efficiency (by reducing network communication) and statistical efficiency (by reducing the staleness of the FC model). However it means that only one gradient computation (batch) is processed by the FC at a time, and so it may become a computational bottleneck. 

Many optimizations exist for both of these observations. They are presented after our derivation of the model.

\subsubsection{Full Derivation of Analytic Model}

Formally, 
let there be $N+1$ machines. 1 machine is allocated to the FC phase, and $N$ machines (e.g., 32) to the conv phase. An \textit{execution strategy} partitions the $N$ conv machines into $g$ compute groups. Each group contains $k$ machines, and the $k$ machines in a group compute the conv phase with data-parallelism. Therefore, there will be $g = N/k$ compute groups sharing the single FC server machine.
In addition, let $t_{conv}(k)$ be a function that returns the time that a group of size $k$ needs to compute the convolution phase (forwards and backwards for only the conv phase), and $t_{fc}$ be the time that an FC server needs to serve one group (forwards and backwards for only the FC phase. Note that $t_{fc}$ is independent of $k$, the number of machines used to perform convolution on each batch). We also define that $t_{fc}$ includes the network time to transfer the data from the conv phase to the fc phase and the data gradients from the fc phase back to the conv phase, although we observe that this network time is often small compared to the computation time of $t_{fc}$. Note that $t_{fc}$ is independent of $k$, the number of machines used to perform convolution on each batch. Finally, assume for now as we did above that different groups (batches) cannot be executed in parallel on the FC server (that case is described later). 

\begin{figure}
\centering
\includegraphics[width=0.5\textwidth]{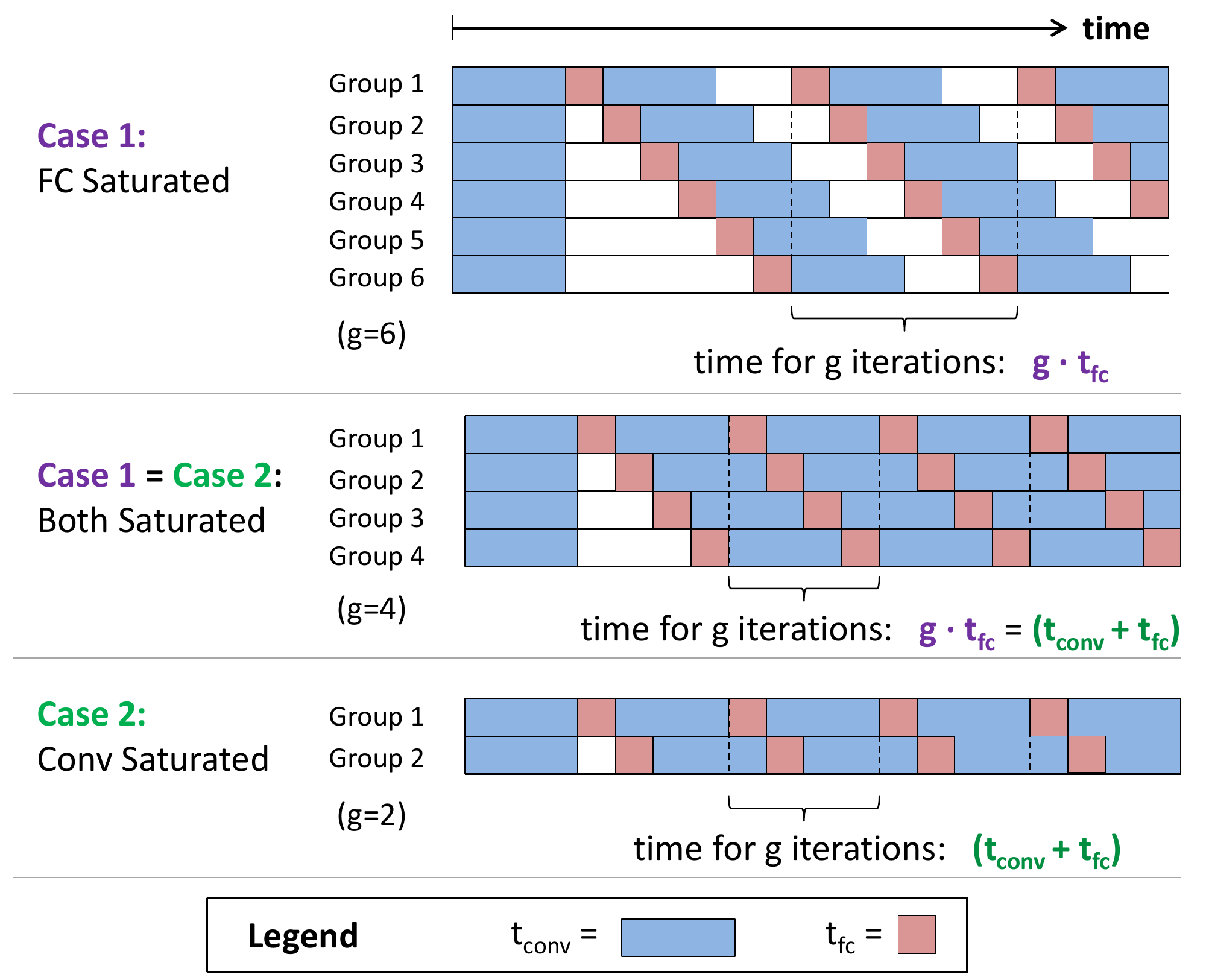}
\caption{Gantt chart illustrating the Hardware Efficiency model for the server architecture of Figure~\ref{fig:arch_cg}. Case 1 is FC saturation (top), Case 2 is Conv Saturation (bottom), and the boundary between the two is shown in the middle.}
\vspace{-1.0em}
\label{fig:HE_Model_Gantt}
\end{figure}

Given $k$, $g$, $t_{conv}(k)$ and $t_{fc}$, our goal is to create a hardware efficiency model which predicts the time per iteration or, equivalently, which given a time interval $T$ predicts how many batches will be processed by the system.
Because each batch must be processed by the FC server, this is equivalent to determining the number of requests that the FC server can process from the $g$ convolution groups in time $T$. There are two cases, also illustrated in Figure~\ref{fig:HE_Model_Gantt}.

\textbf{Case 1: Saturated FC Server}
The first case is when FC server is saturated, i.e. it starts to serve the next request immediately after the previous request finishes. In this case, the hardware efficiency is straightforward. The server will serve $T / t_{fc}$ requests in time T, or equivalently,
\begin{equation*}
\text{Time per iteration}_{\text{saturated fc}} = t_{fc}
\end{equation*}

\textbf{Case 2: Saturated Conv Server}
When the FC Server is not saturated, each conv server becomes the bottleneck. In this case, the FC server serves $T g/(t_{conv}(k) + t_{fc})$ requests in time T, or equivalently,
\begin{equation*}
\text{Time per iteration}_{\text{saturated conv}} = (t_{conv}(k) + t_{fc}) / g
\end{equation*}
which is the total time for a single iteration divided by the number of parallel groups. This is because the groups all are computed in parallel, with the exception of the FC server which is serial, but the FC server is never saturated so it can also be seen as being part of each parallel group. Refer to Figure~\ref{fig:HE_Model_Gantt} for an illustration of this case.
To understand this case, note that:
\begin{enumerate}
  \item When each conv server is fast ($t_{conv}(k)$ is small), the FC server serves more requests in time T
  \item When the FC server is fast ($t_{fc}$ is small), the FC server serves more requests in time T
  \item When the number of concurrent group is large ($g$ is large), the FC server serves more requests in time T
\end{enumerate}

\textbf{Determining Saturation}
Finally, the model needs to predict when the FC server will saturate. This occurs at the boundary of the times above, specifically the FC server saturates (case 1) when:
\begin{equation*}
t_{conv}(k) + t_{fc} < g t_{fc}
\end{equation*}
Intuitively, if the combined FC time to process all groups ($g t_{fc}$) exceeds the time for a single group's iteration ($t_{conv}(k) + t_{fc}$), then the FC server will always be saturated. Note that:
\begin{enumerate}
  \item When each conv server is fast ($t_{conv}(k)$ is small), it is easier to saturate
the FC server
  \item When the FC server is fast ($t_{fc}$ is small), it is harder to saturate the FC server.
  \item When the number of concurrent group is large ($g$ is large), it is easier to saturate the FC server.
\end{enumerate}
We now have an expression for the time per iteration in both cases as well as a condition to decide which case applies. Given the following:
\begin{itemize}
  \item Two of: $N$, $g$ or $k$ (the third can be calculated from the other 2),
  \item $t_{conv}(k)$, the time to complete the convolution portion of the network (forwards and backwards) given the group size, and
  \item $t_{fc}$, the time to complete forwards and backwards on the FC phase,
\end{itemize}
the model can predict the mode of saturation and therefore the time per iteration. 

$t_{conv}(k)$ can be calculated given the throughput of each node and the network speed. It has two components:\\$t_{conv, compute}(k)$ and $t_{conv, network}(k)$.

Let us define $t_{conv, compute}(1) = T_{c,c}$, i.e. $T_{c,c}$ is the time it takes for a single machine ($k=1$) to compute the forward and backward pass of the convolution phase. Similarly, let us define $t_{conv, network}(1) = T_{n,c}$, i.e. $T_{n,c}$ is the time needed for a single copy of all the conv phase's models (forwards pass) and model gradients (backwards pass) to be passed over the network. We will describe how to determine these two quantities later.

The computation time for the convolution phase for $k > 1$ is then $t_{conv, compute}(k) = T_{c,c} / k$, because recall that a single compute group performs computation on a single batch of data (data parallelism), i.e. the amount of data per group is always the same per iteration (e.g., $b$ images) and so if there are $k$ machines in a group, each will process $b/k$ images. We assume a linear speedup.

On the other hand, the time for the network communication increases with $k$. This is because of increased network communication from the conv model server, i.e. the model needs to be sent to $k$ workers simultaneously and gradients will be received from $k$ workers simultaneously (all requests are made at almost the same time, because the workers in the group are synchronous). The network time for the convolution phase for $k > 1$ is then $t_{conv, network}(k) = T_{n,c} * k$. Here, we assume a linear slowdown.

So while the compute time decreases with $k$, the network time increases with $k$. We assume that both of these are linear. Empirically we notice that the convolution computation does not scale exactly linearly with $k$: on 8 c4.4xlarge machines in a single group, the forward pass of the convolution becomes $7.2\times$ faster and the backwards pass $6.6\times$ faster.
Similarly, we observe that the network slowdown is usually linear but can be super-linear, which we attribute to thrashing.

Given $t_{conv, compute}(k)$ and $t_{conv, network}(k)$, we can naively approximate
\begin{equation*}
t_{conv}(k) = t_{conv, compute}(k) + t_{conv, network}(k)
\end{equation*}
However, these two can be done in parallel, i.e. while one layer is computing its forwards pass, the model for the next layer is being sent over the network. This does not entirely overlap because the first layer needs to complete its backwards pass before requesting the model for its next forwards pass, but we can approximate the total convolution phase time as:
\begin{equation*}
t_{conv}(k) = max( t_{conv, compute}(k), t_{conv, network}(k) )
\end{equation*}
Finally, it is necessary to obtain $T_{c,c}$, $T_{n,c}$ and $t_{fc}$. We measured these because they only need to be measured once (not for each $k$), but they can be calculated using the node throughput and network throughput:
$T_{c,c}$ and $t_{fc}$ can be approximated by counting the total number of operation from each GEMM operation in the conv and fc phases and assuming that BLAS achieves the device peak. $T_{n,c}$ can be approximated by counting the total number of bytes in the conv models and assuming the peak network throughput is achieved. These assumptions are justified because the matrices and models are large.

\begin{figure}
\centering
\includegraphics[width=0.5\textwidth]{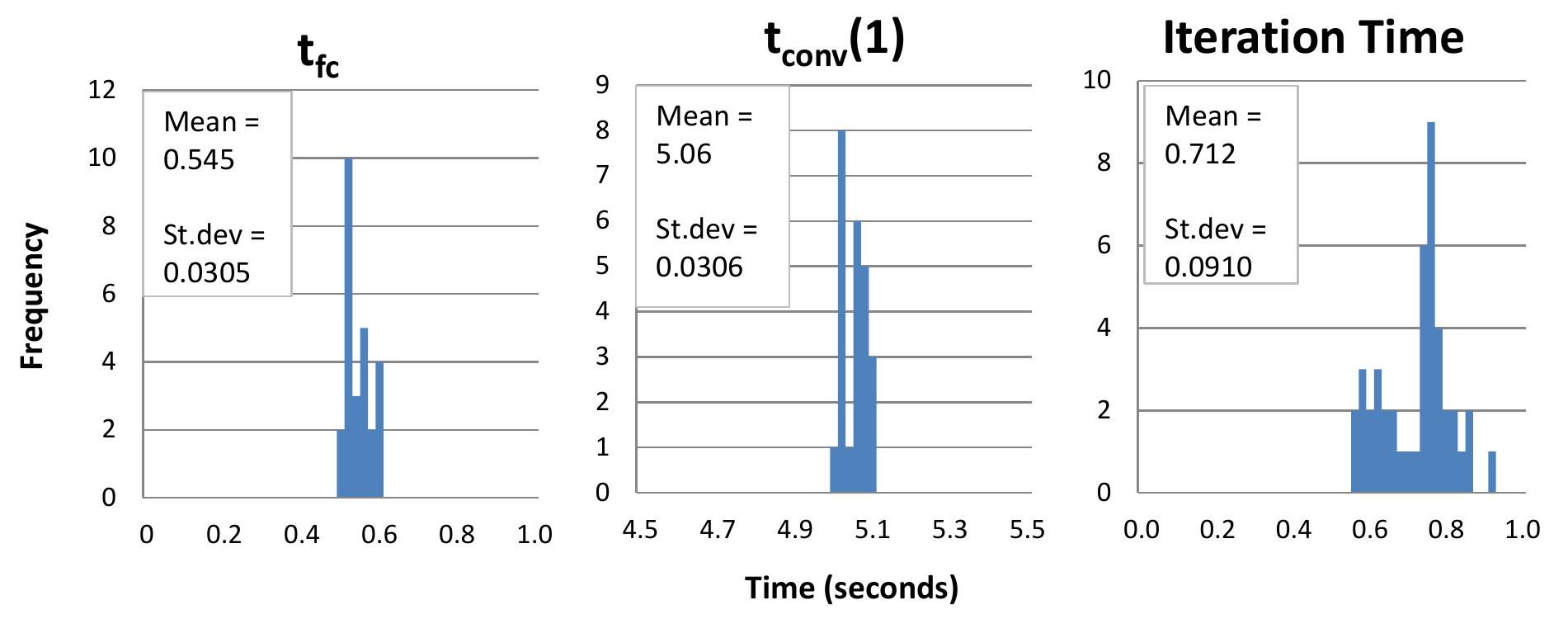}
\caption{Variance of \HE times. These are on a cluster of 9 CPU machines, 8 Conv compute groups, 1 machine per group}
\label{fig:HE_hisogram}
\end{figure}

Also note that measurements of these quantities are accurate for all iterations because deep learning computations are dense and so there is little variation in the computation time across iterations, as shown in Figure~\ref{fig:HE_hisogram}. Note that there is a standard deviation of than $6\%$ for $t_{conv}(1)$ and $t_{fc}$, and a standard deviation of $8\%$ for the total iteration time. For CIFAR-10, the standard deviation for total iteration time was $1.5\%$. We also observed similar variances on a GPU cluster.

Using measurements of $T_{c,c}$, $T_{n,c}$ and $t_{fc}$, Figure~\ref{fig:HE_predict} shows that the analytic model of hardware efficiency is accurate. When the FC server is saturated (right side of the graph), the model is almost exact. When the FC server is not saturated (left side of the graph), the slowdown and speedups are not exactly linear, and we under-estimate the time per iteration.\\

While this analytic model may seem specific to CNNs, it extends to any deep learning model because its derivation relies only on queuing theory, not any specific properties of CNNs.

\subsubsection{Further Optimizations}

A primary goal for understanding the hardware efficiency above is to determine possible optimizations.

\paragraph*{\textbf{Saturated Conv Server}}

We showed above that as the group size ($k$) increases, there is no longer FC saturation, because 
\begin{equation*}
t_{conv}(k) + t_{fc} > g t_{fc}
\end{equation*}
and recall
\begin{equation*}
t_{conv}(k) = max( t_{conv, compute}(k), t_{conv, network}(k) )
\end{equation*}
Specifically, in Figure~\ref{fig:HE_predict} the case of a single group (left side of the graph) is so much slower than FC saturation (right side of the graph) because 
$ t_{conv, network}$ becomes
very large, i.e. the time it takes to send the conv model to all 32 machines in the group is significantly greater than the computation time, which is small due to data parallelism across 32 machines.

In particular, note that a single, large group is slower than many small groups not because of synchronization delays due to machine variance exists, but because of increased network congestion, although some variance does exist in the computation time across machines.

Microsoft's Adam~\cite{Chilimbi:2014:OSDI} discusses techniques to mitigate both of these problems, from not requiring each worker in a group to finish processing all of its images to adding multiple NIC cards on the parameter server machines. FireCaffe~\cite{iandola15firecaf} uses the technique of reduction tress for their parameter servers to reduce communication overhead, which allows them to reduce this network congestion and scale to many machines in the synchronous case using a larger batch size.

\paragraph*{\textbf{Saturated FC Server}}

The optimizations above improve the hardware efficiency for the synchronous case, or generally for small $g$ and large $k$. On the right side of Figure~\ref{fig:HE_predict} (small $k$, large $g$), the FC server becomes saturated and no further speedups are possible. 
Recall that this is because the FC compute and FC model servers are mapped to the same physical machine. Also recall that this has benefits for both hardware efficiency (by reducing network communication) as well as statistical efficiency (by reducing the staleness of the FC model). However it means that there is only one FC compute server, and so it may become a computational bottleneck.
This is seen in the top Gantt chart of Figure~\ref{fig:HE_Model_Gantt} in which there are blank spaces which indicate un-utilized machines.

A simple way to fix this is to make the FC machine faster, for example if there is limited access to GPU hardware, it it best to use them on this machine (indeed, we see in Section~\ref{sec:Experiments} that the GPU cluster does not saturate FC). Another technique is to use multiple machines for the FC phase, for example using model parallelism across multiple machines such that each machine stores a portion of the FC model (this still has a staleness of 0 for the FC phase).

Another solution is to remove this physical mapping, i.e. rather than have a single FC compute server mapped to the FC model server, to have a separate FC compute server per Conv compute server. This removes the bottleneck of a single FC compute server, although it also sacrifices the benefits mentioned above. Section~\ref{sec:Experiments} demonstrates the consequences of this decision experimentally.

\paragraph*{\textbf{Physical Mapping Details}}

In addition to mapping the FC compute and model servers to the same physical machine, note that the conv model server does minimal computation and can also be mapped physically to the same machine as the FC compute/model server or one of the conv compute machines. The primary concern with this server is network congestion, so it makes more sense to map to one of the conv compute server's machines.

In addition to multiple servers physically mapping to a single machine, it is also possible for a server to physically map to many machines, for example using multiple machines in a model server to implement a reduction tree as in FireCaffe, or mapping 4 FC Compute servers to a machine that contains 4 GPUs, etc. Another example is merging the FC compute and FC model server and mapping them to the same physical hardware as in Figure~\ref{fig:arch_simple} (b), but where that hardware is not a single machine 
as shown in Figure~\ref{fig:arch_simple} (b)
but multiple machines e.g., using model parallelism.

Finally, a common technique is to ``pipeline'' the servers by mapping multiple conv compute servers to the same physical machine.
For instance in the synchronous case (1 group of $N$ machines), during the FC computation all $N$ machines are idle (because they are waiting for the FC to return data gradients before they can begin the backwards pass of the conv phase). During this idle time those machines can be processing a different batch, i.e. $N=32$, but there are two groups of size 32. Note that in this example, this pipelining increases staleness from 0 to 1. Using this pipelining, now the time per iteration in conv saturation becomes:
\begin{equation*}
\text{Time per iteration}_{\text{saturated conv}} = t_{conv}(k) / g
\end{equation*}
i.e. the FC time is completely hidden.

\subsection{Statistical Efficiency Model}

Because asynchrony can be viewed as increasing implicit momentum, asynchrony can be made equivalent to synchronous execution by properly reducing the explicit momentum in order for the total explicit + implicit momentum to stay the same as the optimal momentum of the synchronous case.
This is true as long as the implicit momentum is less than the optimal momentum of the synchronous case.
This is a key discovery because it means that staleness can exist in the system without incurring a statistical penalty, which is advantageous for hardware efficiency.
Also, making the momentum stay the same (rather than just ignoring this result and letting there be extra momentum) is important because a total momentum that is too high will diverge, which we show experimentally in Appendix~\ref{sec:app:optimizer}.
This theory also successfully predicts measured system behavior: Figure~\ref{fig:momentumstaleness} shows the predicted and actual measured momentum for several popular deep learning problems.
In both cases, upon reducing the explicit momentum to compensate for implicit momentum, we observe no SE penalty.
Moreover, the momentum increase due to asynchrony closely matches the theoretical curve for both datasets.
Our study is the first to identify a relationship between hyper-parameters and asynchrony, and next these results are used to design the optimizer in Section~\ref{sec:Optimizer}.

\section{Appendix for Distributed Optimizer (Section~\ref{sec:Optimizer})}

\label{sec:app:optimizer}

This section describes how to use the models from the previous two sections to choose (1) a physical mapping which maps each server to a machine, and (2) an execution strategy which defines the number of compute groups by allocating data batches to each server. As in previous sections we assume that the number of machines are fixed.

\subsection{Selecting the Batch Size}
\label{sec:app:optimizer:batchsize}

\begin{figure}
\centering
\includegraphics[width=0.5\textwidth]{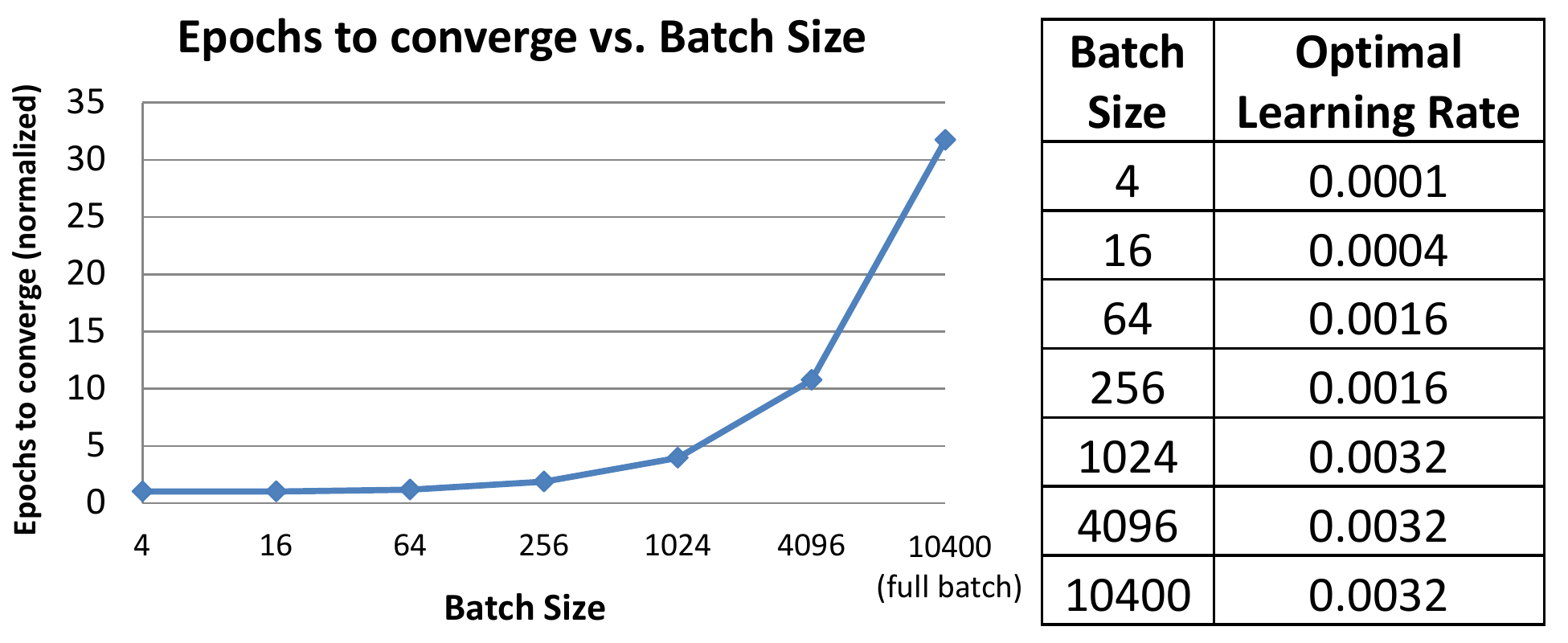}
\caption{Using a batch size that is too large greatly increases the \# epochs to converge (Imagenet 8-class).
This slowdown begins when the optimal learning rate no longer scales with the batch size.
}
\label{fig:batch_size}
\end{figure}

We first study the batch size in Figure~\ref{fig:batch_size}, which uses the imagenet-8 dataset with $S=0$ and momentum $\mu=0.9$. 
The x axis varies the batch size 
and the y axis plots the \# passes over the dataset (or \textit{epochs}) until convergence.
For each batch size we used
an oracle to find the optimal learning rate, $\eta^{*}$. $\eta > \eta^{*}$ diverged.

We see that as long as $\eta^{*}$ increases with the batch size, there is little penalty for larger batch sizes.
This is because larger batch sizes provide a
truer gradient each iteration and permit a larger $\eta$ before divergence,
therefore while a larger batch consumes more of the dataset, the progress made by each step is greater.
$\eta^{*}$ cannot scale infinitely however, and plateaus beyond $\eta^{*}=0.0032$.
As a result, larger batch sizes make no more progress per iteration than smaller batch sizes,
but consume much more of the dataset each iteration. This is catastrophic for performance
(it can take $30\times$ more epochs to converge) as computation is effectively ``wasted'',
which is why neural networks have been trained with SGD rather than batch gradient
descent since the early days.
This also greatly exceeds the staleness cost 
incurred of ``splitting'' a large batch into smaller, asynchronous batches (which we show is nearly flat), which is why asynchrony 
is necessary for systems to scale to very large clusters.

This suggests that the optimizer needs to tune batch size, however for imagenet-8 and other datasets
we observe that this performance penalty is negligible around 256 (specifically we use 256 for Imagenet, 128 for
CIFAR-10 and 64 for MNIST, based on published results for these datasets).
In principle the optimizer
could tune $b$ as well, but we observe that unless $b$ is too large the penalty is small so we don't
study this in more detail.  

\subsection{Physical Mapping}
\label{sec:app:optimizer:pm}

As discussed in the text, for the physical map which maps servers to machines, we map the FC compute and model servers to the same machine (i.e. ``merge'' the FC servers, which as~\cite{Chilimbi:2014:OSDI} argues reduces communication overhead because it avoids communicating the large FC model) and use one compute group for the FC phase. The rest of the machines are used for the conv compute servers. The conv model servers are mapped to one of the conv compute machines.

Empirically we show in Appendix~\ref{sec:app:distexperiments:largecluster} that this mapping is best for both hardware and statistical efficiency:
on a cluster of 33 EC2 c4.4xlarge machines, not merging the FC servers incurs an additional hardware efficiency penalty of $1.2\times$
due to increased communication
as well as a statistical efficiency penalty of $2.5\times$
because of staleness in the FC model.
Our current optimizer therefore always chooses this server architecture,
although Appendix~\ref{sec:app:distributed:hemodel} (for Section~\ref{sec:HE_Model}) described other scenarios in which these penalties are justified to eliminate FC saturation,
as well as additional optimizations within this server architecture (such as reduction trees
or multi-machine model parallelism for the FC phase).

\subsection{Optimizer Details}
\label{sec:app:optimizer:details}

\label{sec:app:optimizer:optimizer}
For each epoch, Algorithm~\ref{alg:optimizer}
performs an adaptive grid search over both the learning rate and
the momentum starting at the current value of $g$. Specifically, we
  run each learning rate and momentum (see below) for one minute and select the configuration with
  lowest loss after 1 minute of execution. If after 1 minute all these
  configurations have the same loss, we continue to run another minute
  until there is a clear winner in terms of loss (we determine this using a threshold of 5\% from the loss of the past 50 iterations, although a statistical test can be used as well). 
We then run this best $(\mu^{*}, \eta^{*})$ for an hour and then rerun the optimizer.
  
  One could use more sophisticated
  parameter search routines, but this took less than 10\% of the
  execution time.

We search the learning rate as follows. Let the learning rate and momentum
used in the previous 1 hour epoch (i.e. the result of the grid search from that
epoch) be $(\mu^{*}_{last}, \eta^{*}_{last})$. 
For the current epoch, we then search $\eta \in \{\eta^{*}_{last}, \eta^{*}_{last}/10\}$,
and $\mu \in \{0.0, 0.3, 0.6, 0.9\}$. As an optimization to prune the search space, we do
not search $\mu > \mu^{*}_{last}$ when $\eta = \eta^{*}_{last}$, as we notice empirically
that as the run progresses, the optimal total momentum decreases. 

If the optimal $\mu^{*} = 0.0$, we try $\mu^{*} = 0.1$ and $\mu^{*} = 0.2$ as well, 
and if the lowest loss is still achieved at $\mu^{*} = 0.0$, we decrease $g$ and repeat the search.

$(\mu^{*}, \eta^{*})$ in the initial (cold-start) phase are selected using a similar
procedure, described in Appendix~\ref{sec:app:optimizer:coldstart}.

\subsubsection{Empirical Validation}

\begin{figure}
\centering
\includegraphics[width=0.5\textwidth]{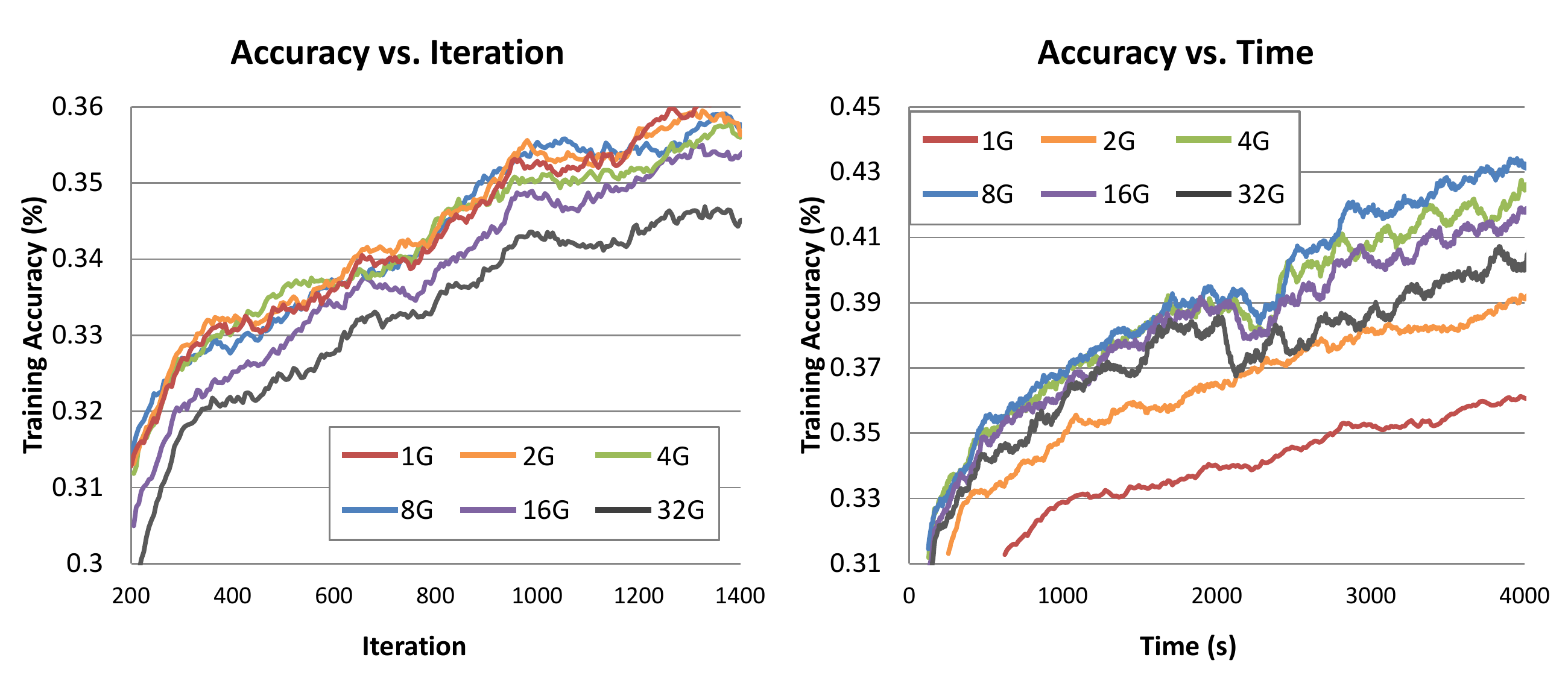}
\caption{Accuracy vs. iteration and Accuracy vs. Time for each execution strategy shown in Figure~\ref{fig:sec4full}}
\vspace{-1.0em}
\label{fig:CG_Opt}
\end{figure}

We now justify the theoretical results of Section~\ref{sec:SE_Model} experimentally.
We use a cluster of 33 EC2 CPU machines (CPU-L in Figure~\ref{tab:machine}),
the Imagenet 1000-class dataset, and the AlexNet CNN. 
We run each execution strategy from sync ($g = 1$ conv compute group) to async ($g = 32$), each for a single epoch (1 hour).
Specifically, we plot 6 strategies: $g = \{1, 2, 4, 8, 16, 32\}$, where the
number of conv compute machines is fixed to 32.

Each strategy starts from the same checkpoint (i.e. same loss), but achieves a different
final loss by the end of the epoch. We select the lowest final training loss achieved by 
all strategies, $\ell_{F}$, and plot three measures in Figure~\ref{fig:sec4full}: (a) the time
per iteration (hardware efficiency, \HE), (b) the number of iterations to reach $\ell_{F}$ (statistical
efficiency, \SE), and (c) their product, which
is the total time to reach $\ell_{F}$. For completeness, each 
strategy uses an oracle to exhaustively find its optimal explicit momentum, $\mu^{*}$ (within $0.3$) and 
optimal learning rate $\eta^{*}$ (within an order of magnitude. 
For all strategies $\eta^{*} = 0.01$ was optimal).
The \HE curve in (a) is the same as in Figure~\ref{fig:HE_predict} (b). 

We see in (c) that $g=32$ (fully asynchronous) is $3.7\times$ faster to reach $\ell_{F}$ than $g=1$ (synchronous). 
This is due to its faster iteration time (\HE) in (a), although it requires $1.8\times$ more iterations (\SE) to reach $\ell_{F}$ in (b).
This matches the theory's prediction: increasing $g$ decreases the explicit
momentum $\mu^{*}$, which falls to 0 at $g=32$ (see Figure~\ref{fig:momentumstaleness} (right)) , and consequently there is a penalty in \SE.
The optimizer of Algorithm~\ref{alg:optimizer} would therefore select $g=16$, which is near-optimal.

However, our optimizer additionally
employs an optimization to leverage the \HE model from Section~\ref{sec:HE_Model}:
because the FC server saturates at $g=4$ (determined analytically or through measurements
during the cold start phase), the optimizer will ``short-circuit'' Algorithm~\ref{alg:optimizer}
to begin at $g=4$ instead of $g=32$, and ends up selecting $g=4$, which is $5.3\times$ faster than sync.

Figure~\ref{fig:CG_Opt} shows the accuracy vs. time (and for reference accuracy vs. iter, i.e. statistical efficiency) for each configuration (\# groups) in Figure~\ref{fig:sec4full}. Recall that the optimizer selected 4 groups because with proper momentum tuning the statistical efficiency was nearly the same for all curves, but 4 or more groups had the best hardware efficiency.
Figures~\ref{fig:sec4full} and~\ref{fig:CG_Opt} use ImageNet 1000 class (1 hour of training) as described above with 33 EC2 c4.4xlarge machines.

\subsection{Cold Start Phase}
\label{sec:app:optimizer:coldstart}

The model is trained synchronously before beginning asynchronous execution in
Algorithm~\ref{alg:optimizer}. This is needed in order to set the appropriate scale for the weights.
However fully synchronous execution may be slow, and just as an optimization to Algorithm~\ref{alg:optimizer}
was to run asynchronously only up to FC saturation, similarly this section focuses on accelerating training
during the cold-start phase.
In particular, a fully synchronous execution may significantly increase the duration of the cold-start phase due to 
poor hardware efficiency, and so a cold-start run with slight asynchrony may more quickly terminate the cold start phase.
Therefore, tuning the number of compute groups is also important for the cold-start phase.

\paragraph*{Cold Start Grid Search} 
To do this, as in Algorithm~\ref{alg:optimizer}, we grid search hyper-parameters for each number of groups
($1, 2, 4, \ldots, N$, for $N$ machines). For each we also grid search learning rate and momentum. 
We use a standard adaptive grid search algorithm for its simplicity. For each staleness, the algorithm searches for optimal settings of momentum and learning rate by running each configuration of parameters for 1 minute, and selecting the parameters with the lowest loss after 1 minute. 

The search happens as follows, and is similar to the search in the steady-state execution of Algorithm~\ref{alg:optimizer}. We start with $S=0$, fix the momentum to 0.9, and run 1 minute for each learning rate $\eta \in \{0.1, 0.01, 0.001, 0.0001, 0.00001\}$. We search from lowest to highest and stop early if a learning rate produces a final loss worse than the previous learning rate (or if a learning rate causes divergence). We select the learning rate which has the lowest loss after 1 minute. Call this $\eta^{*}_{sync}$. Therefore for $S=0$, the optimal configuration $(\mu^{*}, \eta^{*}) = (0.9, \eta^{*}_{sync})$. We do not tune momentum for sync because 0.9 is standard~\cite{Krizhevsky:2012:NIPS}, because this saves optimizer time, and because there is no implicitly induced momentum due to asynchrony for $S=0$.

For the remaining $S$ after $S=0$, we perform the following iteration: We increase the number of groups to the next highest power of 2 (after sync, this is $g=2$, then $g=4$, $g=8$, etc, i.e. $S=1,3,7,\ldots$) Let the optimal configuration from the previous $S$ be $(\mu^{*}_{last}, \eta^{*}_{last})$. For the current $S$, we run a grid search for $(\mu, \eta) | \mu \in \{0.0, 0.3, 0.6, 0.9\}, \eta \in \{\eta^{*}_{last}, \eta^{*}_{last} / 10\}$. I.e., $\eta^{*}$ defines the search range for the next $S$. In addition, $\mu^{*}$ reduces the search space for the next $S$: we do not search a higher momentum than $\mu^{*}_{last}$ while searching $\eta = \eta^{*}_{last}$.

We notice empirically that there is not a large impact of running a finer grid for momentum (although this can be done by adding a second-phase of search which fixes $\eta^{*}$ and searches $\mu$ around $\mu^{*})$.
Running multiple random seeds (network weight initializations) can also be used to find a good starting point for the SGD algorithm (this is a known technique).
Tuning parameters is not a novel idea in machine learning, but unlike existing work our 
problem is more sophisticated as we are coupling tuning hyper-parameters and execution strategies (staleness).
Our work is the first to show that hyper-parameters and execution strategies need to be tuned jointly to avoid divergence as staleness increases. 

Once we obtain $(\mu^{*}, \eta^{*})$ for each $S$, we then run each $S$ for one minute at a time until there is a clear winner in terms of loss (we determine this using a threshold of 5\% from the loss of the past 50 iterations, although a statistical test can be used as well). We then run this best $S$ with its $(\mu^{*}, \eta^{*})$ for an hour (the cold-start period).

\paragraph*{Parameter Search Experiments} 
The remainder of this section describes experiments motivating the pruning above, in particular why a larger staleness does not need to try larger learning rates or larger momentum values at the same learning rate.
There are a number of insights which allow us to prune the search space for the cold-start phase and
reduce the total search time. 
We discovered that as staleness increases, the optimal learning rate and momentum parameters when $S=0$ cause divergence (loss goes to infinity) for larger staleness values, e.g., $S=31$. This makes sense given our theoretical foundation from the steady-state optimizer: staleness induces implicit momentum, hence if explicit momentum is not decreased as $S$ increases, total momentum can be $> 1$ and cause divergence. As $S$ increases, we note that one or both of $\eta$ and $\mu$ need to be reduced otherwise SGD will diverge (loss goes to infinity).

\begin{table}[]
\centering
\caption{(Cold Start) Optimal momentum and learning rate during the cold start for various datasets/networks and various amounts of staleness in the conv model}
\begin{tabular}{|l|l|l|l|}
\hline
\textbf{\begin{tabular}[c]{@{}l@{}}Dataset\\ (Network)\end{tabular}}               & \textbf{Staleness} & \textbf{\begin{tabular}[c]{@{}l@{}}Optimal\\ Momentum\end{tabular}} & \textbf{\begin{tabular}[c]{@{}l@{}}Optimal\\ Learning\\ Rate\end{tabular}} \\ \hline
\multirow{3}{*}{\begin{tabular}[c]{@{}l@{}}MNIST\\ (LeNet)\end{tabular}}           & 0                  & 0.6                                                                 & 0.1                                                                        \\ \cline{2-4} 
                                                                                   & 31                 & 0.0                                                                 & 0.1                                                                        \\ \cline{2-4} 
                                                                                   & 127                & 0.8                                                                 & 0.01                                                                       \\ \hline
\multirow{3}{*}{\begin{tabular}[c]{@{}l@{}}CIFAR-10\\ (Krizhevsky)\end{tabular}} & 0                  & 0.9                                                                 & 0.001                                                                      \\ \cline{2-4} 
                                                                                   & 31                 & 0.7                                                                 & 0.0001                                                                     \\ \cline{2-4} 
                                                                                   & 127                & 0.1                                                                 & 0.0001                                                                     \\ \hline
\multirow{2}{*}{\begin{tabular}[c]{@{}l@{}}ImageNet-8\\ (CaffeNet)\end{tabular}}   & 0                  & 0.6                                                                 & 0.01                                                                       \\ \cline{2-4} 
                                                                                   & 31                 & 0.0                                                                 & 0.01                                                                       \\ \hline
\end{tabular}
\label{table:staleness_tune}
\end{table}

Table~\ref{table:staleness_tune} shows the optimal parameters for the same datasets and networks used in Figure~\ref{fig:Phe}, at different staleness values. Here the optimal parameter settings are defined as the parameter settings with which the training converges in the fewest number of iterations. We say that a model has converged once the training accuracy reaches 99.5\%. Recall that a staleness value of $S$ corresponds to $N = S+1$ parallel groups updating the model asynchronously. In these experiments, the staleness of the fully-connected models was zero, and so only the conv model had staleness.
Note that for imagenet 8-class there are only 10400 examples and batch size is 256 so 128 parallel workers was not possible (there is insufficient data).
Note that these small datasets all converge in under an hour and therefore consist entirely of the cold-start
phase in our implementation. 
While the cold-start phase of these datasets is less than an hour
(e.g., MNIST converges in seconds),
and therefore Algorithm~\ref{alg:optimizer} could be run part-way during 
execution to select a better strategy for the remainder of the execution (e.g., asynchronous), 
the overhead of re-running Algorithm~\ref{alg:optimizer} is not
justified for these small datasets, i.e. it is faster to treat the entire run as the cold-start phase.
Therefore we use these smaller datasets to study the cold-start phase.

The table shows that, with a fixed batch size, as staleness increases the optimal momentum and/or learning rate decreases, and in some cases not decreasing these parameters and reusing the parameters for $S=0$ causes divergence. 
Also, we see that decreasing the learning rate means momentum can increase again. Intuitively this is because momentum can be viewed as increasing the learning rate (larger SGD steps), and so if the learning rate is decreased too much, momentum increases to compensate for this decrease. Our grid search searches orders of magnitude for the learning rate, following from previous work~\cite{Krizhevsky:2012:NIPS}, but decreasing the learning rate by a smaller factor may avoid the need for momentum to increase and provide faster overall convergence. We leave this exploration to future work.

\section{Appendix for Distributed Experiments (Section~\ref{sec:Experiments})}
\label{sec:app:distexperiments}

\subsection{Single-Node Experiments}

See Appendix~\ref{sec:app:singlenode::experiments}.

\subsection{Small Cluster Experiments}
\label{sec:app:distexperiments:smallcluster}


This section provides additional details of the experimental setup. For the end-to-end experiment on ImageNet 1000, see Appendix~\ref{sec:app:distexperiments:endtoend}.

We ran all systems to $99\%$ accuracy (which we define as convergence) with a timeout of 2 hours.
For each system we used the same CPU and GPU external libraries as discussed in Appendix~\ref{sec:app:singlenode::experiments}.

We further sped up other tools by applying our optimizer to the extent that no code change was required.
MXNet offers both the sync and async strategy. Given our observation that async requires tuning parameters, to ensure training did not diverge we ran each strategy of MXNet with 4 orders of magnitude of the learning rate for 10 minutes each. We then selected the best strategy (as the static optimizer would) and ran it until convergence or timeout.
For SINGA we followed the same procedure and tried all available configurations. SINGA supports not only sync (1 group) and async (1 machine per group) strategies, but also intermediate group sizes. We ran SINGA with 1, 2, 4, and 8 machines per group, and also 4 orders of magnitude for the learning rate $\eta$ in each case. All runs were also for 10 minutes, and then as with MXNet the best one was run to convergence.

For Omnivore we ran our optimizer, which merged the FC compute and model servers to one machine and used the other 8 machines as conv compute machines. As with SINGA, the optimizer searched statically among 1, 2, 4 and 8 machines per group, but for 1 minute per execution setting. Overall the optimizer ran for less time than the tuning we did to ensure no divergence for MXNet and SINGA.

We followed the tutorials for each system and also ensured that all three systems used identical networks and parameters, including weight initializations and data preprocessing.

The network we use is CaffeNet, which is Caffe's version of Alexnet\footnote{\scriptsize{\url{https://github.com/BVLC/caffe/blob/master/models/bvlc_reference_caffenet}}}. AlexNet is the standard network for ImageNet and Caffe is the most widely used CNN framework, so this ensures reproducibility.
The weight initializations, batch size, regularization, and other hyperparameters are the same as CaffeNet, with a few minor differences:

Unlike Caffe and SINGA, MXNet does not easily support learning rate and weight decay multipliers, or different initializations for each model and bias. For consistency across tools, we therefore just made all 3 tools use the same weight initialization scheme, which is Gaussian with mean 0 and standard deviation 0.01, and no multipliers.

MXNet and SINGA do not support the grouping in AlexNet (which was done in 2012 to save GPU memory), so this is disabled from CaffeNet (and also is not important anymore as GPUs have more memory)

No random data preprocessing was used (crop, mirror), and we show convergence on the training set, not a test or validation set. We do this because these are machine learning concerns/optimizations and our focus is the system. We do subtract the image mean to avoid divergence.

Similarly, we disable the learning rate schedule in all tools and use a constant learning rate. We do this because we only train on a subset of ImageNet and to reduce the search space of the parameter configurations.

The following subsections describe in detail the individual settings used for each system to ensure fairness in our comparison.

\subsubsection{Detailed Settings for Both Systems}

For both systems we built in a wall-clock timer to ensure accurate timing. We created and shuffled the dataset using the tools provided by the systems: MXNet required shuffling beforehand, SINGA provided an \texttt{im2rec} utility. Those tools also were used to calculate the image mean: MXNet automatically generated the mean file when it ran and SINGA as part of \texttt{im2rec}. Because we focus on the training set we removed any validation set from the tools to ensure no time was spent on that. We used the provided AlexNet examples for each system and changed them only as above to ensure identical settings across all three systems (e.g., weight initializations, L2 regularization, etc.)
Accuracy was reported instead of loss for all tools to ensure consistency. The MXNet examples do not report loss so we used their accuracy \texttt{eval\_metric}. Moreover MXNet's \texttt{acc} metric is by default on the entire epoch while SINGA's is since last time printing, so we averaged the logs to ensure consistency across all three tools.

\subsubsection{Detailed MXNET Settings and Results}

We removed all machine learning optimizations from both tools except those described above. For MXNet this meant removing gradient clipping.
Because we ensure the parameters are used for all tools, including the batch size (256 for CaffeNet),
this meant that for MXNet's \texttt{dist\_sync} strategy on 8 machines, a batch size of 32 was used, and for 32 machines, a batch size of 8 was used (other tools partition the batch size for the sync strategy, i.e. they partition $b$ images by sending $b/N$ to each sync worker, but MXNet uses that batch per worker, i.e. they use a batch size of $b \times N$).
We fixed the random seed to 1 so the initialization is always the same.

We created a single data file and ensured that each worker read it from different location (using \texttt{ImageRecordIter} as in the AlexNet example). The timer was added as modified version of the \texttt{speedometer} callback (using \texttt{time.time()}, which is wall-clock time in python).

We used a cluster of 9 machines in these experiments because MXNet's AWS documentation instructs to ``Prepare a host file with all slaves's private IPs''. Therefore in order to test parallelism across 8 machines, we opened 9 EC2 machines, ran MXNet from the master (root) machine, and placed the other 8 machines in the \texttt{hosts} file. 

On the cluster of 9 c4.4xlarge machines we ran the 4 orders of magnitude learning rate for each execution strategy and noticed after 10 minutes that the best was learning rate 0.01 and sync, so we ran until 99\% convergence. We needed 4 orders of magnitude to ensure that the optimal setting was never on the ``boundary'' of the interval, i.e. the optimal $\eta$ we report was superior to an order of magnitude higher $\eta$ and lower $\eta$. For async no parameter setting had high accuracy after 10 minutes: The best sync was 60\% in 10 min and the best async got to 20\% in 10 min, in spite of better hardware efficiency for async (72 s per epoch async, 120 s per epoch sync). The best async was with learning rate 0.0001.

On the cluster of 9 g2.8xlarge machines (again following MXNet's documentation, 1 parameter server machine and 8 compute machines), we tried both cuDNN v3 and v4 and found no speed difference. Again we searched learning rate and found that 0.01, sync was best.

On the c4.4xlargs machines we ensured each worker was using all the cores, and on the g2.8xlarge machines that all 4 GPUs on each worker were utilized (using \texttt{nvidia-smi}).

\subsubsection{Detailed SINGA Settings and Results}

For SINGA, the timer built into \texttt{TrainOneBatch}. It also uses wall-clock time (\texttt{clock\_gettime}, same as Omnivore). Again we use the default AlexNet example with only small changes to make all weight initializations the same (as in MXNet), and to remove learning rate / weight decay multipliers (since not supported easily in MXNet). 

Tuning parallelism within groups: we tried tuning\\\texttt{partition\_dim} for each layer. Specifically we first used the default, i.e. \texttt{partition\_dim} commented out (as in their AlexNet example). We then uncommented those recommended \texttt{partition\_dim} settings in the example (i.e. dim 0 or batch parallelism for convolution, and dim 1 or model parallelism for FC) and found no difference, so we left\\ \texttt{partition\_dim} commented out as in the default AlexNet SINGA example.

Tuning parallelism across groups: To ensure different data for each worker we tried specifying a \texttt{random\_skip} in the data section but this made no difference. Documentation v0.1.0 suggested using \texttt{random\_skip} but in v0.2.0 (which we used) it was deprecated, so as with \texttt{partition\_dim} we left this out and used the default AlexNet SINGA example settings.

Next, we had to select for each machine number of workers per process. For each machine, we tried 1 process of 8, 16, and 32 threads on a single machine. The number of physical cores is 8 on the c4.4xlarge, and virtual cores is 16 (\texttt{nproc} $=$ 16). 16 was fastest so we used 16 workers per process.

As with MXNet we followed the documentation for SINGA and also included 8 machines in the hostfile. We ran 4 orders of magnitude for the learning rate as described above and found that 0.0001 and 4 groups of 2 machines each was best after 10 minutes. The result was noisy however, and looked similar to the distributed results in the ``SINGA: Putting Deep Learning in the Hands of Multimedia Users'' paper. We then ran the best configuration for but it did not converge in 3 hours (got to 70-80\%).

SINGA GPU distributed did not work at the time of this study so it is not included.

\subsubsection{Detailed Omnivore Settings and Results}

Omnivore was run using the same network and parameters as the systems above.
The optimizer was run as described above in Appendix~\ref{sec:app:optimizer}.
Each configuration was searched by the optimizer for 1 minute and 
search time was reduced by pruning the space across staleness values.
The second search phase was skipped (momentum granularity was 0.3).
We fixed the random seed to 1 so the initialization is always the same.
The overall 
optimizer time was less than the search time to avoid divergence in other tools.

On the CPU cluster, the strategy chosen was the same as with MXNet, i.e. sync with $\eta=0.01$. Momentum is untuned for both Omnivore and MXNet, i.e. 0.9, because our contribution is tuning momentum to compensate for staleness and sync has no staleness. Since the parameters and staleness are the same as MXNet, as expected Omnivore achieves the same statistical efficiency. However, it is $2.3 \times$ faster in terms of hardware efficiency, for an overall end-to-end convergence speedup of $2.3\times$. On the GPU cluster, the optimizer chose 2 groups of 4 m/g, and was $5.0\times$ faster to converge. The following section studies the benefits of the optimizer in more detail, and also examines how the tradeoffs change on a large cluster which has more options for execution strategies: for example, the extreme strategies of sync or async may not be sufficient for a larger cluster. This may prevent MXNet, which only supports these strategies, from scaling to a larger cluster.

\subsection{Detailed Tradeoff Space Analysis}
\label{sec:app:distexperiments:detailed}

This section analyzes the speedups observed in the previous section to understand the contribution of each tradeoff selection that the optimizer made. These tradeoffs include (1) execution strategy (number of groups), (2) optimizing hyper-parameters to compensate for staleness, and (3) physical plan (server to machine allocation).

\subsubsection{Penalty Definition}

\begin{figure}[t] 
\centering
\includegraphics[width=0.5\textwidth]{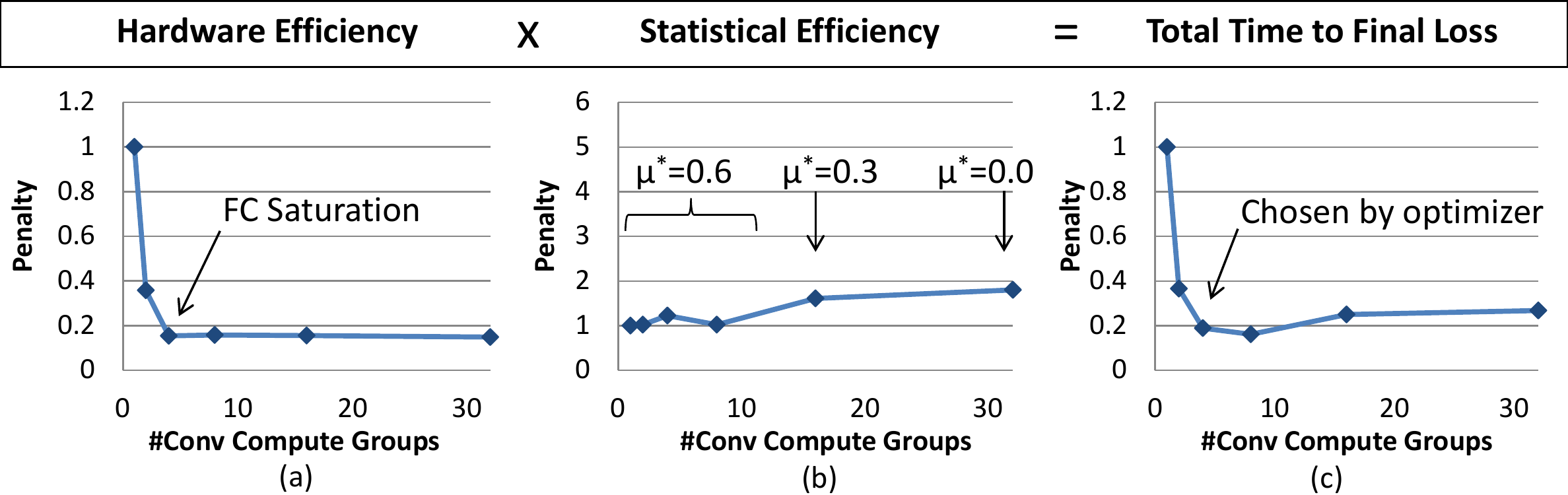} 
\caption{Recall Figure~\ref{fig:sec4full} (replicated here for convenience and also showing momentum values from Figure~\ref{fig:momentumstaleness} (right))}
\label{fig:sec4full2}
\vspace{-1.0em}
\end{figure}

Consider again Figure~\ref{fig:sec4full}, which we've replicated here for convenience in Figure~\ref{fig:sec4full2} and also shown momentum (note that when the optimal explicit momentum is 0, there is an associated \SE penalty). Recall this figure showed the tradeoff for compute groups on the CPU-L cluster for an epoch of ImageNet 1000. The \HE and \SE plots were multiplied to produce the right-most plot of total time to reach a final loss. 

The vertical axis of the \SE figure in Figure~\ref{fig:sec4full2} shows what we call the \SE penalty, $P_{SE}(S)$, which is defined as the ratio of the \# iterations needed to converge relative to the case of no staleness ($S=0$),
\begin{equation}
P_{SE}(S) = \frac{SE(S)}{SE(0)}
\label{eq:SE_Penalty}
\end{equation}
$P_{SE}$ is dimensionless, because it is the ratio of statistical efficiencies (\#iter / \#iter).
The penalty is 1 when the staleness is 0, and should be higher for all $S > 0$. A higher $P_{SE}$ is worse (more iterations to converge).

Recall also that we defined \textit{hardware efficiency penalty} ($P_{HE}$). This is shown in the middle graph of Figure~\ref{fig:sec4full2}.
Since the statistical efficiency penalty is defined as the ratio of the \# iterations to convergence with respect to $S=0$, for consistency $P_{HE}(S)$ is also normalized with respect to $S=0$.
$S=0$ is the case of 1 compute group, also called the synchronous case. The hardware efficiency penalty is defined as the ratio of the time per iteration relative to the time per iteration for this synchronous case,
\begin{equation}
P_{HE}(S) = \frac{HE(S)}{HE(0)}
\label{eq:HE_Penalty}
\end{equation}
As with $P_{SE}$, a higher $P_{HE}$ is worse (more time per iteration).
Whereas $P_{SE}(S)$ increased with staleness, for hardware efficiency this trend is reversed: $P_{HE}$ decreases (iterations become faster) as the number of compute groups increases.

Note in these figures, the staleness is 0 for the FC model, so staleness on the horizontal axis refers only to the conv models (i.e. the number of conv compute groups). 

Finally, recall that the product of hardware and statistical efficiency is the total time to convergence.
Since the horizontal axis (staleness, i.e. \# groups) is the same on both the \SE and \HE plots, these plots can be multiplied, and the resulting vertical axis is the \textit{total penalty}, defined as the ratio of the total time to convergence (normalized to sync, i.e. $S=0$):
\begin{equation}
P_{\text{Total}}(S) = P_{\text{SE}}(S) \cdot P_{\text{HE}}(S)= \frac{SE(S) \cdot HE(S)}{SE(0) \cdot HE(0)}
\label{eq:Total_Penalty}
\end{equation}

We use figures of this format throughout this section to quantify the benefit of the choice of compute groups.

\subsubsection{End-to-End Imagenet 1000}

\begin{figure}[t] 
\centering
\includegraphics[width=0.5\textwidth]{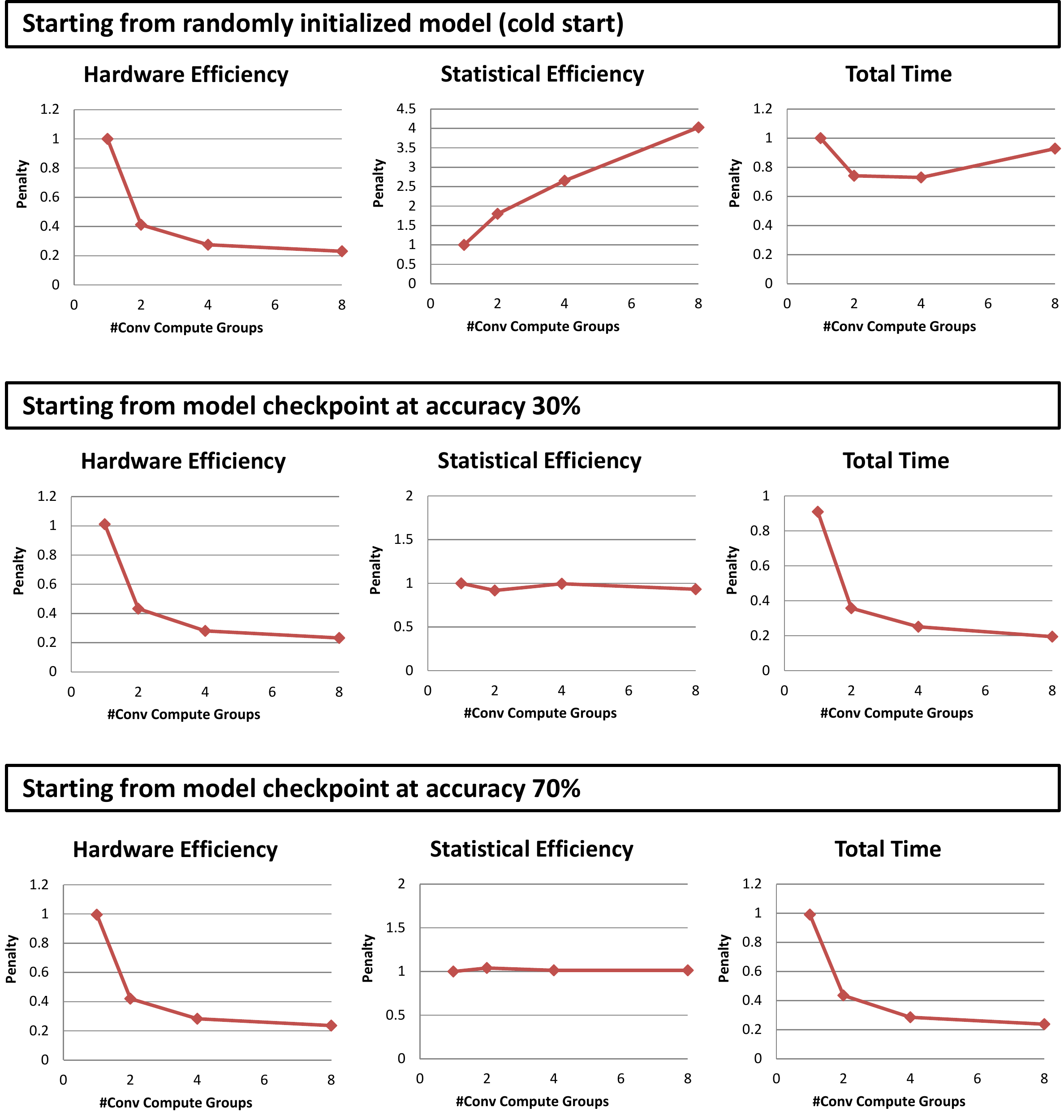}
\caption{Cold start vs Steady state for Imagenet 1000 on GPU-S}
\label{fig:cold_vs_ss_1k_gpu}
\vspace{-1.0em}
\end{figure}

\begin{figure}[t] 
\centering
\includegraphics[width=0.5\textwidth]{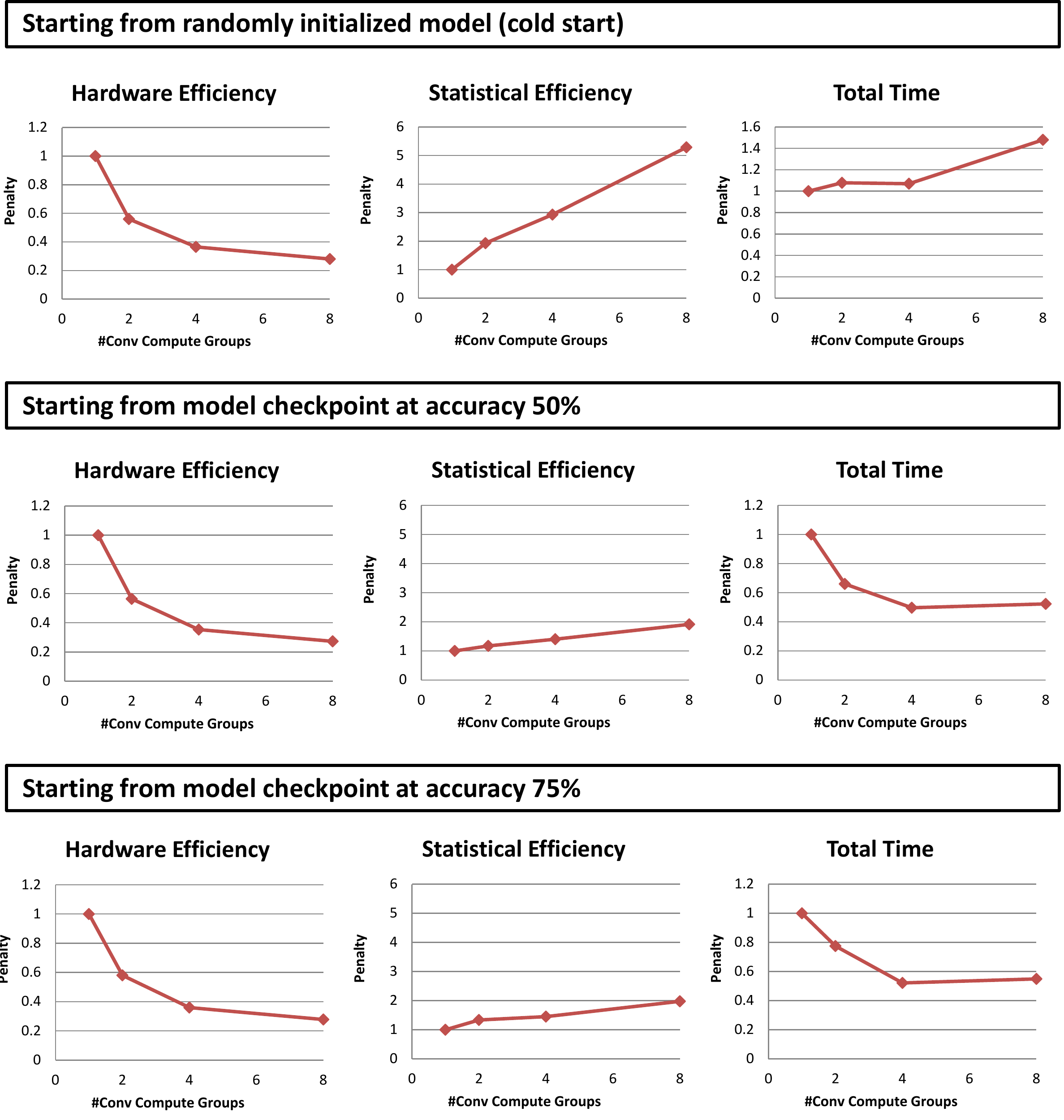}
\caption{Cold start vs Steady state for cifar on GPU-S}
\label{fig:cold_vs_ss_cifar_gpu}
\end{figure}

Recall that Figure~\ref{fig:sec4full2} was run for a 1-hour optimizer epoch of Imagenet 1000, on the CPU-L cluster, as discussed in Appendix~\ref{sec:app:optimizer:details}. 
Figure~\ref{fig:cold_vs_ss_1k_gpu} (bottom 2 sets of figures) shows the same experiment on the GPU-S cluster. Notice again that \SE is flat, i.e. maximum asynchrony is optimal.

These figures show steady-state execution, hence the \SE curves show no penalty (nearly flat). The same curves but for the cold-start epoch of the GPU-S cluster are shown in the top of
Figure~\ref{fig:cold_vs_ss_1k_gpu}.

We also validate this for the small cifar dataset in Figure~\ref{fig:cold_vs_ss_cifar_gpu}. Here for exposition we reduced the epoch size to 2 minutes (otherwise the cold start would converge after only a few minutes).

\subsubsection{Small Clusters}

\begin{figure}[t] 
\centering
\includegraphics[width=0.5\textwidth]{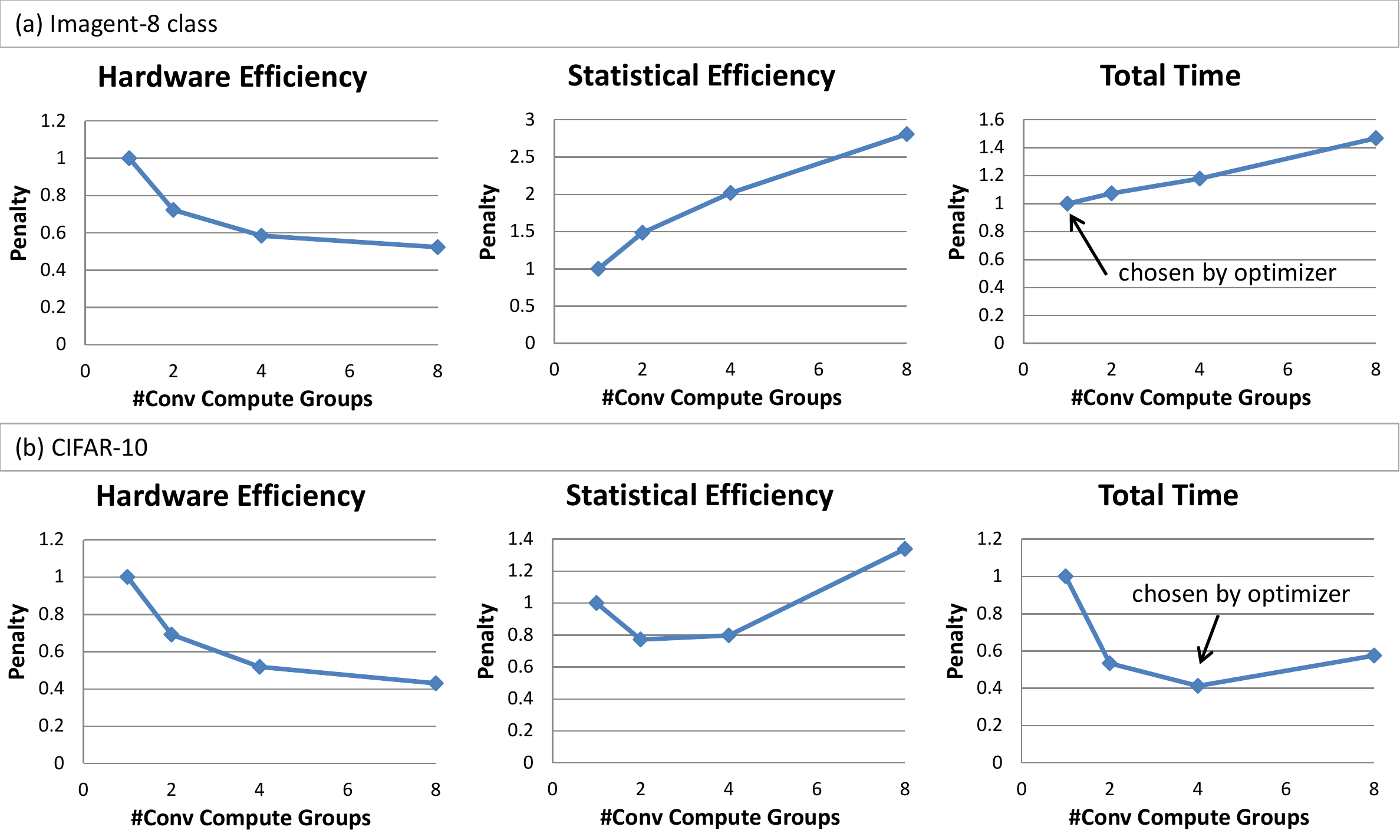} 
\caption{Imagenet 8-class and CIFAR-10 tradeoff on a cluster of 9 EC2 c4.4xlarge CPU machines.}
\label{fig:8cpu_tradeoff}
\vspace{-1.0em}
\end{figure}

\begin{figure}[t] 
\centering
\includegraphics[width=0.5\textwidth]{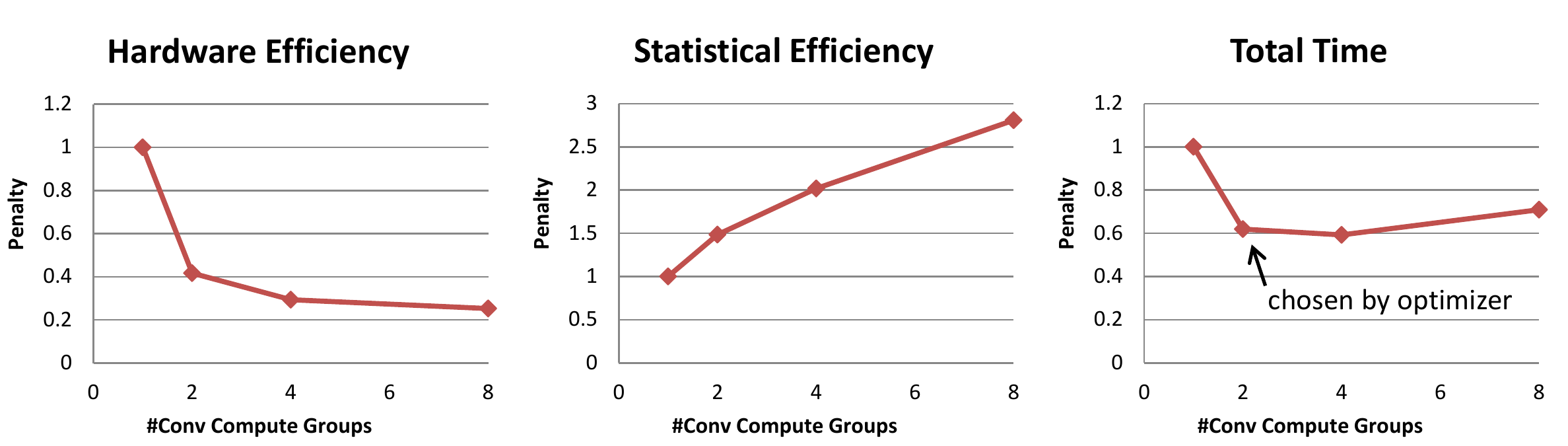} 
\caption{Imagenet 8-class tradeoff on a cluster of 9 EC2 g2.8xlarge GPU machines (36 GPUs total).}
\label{fig:8gpu_tradeoff}
\end{figure}

The optimizer's choice of execution strategy for the small cluster experiments (Figure~\ref{tab:e2e_imagenet8}(a) and (b))
is shown in Figure~\ref{fig:8cpu_tradeoff} (CPU-S) and
Figure~\ref{fig:8gpu_tradeoff} (GPU-S). 
In addition to the imagenet-8 dataset we also 
include CIFAR-10 to show the optimizer is robust across datasets.
We see that for these small
clusters, choosing the execution strategy incorrectly incurs a penalty
of roughly $1.5\times$. 

Note that Figure~\ref{fig:8cpu_tradeoff} and Figure~\ref{fig:8gpu_tradeoff} have the same statistical efficiency curves but different hardware efficiency curves. This is because the difference in throughput between the GPU machines and CPU machines exceeds the difference in network speed between these clusters so there is a higher penalty for the sync case on the GPU cluster. Also, neither of these 9 machine clusters reach FC saturation.

Next we consider the CPU-L cluster.

\subsubsection{Large Cluster}
\label{sec:app:distexperiments:largecluster}

\begin{figure}
\centering
\includegraphics[width=0.5\textwidth]{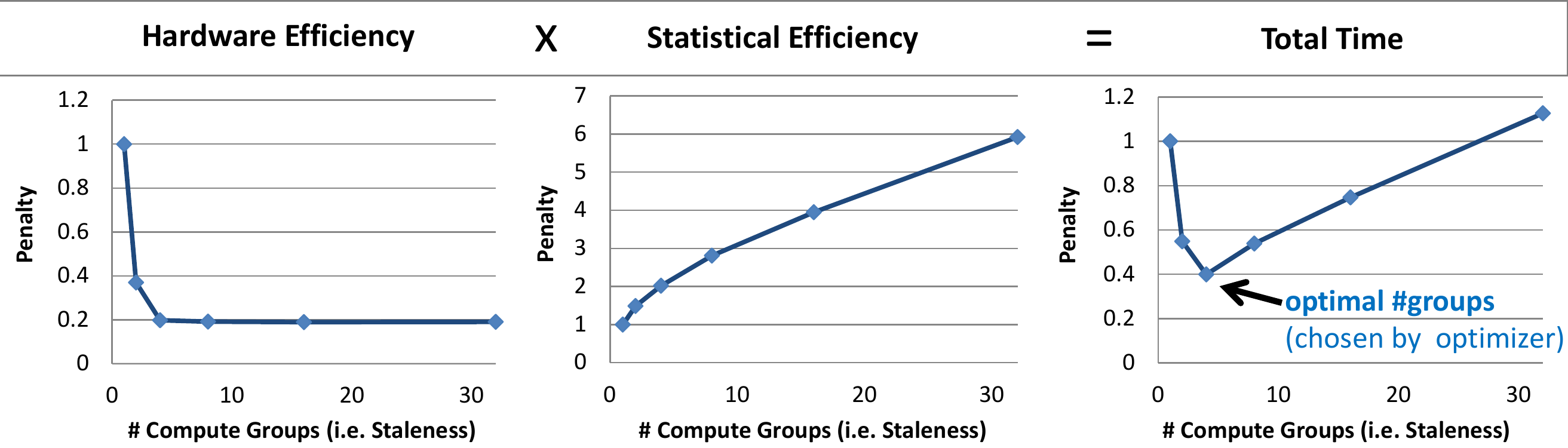}
\caption{Hardware efficiency, statistical efficiency, and total time tradeoff curve (ImageNet 8-class, 33 machines)}
\vspace{-1.0em}
\label{fig:sec4full_8}
\end{figure}

The tradeoff for CPU-L (Figure~\ref{tab:e2e_imagenet8}(c))
is shown in
Figure~\ref{fig:sec4full_8}. 4 groups (8 machines per group)
was the optimal point, and that the optimizer chose this execution
strategy.  The detailed tradeoff space for CPU-L is analyzed in
Figure~\ref{fig:Tradeoff_New}. Each curve, from the bottom up,
represents a selection made by the optimizer. We've isolated each
selection to observe their relative impact.

\begin{figure}[t] 
\centering
\includegraphics[width=0.5\textwidth]{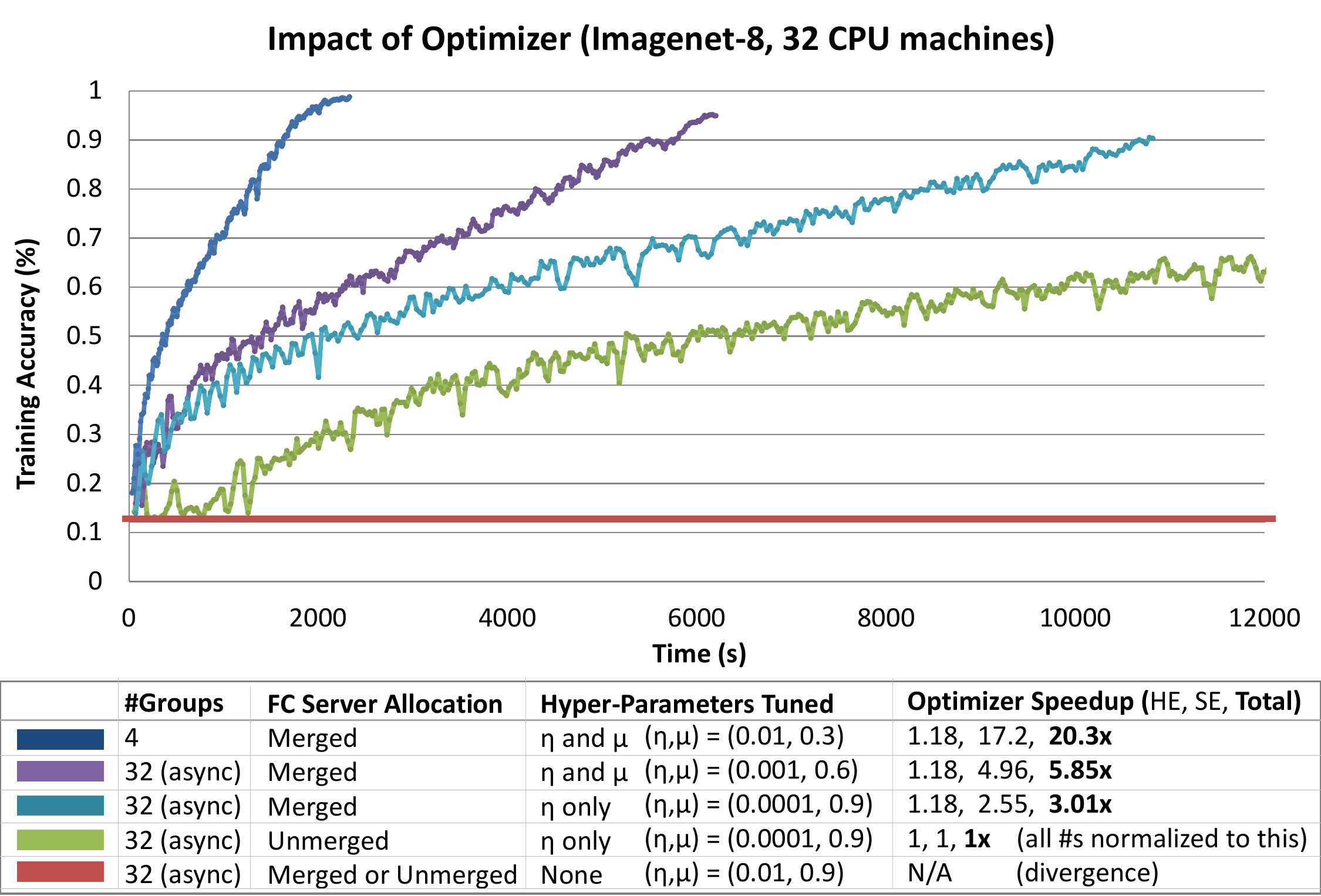} 
\caption{Tradeoff curve showing the impact of each dimension of the optimizer.}
\label{fig:Tradeoff_New}
\vspace{-1.5em}
\end{figure}

\paragraph*{Avoiding Divergence}
First, consider the red line, which represents the default point
chosen by many systems: asynchronous with a large number of
machines. Indeed, statistical efficiency is often ignored by other
systems and so by default, the configuration with the best hardware
efficiency (fastest iteration time) is erroneously selected.
However if the published AlexNet hyper-parameters~\cite{Krizhevsky:2012:NIPS} (which are
optimal for the sync case) are naively used in the async case, there is
divergence. Thus, our tuning approach is
critical.

The green curve shows that if only the \textbf{learning
  rate} $\eta$ is tuned, divergence can be avoided. Tuning $\eta$
is also common practice, although prior work does not do so explicitly
to compensate for staleness as we advocate. 
In the green curve momentum is not been tuned, as many systems always use a momentum of 0.9 (as mentioned in~\cite{Krizhevsky:2012:NIPS}). For example, at the time of this study MXNet hard-codes this momentum into every example \footnote{\scriptsize{\url{https://github.com/dmlc/mxnet/blob/52ea0f0cbbf5eaaf38a2341e57afd6829f88a86d/example/image-classification/train\_model.py\#L77}}}.

Also, the green curve does not merge the FC compute and model servers by physically mapping them to the same machine. Instead, this curve represents the architecture shown in Figure~\ref{fig:arch_simple} (a), i.e. there is an FC compute server for each CONV compute server, and each of these server pairs is mapped to a separate machine. Of the 33 machines therefore, one machine contains the CONV and FC model servers, while each of the other 32 contains a CONV compute and FC compute server. This configuration represents the strategy chosen by MXNet, so we report their async curve as the green line because their system is optimized for this case (note that doing this disadvantages our final speedup figure, i.e. if we had used Omnivore's implementation of this tradeoff point our optimizer's speedup would be $>20\times$).
The remaining curves have their hardware and statistical efficiencies normalized to those of this green curve.

\paragraph*{Device Mapping}
We now examine our choice of merging the FC compute and model servers to
the same machine (the 33rd machine), as Section~\ref{sec:Optimizer} described.
The other systems do not support this merging so we take ``unmerged'' as the
baseline, and use the 33rd machine for the conv and FC model servers
as MXNet and SINGA's documentation suggests.
The remaining curves will have their hardware and statistical efficiencies
normalized to those of this curve.

The turquoise curve merges the \textbf{FC servers}. We see that this
gives a $1.18\times$ improvement to hardware efficiency (due to
reduced communication) and a $2.55\times$ improvement to statistical
efficiency (due to no staleness in the FC model). Overall, this is
$3.01\times$ faster to converge than the baseline.

In the turquoise curve, note that the hardware efficiency improvement is only $1.18\times$ for the merged FC. As discussed in Section~\ref{sec:HE_Model}, this is because while communication is reduced by merging these servers, on the large CPU cluster the FC saturation point is reached at 4 groups, and not mapping the FC servers to the same physical machine as described above eliminates the saturation (but requires more network communication and incurs a statistical efficiency penalty).

\paragraph*{Parameter Tuning for Staleness}
Divergence can always be avoided by tuning $\eta$ alone, and indeed 
most systems always use a momentum of 0.9 (see comment above) which is
standard for the sync case~\cite{Krizhevsky:2012:NIPS}.
However,
we show in the purple curve that at larger staleness values,
additionally tuning the \textbf{momentum}
$\mu$ permits using a higher $\eta$. This does not change hardware efficiency, but now gives an
overall $5.85\times$ speedup over the baseline due to improved statistical
efficiency.

\paragraph*{Execution Strategy}
Finally, the blue
line represents the actual choice made by the optimizer. In addition
to the selections above, recall that the optimizer did not choose 32 groups but
\textbf{4 groups} (Figure~\ref{fig:sec4full_8}), which further improves the statistical efficiency
to give an overall speedup of $>20\times$ compared to standard choices
traditionally made by deep learning systems. Note that changing the
number of groups to 4 did not hurt hardware efficiency because the FC
server is already saturated (see Section~\ref{sec:HE_Model}).

\subsection{Scalability to a Larger Cluster}
\label{sec:app:distexperiments:scalability}

The previous section (Appendix~\ref{sec:app:distexperiments:detailed}) showed that as a result of the optimizer's tradeoffs Omnivore is able to scale to 32 machines. In this final experiment we compare Omnivore to the best competitor from the small clusters, MXNet, on this larger cluster. We use the same dataset and network as the small cluster experiments, and define convergence the same way (99\% accuracy). 

We attemped to open a cluster of 33 g2.8xlarge instances but continuously ran into EC2 errors related to not enough machines available (InsufficientInstanceCapacity).

As in Section~\ref{sec:small_cluster}, we repeat the same procedure to apply our optimizer to MXNet, i.e. we run each configuration for 10 minutes, select the best execution strategy and learning rate, and run that to convergence. The best strategy was once again sync with $\eta=0.01$.
Specifically, as in the previous experiments, we followed MXNet's documentation and used the EC2 master machine as the root, and put the other 32 workers in the hostfile.
We ran MXNet for 10 minutes with both sync/async and 4 orders of magnitude learning rate as described above. This time all 4 orders of magnitude for $\eta$ were needed because for sync with 32 machines, after 10 minutes 0.001 and 0.01 were almost the same, although 0.001 was better by $\sim5\%$ points. Since this differed from the optimal $\eta$ from the 8 machine case, which was 0.01 (i.e. statistical efficiency changed for the larger cluster), to be sure we ran MXNet with each $\eta$ to convergence and noticed that in fact 0.01 was significantly faster to converge in the end, as was true on the smaller clusters. Similarly, for async after 10 minutes the best $\eta$ was $0.00001$, although this was close to 
$0.0001$, therefore once again we ran both to convergence and noticed that $0.0001$ was faster to converge for MXNet. We did not do this extended parameter tuning for Omnivore, and only ran the 1 minute static runs described above, to make sure that we were getting the best possible performance from MXNet for our comparison.

For Omnivore the optimizer was used as described in the previous section.
The best result of MXNet and Omnivore is shown in Figure~\ref{tab:e2e_imagenet8}(c). We see that Omnivore is $3.2\times$ faster than MXNet to converge now (it was $2.3\times$ faster on the 9 CPU cluster). In addition, we see that compared to Figure~\ref{tab:e2e_imagenet8}(a), Omnivore sped up on the larger cluster but MXNet did not. Therefore not only does the optimizer give speedups by not relying solely on the sync strategy and by merging the FC servers, but also enables scalability to more machines.

If we do not apply our optimizer to MXNet, Omnivore now converges $20\times$ faster. This is compared to MXNet's async strategy, which has poor statistical efficiency for 32 machines. This $20\times$ corresponds exactly to the speedup in the previous section because the green curve in that section is MXNet using the async strategy (i.e. they are the same point in the tradeoff, see the discussion in Appendix~\ref{sec:app:distexperiments:largecluster}) By applying our optimizer to MXNet we select the sync strategy instead, which lowers the gap with Omnivore to $3\times$ on this cluster. Therefore this section shows not only that the optimizer gives speedups of  more than an order of magnitude, but that it is versatile and can be applied to existing tools.

\subsection{End-To-End Experiments}
\label{sec:app:distexperiments:endtoend}

The end-to-end result is in Figure~\ref{fig:flagship}.
We trained
the standard CaffeNet (same setup as in Appendix~\ref{sec:app:distexperiments:smallcluster}) using ImageNet-1000 on both systems using both the CPU-L and
GPU-S clusters. We time out each run after 8 hours and report
the training accuracy vs. time.

According to MXNet's official performance tuning guideline \footnote{\scriptsize{\url{https://github.com/dmlc/mxnet/tree/db6f6a9418e5696b04be741a78a47ae877bb5505/example/image-classification}}},
they recommend trying the sync strategy, but also state that ``if the 
model size is quite large or you use a large number of machines, you may want to use dist\_async''.
Immediately above, they describe large as ``models with size $>>$ 100MB such as AlexNet and VGG.''
Because we are training AlexNet and use up to 33 machines, which may be considered large, then according to these instructions async
could be the best choice.
Because they do not provide an automatic mechanism to make this decision we followed this advice
and tried both strategies, as we did above in section F.2. 
This required tuning the learning rate for each strategy.
We used the optimal learning rate obtained for each strategy on ImageNet-8, as
recommended by~\cite{Bottou:2012:Tricks} which states that ``the best way to determine the correct learning rates is to perform experiments
using a small but representative sample of the training set''.
In addition, MXNet does not provide a learning rate schedule for their AlexNet example (as of the writing of this study) so we use the standard learning rate schedule of~\cite{Krizhevsky:2012:NIPS}
which decreased the learning rate by $10\times$ when training plateaued.

\begin{table}[]
\centering
\caption{Grid search parameters from Figure~\ref{fig:flagship} (a) on GPU-S}
\label{table:grid_search_GPUS}
\begin{tabular}{|l|l|l|l|}
\hline
Phase   & $\mu$  & $\eta$    & $g$ \\ \hline
cold    & 0.6 & 0.01  & 4 \\ \hline
phase 1 & 0.6 & 0.001 & 8 \\ \hline
phase 2 & 0.6 & 0.001 & 8 \\ \hline
\end{tabular}
\end{table}

\begin{table}[]
\centering
\caption{Grid search parameters from Figure~\ref{fig:flagship} (b) on CPU-L}
\label{table:grid_search_CPUL}
\begin{tabular}{|l|l|l|l|}
\hline
Phase   & $\mu$  & $\eta$    & $g$ \\ \hline
cold    & 0.6 & 0.01  & 2 \\ \hline
phase 1 & 0.6 & 0.001 & 4 \\ \hline
phase 2 & 0.3 & 0.001 & 4 \\ \hline
\end{tabular}
\end{table}

For \oursystem, we ran the optimizer end-to-end. We ensure a $10\%$ overhead by running the optimizer and then training for $10\times$ the optimizer time before rerunning the optimizer.
Each time the optimizer runs, it searches momentum, $\mu$ and learning rate, $\eta$, either reducing one, reducing both, or keeping them the same. The grid search results for each phase in Figure~\ref{fig:flagship} (a) and (b) are shown in Table~\ref{table:grid_search_GPUS} and Table~\ref{table:grid_search_CPUL}.
On the GPU-S cluster, the first time the optimizer ran $\mu$ remained at $0.6$ but $\eta$ decreased from $0.01$ to $0.001$. The second time, the parameters did not change. On the CPU-L cluster, the first optimizer run made the same choices: $\mu$ also began at $0.6$, and remained  $0.6$, while  $\eta$ decreased from $0.01$ to $0.001$. Following the second optimizer run, $\mu$ reduced from $0.6$ to $0.3$, and $\eta$ remained unchanged. Note that the cold-start (first) phase of the CPU-L run in Figure~\ref{fig:flagship} shows a very large slope, and then after the optimizer run this slope decreases. In ongoing work we are exploring a slope-aware optimizer which would not rerun the optimizer at this point but continue the execution. With this change we expect the gap against competitor systems to increase significantly.

\begin{figure}[t] 
\centering
\includegraphics[width=0.5\textwidth]{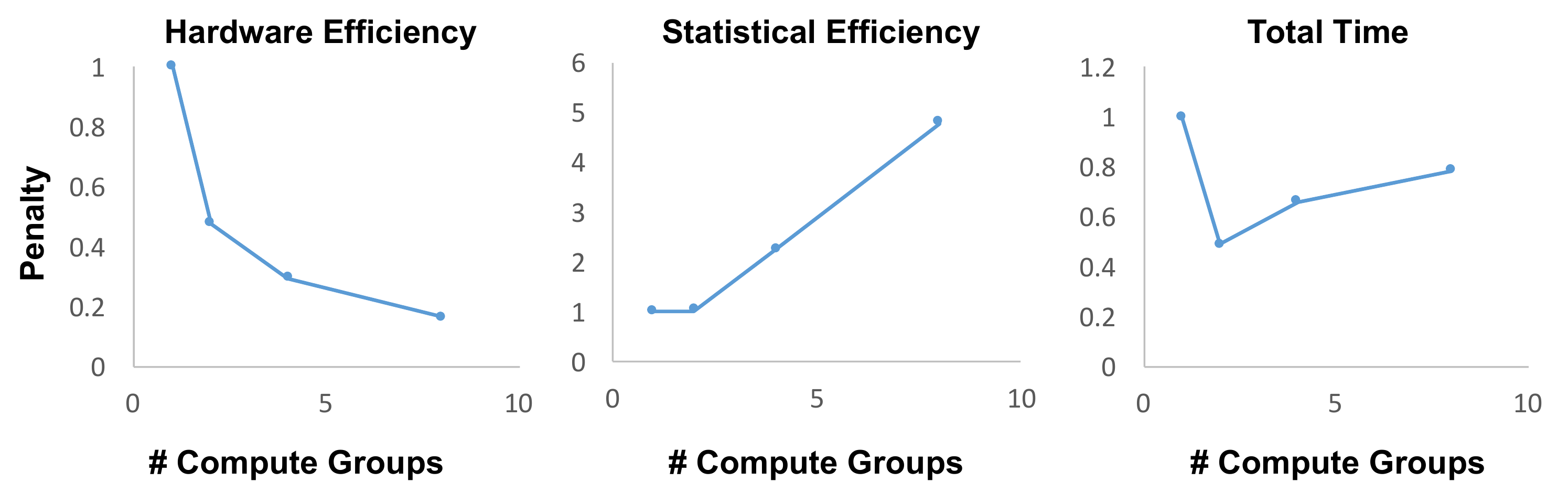} 
\caption{Recurrent Neural Network using 9 EC2 c4.4xlarge CPU machines.}
\label{fig:rnn_tradeoff}
\vspace{-1.0em}
\end{figure}

\subsection{Preliminary RNN/LSTM Result}
\label{sec:app:distexperiments:rnn}

To understand if our tradeoff applies more broadly, we implemented the
Recurrent Neural Network model and LSTM proposed by
Graves~\cite{Graves:2013}. 
Following the same protocol as Figure~\ref{fig:8cpu_tradeoff}
and using the CPU-S cluster,
we see in Figure~\ref{fig:rnn_tradeoff} that the tradeoff between
statistical efficiency and hardware efficiency is comparable, and choosing a completely synchronous or
asynchronous configuration can be up to 2$\times$ slower than the
optimal configuration. 
\footnote{
\scriptsize{
We also see a similar tradeoff in the LSTM variant proposed by
Graves~\cite{Graves:2013}.
}
}

\subsection{Comparison to Standard Schedules}
\label{sec:app:distexperiments:standardschedules}

\begin{figure}[t] 
\centering
\includegraphics[width=0.45\textwidth]{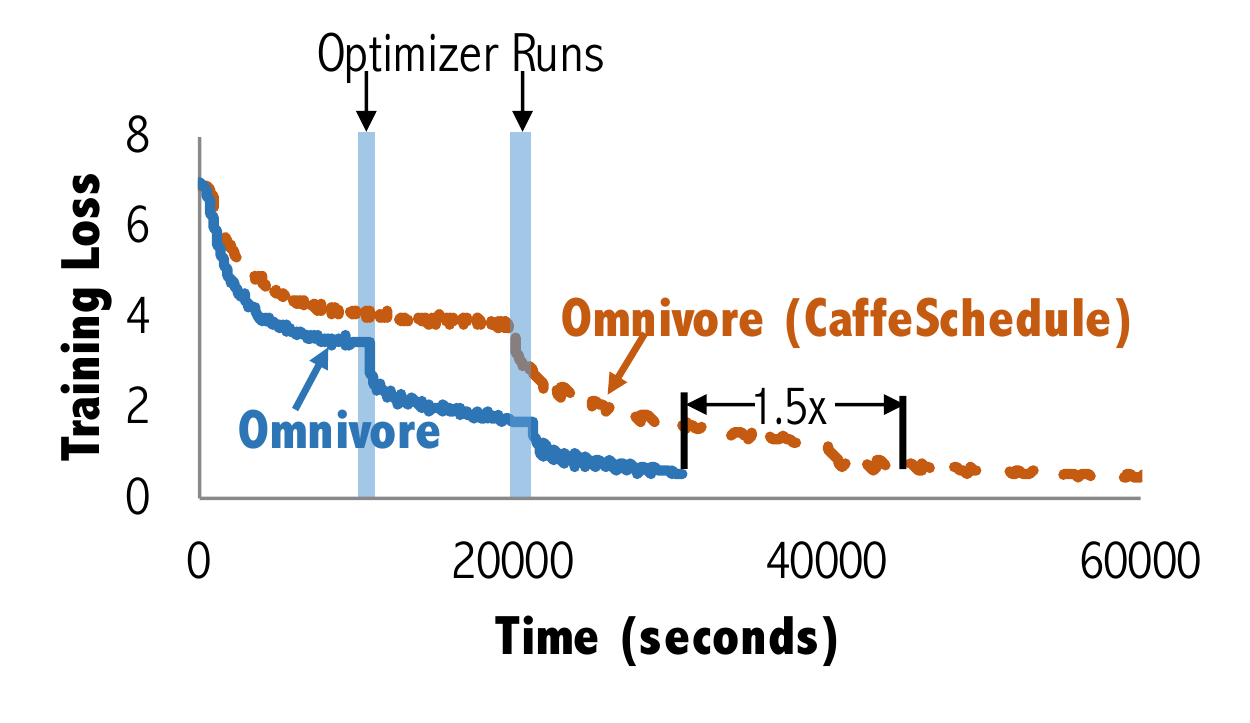} 
\vspace{-2.5em}
\caption{Comparison of Omnivore's optimizer to CaffeNet's default learning rate schedule on Full ImageNet with AlexNet.}
\label{fig:OptimizerValidationCaffe}
\end{figure}

\label{sec:comparisoncaffeschedule}
We have shown that tuning is critical for good performance.
We next validate the hypothesis that 
Omnivore's optimizer outperforms
standard tuning and parameter scheduling methods. 
To validate this, we run Omnivore on the full
ImageNet using the standard
CaffeNet. We run two versions of Omnivore:
(1) \textbf{Omnivore (Default Schedule)},
which uses CaffeNet's default learning
rate schedule that decreases the 
learning rate by $10\times$ every 100,000 
iterations; and (2) \textbf{Omnivore},
which uses the standard Omnivore optimizer.
To be fair, both versions use the same
grid search strategies to select the
optimal learning rate, momentum, and 
number of compute groups at the beginning.
In addition, we run \textbf{Omnivore} 
for $10\times$ the optimizer time 
before it re-optimizes the parameters.

Figure~\ref{fig:OptimizerValidationCaffe} shows the training loss vs. wall-clock time. 
The two plateaus shown in \textbf{\oursystem} correspond to the times Omnivore re-optimizes the parameters.
The losses
of both \textbf{Omnivore (Default Schedule)} and 
\textbf{Omnivore} decrease over time. 
However, after the first parameter re-tuning, \textbf{Omnivore}'s
loss starts to decrease more rapidly. 
Finishing at $36$K seconds, \textbf{Omnivore} 
is $1.5\times$ faster to achieve the same
loss as \textbf{Omnivore (Default Schedule)}.
\oursystem does not
require the user to specify the
number of iterations to run before
re-optimize the parameters.

\subsection{Comparison to Bayesian Optimizer}
\label{sec:app:distexperiments:bayesian}

\begin{figure}[t] 
\centering
\includegraphics[width=0.5\textwidth]{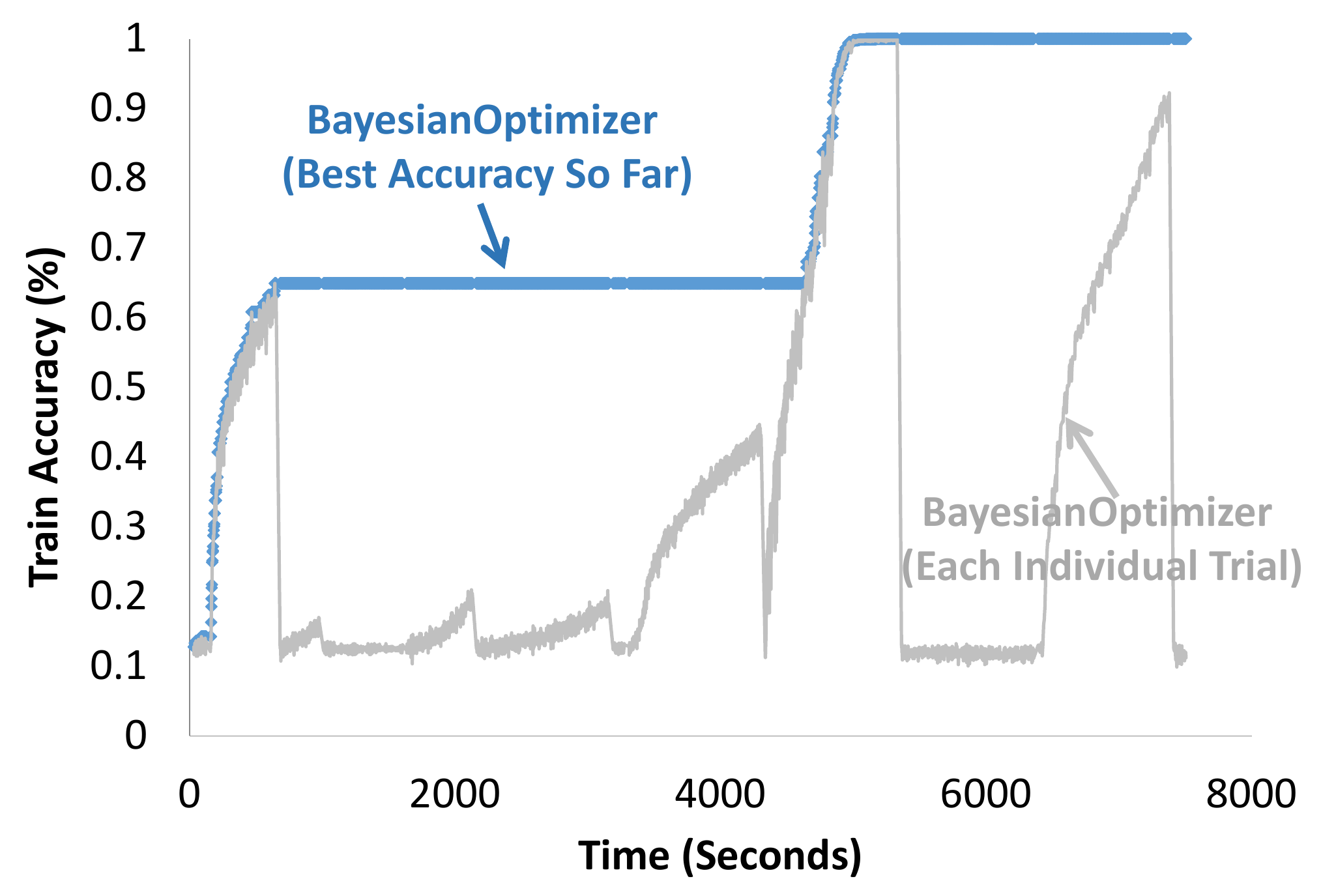} 
\caption{Bayesian optimizer run on Imagenet-8 using the GPU-S cluster.}
\label{fig:bayesian_fig}
\vspace{-1.0em}
\end{figure}

We compare
our simple optimizer with the state-of-the-art Bayesian
optimization approach that explores
the parameter space of CNNs. 
The results are shown in Figure~\ref{fig:bayesian_fig}.
We follow Snoek et al.~\cite{Snoek:NIPS:2012}
to model the search space as $(\eta, \mu, S, N)$
where $N$ is the number of epochs to run. We
use the same search space for $(\eta, \mu, S)$
as in our optimizer and measure both the
number of configurations and the total number of epochs 
that the Bayesian optimizer needs to run before finding
a run that achieves an accuracy within $1\%$ of 
\oursystem's best accuracy. 

Our procedure is as follows. We first run Omnivore to obtain
a run which reaches $99\%$ convergence using 
the same dataset and cluster as in 
Figure~\ref{tab:e2e_imagenet8} (b).
This took 80 epochs and 680 seconds.
We then give the Bayesian optimizer $N=80$
and it tries to fit $\eta, \mu$ and $S$ in order to reach the
lowest loss (highest accuracy) by a timeout of 1000 seconds. It searches
$N$ in the range $1, \ldots, 80$, 
$S$ in the range $1, 2, 4, 8$, 
$\mu$ in the range $0.0, 0.3, 0.6, 0.9$,
and learning rates in the range $0.1, 0.01, 0.001, 0.0001, 0.00001$, i.e.
the same as \oursystem searches.

It took the Bayesian optimizer on average 12
runs before finding a strategy which achieves accuracy
within $1\%$ of \oursystem's run.
On average this takes $6\times$ more epochs than 
just training that strategy to convergence, which makes the Bayesian approach
infeasible to run on Imagenet 1000 (whereas \oursystem's optimizer incurred only
a $10\%$ overhead).

Compared with our optimizer, one difference is that
we are using the first minute's execution as a
proxy for a longer run, while on the other hand,
Snoek et al. have the number of epochs to run
as a parameter to explore and do not share information
across runs. It is of course possible to use Bayesian optimization
to guide our grid search for the first minute, however,
it is future work to integrate this heuristic into
the Bayesian optimization framework in a principled way.

\section{TensorFlow Experiments}
\label{sec:tensorflow}
\label{sec:app:tensorflow}





In this section we see that tuning momentum can significantly improve the performance on platforms other than \oursystem by running experiments on TensorFlow  \cite{tensorflow}.
We use TensorFlow r0.9 starting with the fully synchronous and fully asynchronous implementations of Inception-v3 for ImageNet  \cite{chen2016revisiting} found in \cite{tfinception}.
Synchronous training has each worker send its gradients to the parameter server.
The parameter server aggregates all the gradients, applies them to update the weights and then sends the new weights to all the workers.
In the asynchronous configuration, each worker sends its gradients to the parameter server. 
The gradients are immediately applied to the weights and the new model is returned to the worker.  

On 32 GPU workers, asynchronous training reaches the same loss as its synchronous counterpart $1.5\times$ faster, when properly tuned. In contrast, when momentum is fixed to $0.9$ for both, synchronous training is faster.
We also implement compute groups on top of the code from \cite{chen2016revisiting} and report the tradeoff figures for a $32$ worker configuration. We discuss the experimental setup and discuss results on momentum tuning and compute groups.

\subsection{Experimental Setup}

We deployed 8 g2.8xlarge instances on AWS, each equipped with 4 GPUs. We allocate 1 GPU per worker node, meaning there are 4 workers per instance and a total of 32 worker nodes. There is a single parameter server that runs on one of the instances. Each worker uses a batch size of $32$ images,
as suggested in \cite{tfinception}.\footnote{
This is also the largest batch size that can fit on a single GPU of the g2.8xlarge instances.}
Notice that the suggested setup uses a per-worker batch-size as opposed to the per-group batch-size we used on \oursystem.
We used the SGD with momentum optimizer for Inception-v3 as opposed to RMSProp with momentum used in \cite{chen2016revisiting}.

We perform our experiments after a warm start:
we train the Inception-v3 network synchronously on ImageNet, until it reaches 50\% training accuracy and take a snapshot. This snapshot is used as the starting point for all measured runs.  We run each set of parameter values for 1 hour. We grid seach momentum values in $\{0.0, 0.3, 0.6, 0.9\}$ and learning rates in $\{0.005, 0.01, 0.05\}$. We experimented with other values and found these to capture the range of optimal learning rates. We measure the loss achieved by each configuration after $1$ hour of execution.

\subsection{Results}

\begin{figure}[t] 
	\centering
	\includegraphics[width=0.5\textwidth]{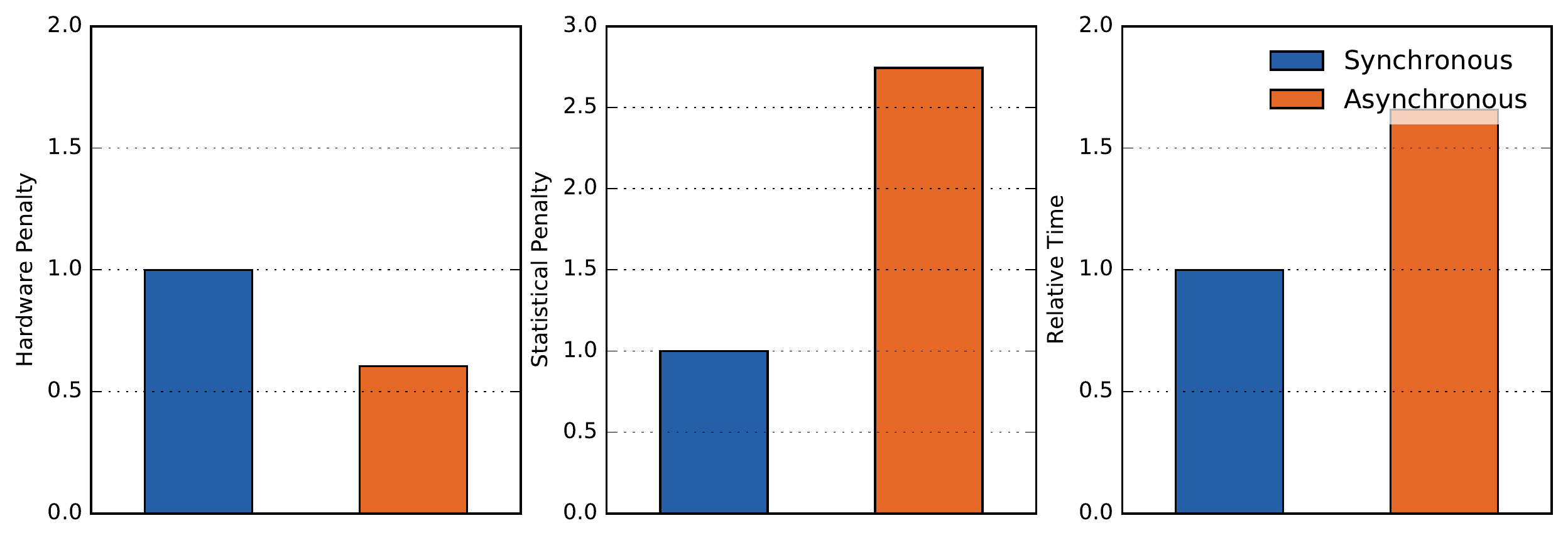} 
	\vspace{-1.0em}
	\caption{Hardware and statistical penalty without tuning (momentum $0.9$)}
	\label{fig:tf_09}
\end{figure}

\begin{figure}[t] 
\centering
\includegraphics[width=0.5\textwidth]{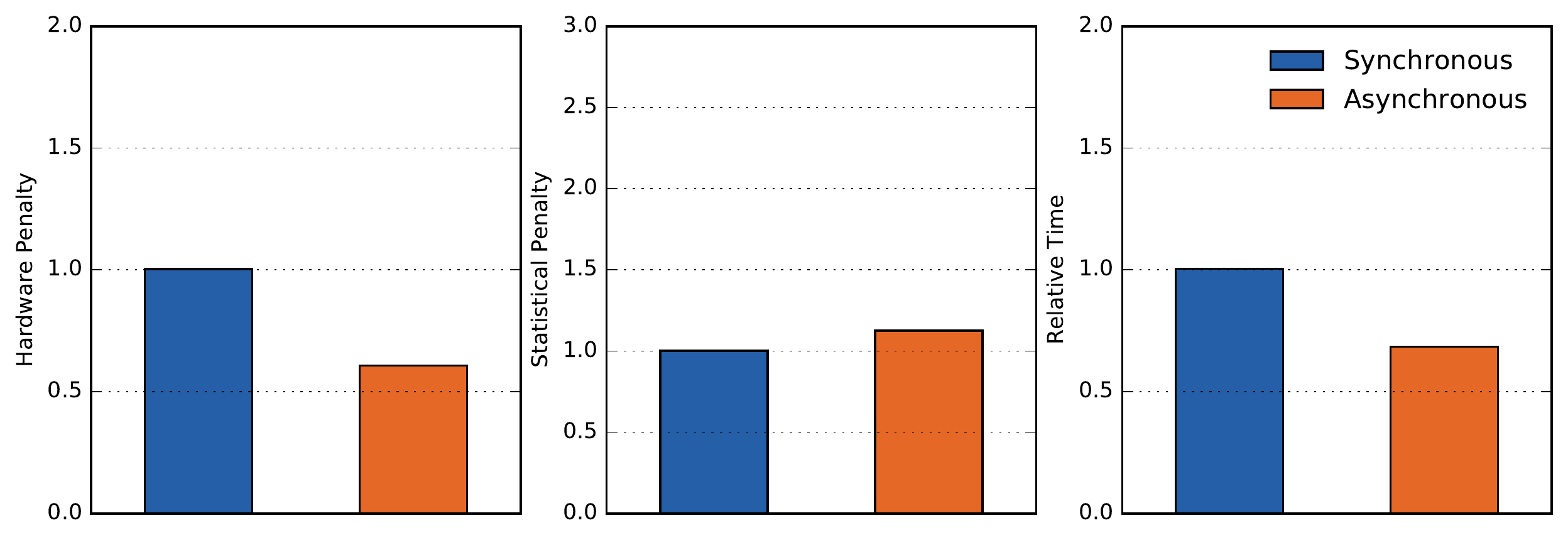} 
	\vspace{-1.0em}
\caption{Hardware and statistical penalty with tuning}
\label{fig:tf_bars}
\end{figure}


We measure the time and number of iterations it takes to reach a target loss and then perform hardware and software efficiency analysis.
Figure~\ref{fig:tf_09} shows the normalized statistical penalty, hardware penalty and wall clock time of training to the target loss for both the synchronous and asynchronous configurations, starting from the same snapshot. In this figure, the momentum parameter is set to the standard 0.9 value (the value used in \cite{chen2016revisiting} is not reported in the paper, but available code \cite{tfinception} suggests it was $0.9$).
The learning rate is tuned using the grid described above.
The statistical penalty is high for asynchronous training, which results in longer wall clock time to reach the same loss compared to synchronous training.
As expected, the hardware penalty is lower for asynchronous training. Figure~\ref{fig:tf_bars} shows the results of the same experiments, but using a grid search to tune momentum and learning rate.
The statistical penalty for asynchronous training relative to synchronous is now improved by a factor of about $2.4\times$.
Tuning in this case is result-changing.
The asynchronous configuration reaches the same loss in less time compared to synchronous training. 

\subsection{Compute Groups}
\begin{figure}[t] 
\centering
\includegraphics[width=0.5\textwidth]{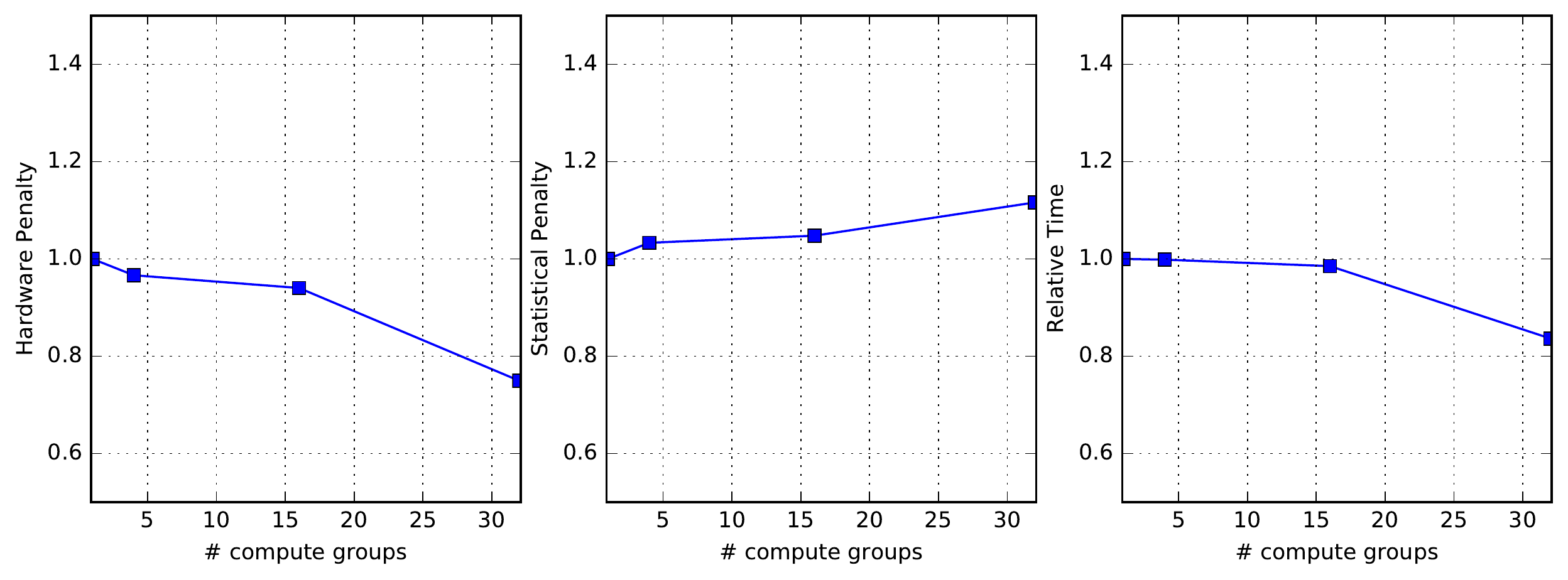} 
\vspace{-1.0em}
\caption{Hardware efficiency, statistical efficiency and wall clock time when tuning momentum and learning rate.}
\label{fig:tf_computegroups}
\end{figure}
We implemented compute groups in TensorFlow r0.9, used the same experimental setup as before and report results on Inception-v3. 
Our goal is to understand the tradeoffs on this different platform.
Figure ~\ref{fig:tf_computegroups} reports the performance curves for this setup.
As in some of our \oursystem experiments, the statistical efficiency remains nearly flat when tuning. This allows us to take advantage of the better hardware efficiency of asynchronous settings. 
We see that, in this case, $32$ nodes are not enough for the limits of asynchrony to start showing.
We expect that given a larger number of nodes, the optimal configuration will not be fully asynchronous, but rather some intermediate compute groups setting. This result, though under a different setup, contradicts the---reported but not demonstrated---claim that hybrid configurations do not perform better than fully synchronous training in TensorFlow. We attribute this to the fact that experiments in \cite{chen2016revisiting} do not involve any momentum tuning.

\section{Appendix Studying Total Cost of Ownership (TCO)}
\label{sec:app:tco}

Our study showed that CNN training is compute-bound regardless of the
compute device used. Given that we can now train CNNs on CPUs
proportional to the CPU FLOPS, this opens new questions in total cost
of ownership (TCO) for running CNN systems. We discuss those trends
and changes.

\textbf{Single Node:} We compare the price of running Omnivore on a
GPU instance (g2.2xlarge, \$0.65/hr, 1.2 TFLOPS) and a CPU instance
(c4.4xlarge, \$0.838/hr, 0.7 TFLOPS) for the same number of
iterations.  As Figure~\ref{fig:e2e} showed, Omnivore's speed on the
c4.4xlarge instance is $0.57\times$ the speed of the g2.2xlarge
instance. This ratio closely matches the FLOPS ratio
$0.7/1.2$. Therefore we observe that running on a CPU instance is
2.1$\times$ more expensive than a GPU instance, due to the difference
in the FLOPS/dollar ratio for these instances.  This suggests that on
cloud services such as Google Compute which do not have GPU instances,
CPU-based deep learning is a viable and cost-effective option when
using Omnivore. Moreover, organizations that can amortize the cost of
CPUs in more ways than GPUs may find them to be a cheaper alternative.

\textbf{Distributed:} In the distributed setting we consider again the
case of 9 machines and compare Omnivore running on the GPU cluster
(g2.8xlarge, \$2.60 per machine-hr, 4.8 TFLOPS per machine) and CPU
cluster (c4.4xlarge, \$0.838 per machine-hr, 0.7 TFLOPS per
machine). The difference in peak FLOPS between these clusters is
$6.8\times$, and the speedup to convergence obtained by Omnivore on
the GPU cluster compared to the CPU cluster is $5\times$--note it is
not quite $6.8\times$ because network speed does not scale with the
node throughput. If we consider only hardware efficiency (since
statistical efficiency is unrelated to the underlying hardware), the
GPU cluster is $5.6\times$ faster than the CPU cluster, which is
nearly the FLOPS ratio. As in the single-machine case therefore it is
only the FLOPS/dollar ratio which matters. The GPU cluster is more
cost-effective, now by a factor of $1.8\times$.

These results show that CPU deep learning is not significantly
different from GPU in terms of consumer cost, and it will be exciting
to see how these trends change for future CPUs which have increased
SIMD parallelism as well as newer GPUs which optimize for lower
power. As SIMD processor bandwidth has been doubling in each
generation, it seems that CPU training may indeed catch GPUs
relatively soon.

\subsection{Distributed Calculation}

The ratio of peak FLOPS of the GPU cluster / CPU cluster is $4.9/0.74 = 6.6$. Considering the optimal points chosen by our optimizer, the CPU / GPU time to convergence is $5\times$. If statistical efficiency is ignored, and we compare only the speeds of the async cases on each cluster, the ratio is now  34s/iter for the CPU cluster and  6 s/iter for the GPU, or $5.6\times$, which almost matches the ratio in device FLOPS. Given that GPU cluster is $\$2.6 / \$0.838 = 3.1\times$more expensive, the GPU cluster is $1.8\times$ cheaper per iteration which matches closely with the FLOPS/dollar ratio.

\section{Appendix for Conclusions (Section~\ref{sec:conclusion})}
\label{sec:app:conclusions}

Our study first demonstrated that on a single machine we could achieve CPU speeds proportional to the device FLOPS, showing end-to-end speedups of more than $5.5\times$ on EC2 CPU instances over state-of-the-art tools. With this improved CPU speed we showed that CNN computations are compute-bound. This allows the underlying hardware in a machine to be treated as a black-box, and we are $2.7\times$ faster than other systems on 4 GPUs and also $15\%$ faster on a single GPU by using the weak CPU alongside the EC2 instance's GPU. More generally, we show that each device or node in a cluster can be treated as a black-box that is characterized only by the throughput which it provides and is irrelevant to the type of hardware on that node (e.g., CPUs or GPUs).

Our second contribution was an empirical study of the factors affecting time to convergence for distributed deep learning training, and a novel, theoretical characterization of asynchrony which demonstrates that by tuning algorithmic (explicit) momentum in SGD there is no statistical penalty associated with asynchronous execution. We justified this empirically. We defined a tradeoff space and demonstrated that the execution strategy and server architecture were key in reducing the total time to convergence. We further showed that all existing distributed deep learning systems fall somewhere along this tradeoff space, but do not optimize within the space.

Finally, we studied each of these tradeoffs by decoupling their impact on hardware and statistical efficiency. This made it possible to study these factors in isolation and build an optimizer which optimizes within the tradeoff space. We showed both theoretically and empirically the need to jointly tune hyper-parameters with execution strategies in order to avoid slower convergence or divergence. We show that our optimizer provides a $>20\times$ reduction in time to convergence compared to other systems which select sub-optimal points in the space, and we also show that our optimizer is versatile by applying it to existing tools. In doing so, we close the gap between our system to $3\times$ faster than other systems, in some cases also preventing divergence in those other tools.


\end{appendices}
\fi

\end{document}